\documentclass[a4paper,11pt]{article}
\pdfoutput=1
\usepackage[utf8]{inputenc}
\usepackage{jheppub}
\usepackage{comment}
\usepackage{bm}
\usepackage[all]{xy}
\usepackage[dvipsnames]{xcolor}

\newcommand{\I}{\mathrm{i}}
\newcommand{\D}{\mathrm{d}}

\renewcommand{\O}{\mathcal{O}}
\newcommand{\<}{\langle}
\renewcommand{\>}{\rangle}
\newcommand{\bs}[1]{\boldsymbol{#1}}
\newcommand{\nn}{\nonumber}
\newcommand{\lla}{\langle \! \langle}
\newcommand{\rra}{\rangle \! \rangle}
\renewcommand{\a}{\alpha}

\newcommand{\gast}{\gamma_{5}}
\newcommand{\FVP}{\text{FVP}}
\newcommand{\WB}{\text{WB}}

\newcommand{\lwick}{:\!}
\newcommand{\rwick}{\!:}

\DeclareMathOperator{\Tr}{Tr}

\newcommand{\blue}[1]{{\color{blue} #1}}
\newcommand{\green}[1]{{\color{Green} #1}}

\newcommand{\ep}{\varepsilon}
\newcommand{\bep}{\bar{\varepsilon}}
\newcommand{\btheta}{\bar{\theta}}

\newcommand{\sA}{\bs{A}}
\newcommand{\schi}{\bs{\chi}}
\renewcommand{\sl}{\bs{\lambda}}
\newcommand{\sD}{\bs{D}}
\newcommand{\sC}{\bs{C}}

\newcommand{\smcM}{\bs{\mathcal{M}}}

\renewcommand{\j}{\varphi}
\newcommand{\bj}{\bar{j}}
\newcommand{\bJ}{\bar{J}}

\newcommand{\bl}{\bar{\lambda}}
\newcommand{\bchi}{\bar{\chi}}
\newcommand{\bpsi}{\bar{\psi}}
\newcommand{\bphi}{\bar{\phi}}
\renewcommand{\a}{\alpha}
\renewcommand{\b}{\beta}
\newcommand{\da}{\dot{\alpha}}
\newcommand{\db}{\dot{\beta}}
\newcommand{\bsigma}{\bar{\sigma}}
\newcommand{\bQ}{\bar{Q}}
\newcommand{\An}{\mathcal{A}}
\newcommand{\bAn}{\bar{\mathcal{A}}}

\begin{document}

\title{Consistency of supersymmetric 't Hooft anomalies}
\author[a]{Adam Bzowski,}
\author[b]{Guido Festuccia,}
\author[b]{and Vladim\'{i}r Proch\'{a}zka}
\vskip1.5cm
\affiliation[a]{Crete Center for Theoretical Physics, Department of Physics, University of Crete, 70013 Heraklion, Greece,}
\affiliation[b]{Department of Physics and Astronomy, Uppsala University, 751 08 Uppsala, Sweden.}
\emailAdd{a.bzowski@physics.uoc.gr}
\emailAdd{guido.festuccia@physics.uu.se}
\emailAdd{vladimir.prochazka@physics.uu.se}
\preprint{UUITP 48/20\\\hspace*{\fill}CCTP-2020-12\\\hspace*{\fill}ITCP-IPP 2020/12}

\abstract{We consider recent claims that supersymmetry is anomalous in the presence of a {\it R}-symmetry anomaly. We revisit arguments that such an anomaly in supersymmetry can be removed and write down an explicit counterterm that accomplishes it. Removal of the supersymmetry anomaly requires enlarging the corresponding current multiplet. As a consequence the Ward identities for other symmetries that are already anomalous acquire extra terms. This procedure can only be impeded when the choice of current multiplet is forced. We show how Wess-Zumino consistency conditions are modified when the anomaly is removed. Finally we check that the modified Wess-Zumino consistency conditions are satisfied, and supersymmetry unbroken, in an explicit one loop computation using Pauli-Villars regulators. To this end we comment on how to use Pauli-Villars to regulate correlators of components of (super)current multiplets in a manifestly supersymmetric way.}

\maketitle

\section{Introduction}

Quantum effects can make a classical symmetry anomalous. The interplay between anomalies and supersymmetry has long been studied~\cite{Nielsen:1984nk,Guadagnini:1985ea,Guadagnini:1985ar,Wang:1985dd,Harada:1985wa,Krivoshchekov:1985ep,Ferrara:1985me}. Here we are interested in ${\cal N}=1$ supersymmetric field theories in four dimensions.   The possibility that supersymmetry may be anomalous in these theories has been recently raised in~\cite{Papadimitriou:2017kzw,Papadimitriou:2019yug,Katsianis:2019hhg,Papadimitriou:2019gel,An:2019zok}. These papers considered theories with a classical $U(1)$~{\it R}-symmetry. The claim is that the presence of a 't Hooft anomaly in the {\it R}-current conservation equation, together with Wess-Zumino consistency conditions, implies an anomaly in the conservation equation for the supercurrent. This anomaly would then be present in very  simple theories. For instance it would arise in some Wess-Zumino model. In theories with higher amount of supersymmetry no such anomalies arise since the {\it R}-symmetry is non-anomalous there. The explicit preservation of supersymmetric Ward identities in these theories was shown long time ago~\cite{Howe:1984am}.
 
Supersymmetry implies that conserved currents are part of multiplets. For instance we will consider theories with a $U(1)$ ``flavor" global symmetry (not an {\it R}-symmetry). The corresponding conserved current is part of a linear multiplet which can be coupled to a background vector multiplet. We can gauge the $U(1)$ symmetry preserving supersymmetry by making the vector multiplet dynamical. When the $U(1)$ current is anomalous the symmetry cannot be gauged. 

Similarly the supercurrent is part of a ${\cal N}=1$ multiplet together with the energy momentum tensor and other operators~\cite{Dumitrescu:2011iu}. Generically this multiplet contains $16$ bosonic and $16$ fermionic degrees of freedom~\cite{Komargodski:2010rb}. In special cases the supercurrent multiplet can be improved to a shorter multiplet. For instance the theory could allow for a $(12+12)$ Ferrara-Zumino (FZ) supercurrent~\cite{Ferrara:1974pz}. Another special case arises when the theory possesses a $U(1)$~{\it R}-symmetry in which case there exists a (12+12) {\it R}-multiplet whose lowest component is the {\it R}-current~\cite{Gates:1981yc,Gates:1983nr}. Finally, if the theory is superconformal, the supercurrent multiplet can be improved to the $(8+8)$ superconformal current multiplet. The anomalies in the conservation of these currents also reside in appropriate multiplets. The different supercurrent multiplets we reviewed dictate which supergravity theory can be coupled to. Every ${\cal N}=1$ theory can be coupled to 16-16 supergravity (see for instance~\cite{Gates:1983nr}). Theories with a FZ supercurrent couple to old minimal supergravity (see, \textit{e.g.},~\cite{Wess:1992cp,Gates:1983nr}) while theories with a conserved {\it R}-symmetry can be coupled to new minimal supergravity~\cite{Sohnius:1981tp}. Finally, superconformal theories can be coupled to conformal supergravity (see~\cite{Freedman:2012zz}).
As in the flavor case we can consider coupling a ${\cal N}=1$ theory to background supergravity which can then be made dynamical when the corresponding multiplet is not anomalous. 

If the anomaly described in~\cite{Papadimitriou:2017kzw,Papadimitriou:2019yug,Katsianis:2019hhg,Papadimitriou:2019gel,An:2019zok} cannot be removed, it would imply that supersymmetry cannot be gauged. This claim appears too strong. Firstly, one can regulate an ${\cal N}=1$ theory preserving supersymmetry. For instance at one loop (which is sufficient to address the anomaly at hand) this can be accomplished by Pauli-Villars (PV)~\cite{Krivoshchekov:1978xg,West:1985jx,Gaillard:1994mn,Gaillard:1994sf,Katsianis:2020hzd}.\footnote{Dimensional regularization breaks supersymmetry explicitly. Nevertheless it was shown to preserve supersymmetric Ward identities for one-point insertions of composite operators provided suitable finite counterterms were added to the action~\cite{Howe:1984zt}.  This analysis was extended to multiple correlators of Ferrara-Zumino supercurrent components in~\cite{Eleftheriou:2020ray}.} Secondly from the discussion above it follows that the presence of an {\it R}-symmetry is not required to couple a theory to dynamical supergravity. Indeed the {\it R}-symmetry can be broken classically, in which case the theory is not superconformal nor possesses a {\it R}-multiplet, but can still be coupled to old minimal supergravity or 16-16 supergravity.

Consider using a supersymmetric regulator. In the presence of a $U(1)$~{\it R}-anomaly the regulator breaks the $U(1)$~{\it R}-symmetry explicitly and hence the regulated theory cannot be coupled to background conformal or new minimal supergravity. Nevertheless, the effective action would be invariant under supersymmetry transformations of the appropriate (old minimal or 16-16) background supergravity. In such a scheme correlators of the supercurrent would display no anomaly.

A different perspective can be gained by considering the case of a theory with a $U(1)$ ``flavor" global symmetry coupled to a background vector multiplet. Anomalies in the $U(1)$ symmetry and supersymmetry are then encoded in the effective action behavior under gauge and SUSY transformations of the vector multiplet. The linear multiplet containing a conserved $U(1)$ current has $(4+4)$ components which couple to the component fields of the vector multiplet in Wess-Zumino gauge. A SUSY variation of the vector multiplet brings out of Wess-Zumino gauge, which has to be restored by a (super)-gauge transformation. As a consequence an anomaly in the $U(1)$ symmetry implies an anomaly in supersymmetry.
However we can restore supersymmetry of the effective action by relaxing Wess-Zumino gauge and coupling to all the components of the vector multiplet~\cite{Guadagnini:1985ea,Wang:1985dd,Kuzenko:2019vvi}. 
The case of a theory coupled to background conformal supergravity is a similar. We can regard restricting the background to the field content of conformal SUGRA as fixing a gauge for the larger set of fields in old minimal supergravity (or 16-16 SUGRA). When the $U(1)$ {\it R}-symmetry is anomalous this leads to an anomaly in supersymmetry as above. However supersymmetry can be restored by reintroducing the extra components of the background supergravity~\cite{Gates:1981yc, Kuzenko:2019vvi}. 

The presence of the anomaly should be reflected by the generating functional as well as by the relevant correlation functions. This can be quantified by Wess-Zumino consistency conditions that impose the algebra of global symmetries at the generating functional level. The use of Wess-Zumino consistency conditions is the cornerstone of the original argument presented in \cite{Papadimitriou:2019gel} for the existence of an anomaly in {\it Q}-supersymmetry. The argument relies on the algebraic properties of the supercharges and their commutators with global symmetries. It implies that any admissible counterterm canceling the supersymmetry anomaly must break diffeomorphisms. This purely algebraic argument is then in tension with the existence of a superspace counterterm as proposed in \cite{Kuzenko:2019vvi}. Here we will argue that the Wess-Zumino conditions are modified in the presence of anomalous global symmetries. In particular we will demonstrate the existence of a non-anomalous supersymmetry current that is consistent with diffeomorphism-invariant counterterms. The price to pay are some extra terms in the supersymmetry Ward identities that were not considered before. We will illustrate this mechanism on an explicit free-field example in two renormalization schemes, that either break or preserve supersymmetry explicitly. 

While supersymmetry can be gauged even in the presence of a $U(1)$ {\it R}-anomaly, there are interesting physical consequences of the anomaly of~\cite{Papadimitriou:2017kzw,Papadimitriou:2019yug,Katsianis:2019hhg,Papadimitriou:2019gel,An:2019zok}. These arise when considering supersymmetric field theories on rigid supergravity backgrounds. Some interesting backgrounds are exclusive to new minimal supergravity. For instance one can define a supersymmetric index by placing a theory with a $U(1)$ {\it R}-symmetry on $S^3\times S^1$ preserving supersymmetry and computing its partition function. This is accomplished by coupling the theory to an off shell new minimal supergravity background. Supersymmetry then implies that the index has certain holomorphy properties~\cite{Closset:2013vra} that are found to be violated in explicit diffeomorphism-invariant schemes (see, \textit{e.g.},~\cite{Genolini:2016sxe,Closset:2019ucb} ). Furthermore in two dimensions there are supersymmetry anomalies that lie in the same multiplet as gravitational ones~\cite{Howe:1985uy}. These are physical in the sense that they cannot be removed by a local counterterm since the supercharge squares to a diffeomorphism whose anomaly cannot be removed by any local, two-dimensional counterterm. \\

The paper is structured as follows. In section 2 we consider a ${\cal N}=1$ theory with a chiral $U(1)$ flavor symmetry. We couple the theory to a background vector multiplet. We show that working in Wess-Zumino gauge a 't Hooft anomaly in the $U(1)$ symmetry results in an anomaly in supersymmetry. This supersymmetry anomaly can be removed by going away from Wess-Zumino gauge.

In section 3 we consider a classically superconformal theory coupled to background conformal supergravity. If the $U(1)$ {\it R}-symmetry is anomalous this implies an anomaly in supersymmetry. We show that this anomaly can be removed by introducing a chiral multiplet that plays the role of a compensator for the chiral $U(1)$ {\it R}-symmetry and conformal supersymmetry. Introducing this compensator is interpreted as coupling the theory to old minimal supergravity.

In section 4 we review some aspects of Pauli Villars (PV) regularization. In particular we summarize how the ABJ anomaly arises using PV regulators. We also discuss different schemes to define conserved currents in PV regularization and how this choice is reflected in the corresponding Ward identities. We consider in some detail the example of trace anomalies.

In section 5 we apply a PV regularization scheme to analyze Ward identities for the supercurrent in a simple free theory model. The anomaly discussed in~\cite{Papadimitriou:2017kzw,Papadimitriou:2019yug,Katsianis:2019hhg,Papadimitriou:2019gel,An:2019zok} arises in this model. We show that  the conservation of the supercurrent is not anomalous using PV.

Various appendices  contain a summary of conventions and formulae used throughout. 

~

{\it Note}: prior to submission we received~\cite{Katsianis:2020hzd} reaching similar conclusions to those reported here.  We thank the authors for sharing the draft of their work prior to publication and for illuminating conversations.

\subsection{Some conventions}

In this paper we will consider a dynamical $\mathcal{N} = 1$ supersymmetric theory in $d = 4$ spacetime dimensions coupled to non-dynamical backgrounds. By doing so we have two types of operators and their correlation functions to consider. If $\phi_k$ denotes a source, then we can define the associated operator $j_{k}$ as
\begin{align}
j_k = \left. \frac{\delta S}{\delta \phi_k} \right|_{\text{all } \phi_j = 0}.
\end{align}

In general a given source $\phi_k$ may not only couple linearly to $j_k$, but the action can contain terms with more than a single source. Such terms will be called seagull terms. Hence, with the generating functional of connected diagrams
\begin{align}
W = - \I \log \< e^{\I S} \>
\end{align}
we can define correlation functions
\begin{align}
\< J_{k_1}(\bs{x}_1) \ldots J_{k_n}(\bs{x}_n) \> = \frac{1}{\I^{n-1}} \frac{\delta}{\delta \phi_{k_1}(\bs{x}_1)} \ldots \frac{\delta}{\delta \phi_{k_n}(\bs{x}_{n})} W.
\end{align}
Due to the presence of seagull terms, the correlation functions of the operators $j_k$ will agree with the correlators defined as derivatives of the generating functional only up to lower-point functions, which we also call seagull terms.

In momentum space correlation functions contain the momentum-conserving delta function. For this reason we define the double bracket notation (omitting the momentum-conserving delta function),
\begin{align}
\< \O_1(\bs{p}_1) \ldots \O_n(\bs{p}_n) \> = (2 \pi)^4 \delta(\bs{p}_1 + \ldots + \bs{p}_n) \lla \O_1(\bs{p}_1) \ldots \O_n(\bs{p}_n) \rra
\end{align}
for any operators $\O_1, \ldots, \O_n$.

\section{Flavor symmetry}

The interplay between flavor and supersymmetry anomalies has been studied extensively. 
As early as in \cite{Nielsen:1984nk,Guadagnini:1985ea} it was argued that supersymmetry is non-anomalous in $\mathcal{N}=1$ supersymmetric gauge theories. It was, however, pointed out in \cite{Guadagnini:1985ea,Wang:1985dd} that supersymmetry and the Wess-Zumino gauge condition can be incompatible 
so that a SUSY-anomaly emerges once the Wess-Zumino gauge condition is imposed. This phenomenon underlies the results of \cite{Papadimitriou:2019yug,Papadimitriou:2019gel,Katsianis:2019hhg}.

While supersymmetry is non-anomalous, flavor symmetry exhibits an anomaly. Explicit results on the structure of the chiral anomaly in supersymmetric theories were derived in \cite{Guadagnini:1985ar}. While the flavor anomaly in non-Abelian theories exhibits a non-polynomial structure \cite{Ferrara:1985me} in the Abelian case simple expressions were derived in \cite{Harada:1985wa,Krivoshchekov:1985ep}. Since only the ABJ anomaly \cite{Adler:1969gk,Bell:1969ts} is physical, all other terms depend on the regularization method used.

\subsection{Formulation} \label{sec:flavor_form}

Let us consider a $U(1)$ flavor symmetry sourced by a vector multiplet, \eqref{vector_multi}. Among its components, fields $C, \chi, \mathcal{M}$ are compensators since they can be gauged away by a supergauge transformation. By choosing $\sigma, \Upsilon$ and $\mathfrak{f}$ in the supergauge transformation \eqref{gauge_tran_first} - \eqref{gauge_tran_last} one can fix Wess-Zumino gauge, where
\begin{align} \label{WZ_gauge}
C = \chi = \mathcal{M} = 0.
\end{align}
In Wess-Zumino gauge the supersymmetry transformations of the physical fields are
\begin{align}
\delta_{\ep,\bep}|_{WZ} A_\mu & = \I \ep \sigma_\mu \bl + \I \bep \bsigma_\mu \lambda, \label{susy_tran_WZ_first} \\
\delta_{\ep,\bep}|_{WZ} \lambda^{\alpha} & = - \ep^{\b} \sigma^{\mu \nu \ \a}_{\: \ \ \b} F_{\mu\nu} + \I D \ep^{\a}, \\
\delta_{\ep,\bep}|_{WZ} D & = - \ep \sigma^\mu \partial_\mu \bl + \bep \bsigma^\mu \partial_\mu \lambda. \label{susy_tran_WZ_last}
\end{align}
The residual gauge symmetry is the standard non-supersymmetric gauge symmetry parameterized by a real function $\theta$,
\begin{align}
& \delta_{\theta} A_\mu = \partial_\mu \theta, && \delta_{\theta} \lambda^{\a} = 0, && \delta_{\theta} D = 0. \label{dtheta:V}
\end{align}

Let us review the argument of \cite{Papadimitriou:2019yug} for the appearance of a SUSY anomaly. Let $\mathcal{A}_{\theta}$ denote the flavor anomaly and $\mathcal{A}_{\ep}, \bAn_{\bep}$ SUSY anomalies \textit{i.e.},
\begin{align}
\delta W = \int \D^4 \bs{x} \left[ \theta \mathcal{A}_{\theta} + \ep \An_{\ep} + \bep \bAn_{\bep} \right].
\end{align}
Here we will concentrate on $\bAn_{\bep}$ and keep $\ep$ and $\bep$ constant. The Wess-Zumino consistency condition, \cite{Wess:1971yu}, implies
\begin{align} \label{WZ_cond_a}
[\delta_{\bep}|_{WZ}, \delta_{\theta}] W = \int \D^4 \bs{x} \left[ \theta \delta_{\bep}|_{WZ} \mathcal{A}_{\theta} - \bep \delta_{\theta} \bAn_{\bep} \right].
\end{align}
While most terms in the flavor anomaly are scheme-dependent and removable by counterterms, the ABJ anomaly is physical,
\begin{align} \label{ABJ}
\mathcal{A}_{\theta}^{\text{ABJ}} = \frac{\kappa}{4} \epsilon^{\mu\nu\rho\tau} F_{\mu\nu} F_{\rho\tau}  = \kappa \epsilon^{\mu\nu\rho\tau} \partial_\mu A_\nu \partial_\rho A_\tau,
\end{align}
where the constant $\kappa$ depends on the theory.\footnote{In general one also has a mixed gravitational contribution proportional to the Pontrjagin density. The analysis of this term is almost identical to the corresponding term in the $R$-symmetry anomaly \eqref{eq:ABJSugra} so we will omit it in this section for brevity.} Its supersymmetric variation reads
\begin{align}
\delta_{\bep}|_{WZ} \mathcal{A}_{\theta}^{\text{ABJ}} = \I \kappa \epsilon^{\mu\nu\rho\tau} F_{\rho\tau} \bep \bsigma_\nu \partial_\mu \lambda.
\end{align}
Because the commutator on the left hand side of \eqref{WZ_cond_a} vanishes we have an anomaly in supersymmetry. In order for the Wess-Zumino consistency condition to be satisfied one needs~\footnote{The supersymmetry anomaly also contains terms cubic in $\lambda$ that we do not keep track of in~\eqref{wrong_An}. }
\begin{align} \label{wrong_An}
\bAn_{\bep}^{(?) \da} = - \I \kappa \epsilon^{\mu\nu\rho\tau} F_{\rho \tau} A_\mu \bsigma_\nu^{\da \a} \lambda_{\a}.
\end{align}
It can be argued that \eqref{wrong_An} is not the supersymmetric variation of any local counterterm. Indeed, the only term producing $\bAn_{\bep}^{(?) \da}$ under $\delta_{\bep}|_{WZ}$ would need to have the form $\epsilon^{\mu\nu\rho\tau} F_{\rho\tau} A_\mu A_\nu$ and hence vanishes.

Equivalently, anomalies appear in Ward identities. With the variations in \eqref{susy_tran_WZ_first} - \eqref{susy_tran_WZ_last} as well as \eqref{dtheta:V} the Ward identities read
\begin{align} \label{wrong_WardQ}
\partial_{\mu} \< \bJ^{\mu \da}_{\bQ} \> & = \I \bsigma_\mu^{\da \a} \lambda_{\a} \< J_A^\mu \> - \left[ \bsigma^{\kappa \lambda \da}_{\ \ \ \ \db} F_{\kappa\lambda} + \I D \delta_{\db}^{\da} \right] \< \bJ^{\db}_{\bl} \> \nn\\
& \qquad + \bsigma^{\mu \da \a} \partial_\mu \lambda_\a \, \< J_D \> - \bAn_{\bep}^{\da}, \\
\partial_{\mu} \< J^{\mu}_{A} \> & = - \An_{\theta}. \label{JA_ward}
\end{align}
If we want to see the tentative anomaly of \eqref{wrong_An} in a correlation function, we should take functional derivatives with respect to $A_\mu$ (twice) and $\lambda_\beta$ and analyze the behavior of the 4-point function $\< (\partial \cdot \bJ_{\bQ}^{\da}) J^\mu_A J^\nu_A J_{\lambda \beta} \>$.

\subsection{Wess-Zumino gauge}

Working in Wess-Zumino gauge, in the previous subsection, we have found an anomaly in supersymmetry \eqref{wrong_An}. On the other hand, in a number of papers \cite{Guadagnini:1985ar,Guadagnini:1985ea,Harada:1985wa,Krivoshchekov:1985ep} one can find superspace expressions where the flavor anomaly is explicitly supersymmetric. All such expressions can be regarded as supersymmetric completions of the ABJ anomaly \eqref{ABJ} and are necessarily scheme-dependent. By the anomaly being supersymmetric we mean that it is the gauge variation of a supersymmetric functional. This does not necessarily imply that $\delta_{\ep, \bep} \mathcal{A}_{\theta} = 0$, hence once again a nonzero supersymmetry anomaly would seem to arise from  \eqref{WZ_cond_a}. Nevertheless we will demonstrate that the Wess-Zumino consistency conditions can be made consistent with a vanishing SUSY anomaly, $\An_{\ep} = \bAn_{\bep} = 0$.

The apparent discrepancy follows from the use of Wess-Zumino gauge fixing as discussed in \cite{Guadagnini:1985ea,Wang:1985dd}. In the discussion of the previous section we used Wess-Zumino gauge throughout the calculations. However, the supersymmetry algebra in Wess-Zumino gauge closes only up to a compensating supergauge transformation.  Starting with $C = \chi = \mathcal{M} = 0$ the SUSY transformations \eqref{susy_tran_first} - \eqref{susy_tran_last} of the full multiplet break the gauge condition,
\begin{align} \label{WZ_fail}
& \delta_{\bep} C = 0, && \delta_{\bep} \chi^\alpha = - \I \bep_{\da} \bsigma^{\mu \da \a} A_\mu, && \delta_{\bep} \mathcal{M} = 2 \bep \bl.
\end{align}
In order to keep the Wess-Zumino gauge condition, an appropriate compensating transformation, $\delta_{\Lambda_{\text{comp}}}$, must be added to the SUSY transformation, $\delta_{\ep, \bep}|_{WZ}$. At first order around the Wess-Zumino gauge condition \eqref{WZ_gauge} we have
\begin{align} \label{WZ_compensate}
\delta_{\bep} = \delta_{\bep}|_{WZ} + \delta_{\Lambda_{\text{comp}}} + \ldots,
\end{align}
where $\delta_{\Lambda_{\text{comp}}}$ is the supergauge transformation \eqref{gauge_tran_first} - \eqref{gauge_tran_last} with components,
\begin{align}
\sigma_{\text{comp}} = \theta_{\text{comp}} & = 0, \\
\Upsilon_{\text{comp}}^{\alpha} & = \frac{1}{\sqrt{2}} \bep_{\da} \bsigma^{\mu \da \a} A_{\mu}, \label{WZ_compensate_val} \\
\mathfrak{f}_{\text{comp}} & = \I \bep \bl.
\end{align}
The original SUSY transformation $\delta_{\bep}|_{WZ}$ in \eqref{WZ_compensate} acts on the physical fields $A_\mu, \lambda, D$, while $\delta_{\Lambda_{\text{comp}}}$ acts only on $\chi$ and $\mathcal{M}$. The omitted terms are ``higher order terms" that make the full $\delta_{\bep}$ transformation close. These terms are higher order in the sense that they produce more compensating terms. In total, $\delta_{\bep}$ is the full supersymmetry transformation acting on the full vector multiplet,
\begin{eqnarray}
\delta_{\ep,\bep} A_\mu & = & \I \ep (\sigma_\mu \bl \green{- \I \partial_\mu \chi}) + \I \bep (\bsigma_\mu \lambda \green{- \I \partial_\mu \bchi}), \label{text_susy_tran_first} \\
\delta_{\ep,\bep} \lambda^{\alpha} & = & - \ep^{\b} \sigma^{\mu \nu \ \a}_{\: \ \ \b} F_{\mu\nu} + \I D \ep^{\a}, \\
\delta_{\ep,\bep} D & = & - \ep \sigma^\mu \partial_\mu \bl + \bep \bsigma^\mu \partial_\mu \lambda, \\
\blue{\delta_{\ep,\bep} \chi^\alpha} & = & \blue{ - \I \bep_{\da} \bsigma^{\mu \da \a} (A_\mu} \green{ - \I \partial_\mu C} \blue{)} \green{+ \mathcal{M} \ep^\alpha}, \\
\blue{\delta_{\ep,\bep} \mathcal{M}} & = & \green{2 \I \bep \bsigma^{\mu} \partial_\mu \chi} \blue{ + 2 \bep \bl}, \\
\green{\delta_{\ep,\bep} C} & = & \green{\I (\ep \chi - \bep \bchi)}. \label{text_susy_tran_last}
\end{eqnarray}
The black terms are the SUSY transformations in Wess-Zumino gauge \eqref{susy_tran_WZ_first} - \eqref{susy_tran_WZ_last}. The blue terms are the first order terms added through the compensating gauge transformation $\delta_{\Lambda_{\text{comp}}}$ in \eqref{WZ_compensate}. The green terms are the higher order terms, which produce more compensating fields.

As long as one is interested only in operators sourced by $A_\mu, \lambda, D$ -- for instance to check the Ward identities -- one can drop the higher order (green) terms. In other words, it is enough to think about $\delta_{\Lambda_{\text{comp}}}$ in \eqref{WZ_compensate} as the compensating transformation. This point of view will be useful for the analysis of the case involving coupling to supergravity. Nevertheless, it is essential to keep the first order (blue) terms. These are the physical terms that result from the variation of compensating fields. In the spirit of ``differentiation before evaluation", one should calculate SUSY variations before the Wess-Zumino gauge condition \eqref{WZ_gauge} is fixed.

\subsection{Consequences} \label{sec:consequences}

\paragraph{Existence of a counterterm.} 

The claim of \cite{Papadimitriou:2019yug} is that the supersymmetric anomaly \eqref{wrong_An} is physical and hence it cannot be removed by counterterms. However, since the compensating supergauge transformation has non-vanishing $\Upsilon$ and $\mathfrak{f}$ components, we have to include the corresponding components of the vector multiplet. Using the compensators $\chi$ and $\mathcal{M}$ we can cancel the SUSY anomaly \eqref{wrong_An}. Indeed, the local counterterm
\begin{align} \label{ct}
S_{\text{ct}} & = - \kappa \int \D^4 \bs{x} \left[ A_\mu \, \lambda \sigma^\mu \bl + \I F_{\mu\nu} \left( \chi \sigma^{\mu\nu} \lambda - \bchi \bsigma^{\mu\nu} \bl \right) \right.\nn\\
& \qquad\qquad\qquad \left. + D \left( \chi \lambda + \bchi \bl \right) - \I \left( \mathcal{M} \, \lambda \lambda - \mathcal{M}^{\ast} \, \bl \bl \right) \right]
\end{align}
removes the anomaly, \textit{i.e.},
\begin{align} \label{eq:DbepF}
\delta_{\ep, \bep} S_{\text{ct}} = - \int \D^4 \bs{x} \left[ \bep \bAn^{(?)}_{\bep} + \ep \An^{(?)}_{\bep} \right] + O(C, \chi, \bchi, \mathcal{M}).
\end{align}
We should remark that in general \eqref{ct} is not unique: the addition of an arbitrary manifestly supersymmetric term does not alter the SUSY anomaly.

\paragraph{Wess-Zumino consistency condition.}

While in Wess-Zumino gauge the commutator between SUSY and gauge transformations vanishes, $[\delta_{\bep}|_{WZ}, \delta_{\theta}] = 0$, this is no longer true when $\delta_{\bep}$ is used. Indeed, the compensating transformation $\delta_{\Lambda_{\text{comp}}}$ in \eqref{WZ_compensate} now acts on $\chi$, which gives $\delta_{\theta} \delta_{\Lambda_{\text{comp}}} \chi = \I \sigma^\mu \bep \partial_\mu \theta$. On the other hand $\delta_{\theta} \chi = 0$, which means that we have a non-vanishing commutator,
\begin{align} \label{chi_comm}
[ \delta_{\bep}, \delta_{\theta} ] \chi^{\a} = \I \bep_{\da} \bsigma^{\mu \da \a} \partial_\mu \theta.
\end{align}
This is the only non-vanishing commutator $[ \delta_{\bep}, \delta_{\theta} ]$ when acting on the component fields of the vector multiplet\footnote{While the SUSY variation of the component field $\mathcal{M}$ involves the physical field $\bar{\lambda}$, the commutator $[\delta_{\bep}, \delta_{\theta}] \mathcal{M} = 0$, due to the gauge-invariance of $\bar{\lambda}$.}.

Alternatively, we could simply start with the full set of SUSY transformations \eqref{text_susy_tran_first}-\eqref{text_susy_tran_last} and calculate its commutator with the full supergauge transformations \eqref{gauge_tran_first}-\eqref{gauge_tran_last}. The commutator vanishes, $[\delta_{\ep, \bep}, \delta_{\Lambda} ] = 0$. Nevertheless, this does not imply that the commutator vanishes in Wess-Zumino gauge because the parameters of the supergauge transformations transform under supersymmetry as well. Therefore, if we pick the actual gauge transformation parametrized by $\theta$ while setting $\Upsilon = \mathfrak{f} = 0$ we have $\delta_{\theta} \chi = 0$ but $\delta_{\theta} \delta_{\ep, \bep} \chi = \I \sigma^\mu \bep \partial_\mu \theta$. This results in \eqref{chi_comm}.

The non-vanishing of the commutator \eqref{chi_comm} alters the Wess-Zumino consistency condition. Let $\{ \phi_k \}$ denote a set of sources for the operators $J_k$. By the chain rule and (anti)commutativity of second derivatives
\begin{align} \label{derive_WZ}
[\delta_{\ep, \bep}, \delta_{\theta}] W[\phi_k] = \sum_k \int \D^4 \bs{x} \, \left( [\delta_{\ep, \bep}, \delta_{\theta}] \phi_k \right)  \< J_k \>.
\end{align}
As a consequence the commutator on the left hand side of the Wess-Zumino consistency condition~\eqref{WZ_cond_a} no longer vanishes,
\begin{align} \label{fullWZ}
\theta \delta_{\bep} \mathcal{A}_{\theta} - \bep \delta_{\theta} \bAn_{\bep} = \I \partial_\mu \theta \, \bep_{\da} \bsigma^{\mu \da \a} \< J_{\chi \a} \>.
\end{align}
Hence, the presence of the flavor anomaly, $\An_{\theta} \neq 0$, does not imply a non-vanishing supersymmetric anomaly, $\bAn_{\bep}$.

We claim that there exists a scheme, where the supersymmetric anomaly is absent, $\bAn_{\bep} = 0$. This is supported by the fact that there exists a supersymmetry preserving regularization scheme, such as Pauli-Villars (PV) regularization. Hence, for constant $\ep, \bep$, the (anti-holomorphic part of) Wess-Zumino consistency condition becomes
\begin{align} \label{true_WZ_cond}
\delta_{\bep} \mathcal{A}_{\theta} = - \I \bep_{\da} \bsigma^{\mu \da \a} \partial_\mu \< J_{\chi \a} \>, && \bAn_{\bep} = 0.
\end{align}
In section \ref{sec:WZ_cond} we will verify this new Wess-Zumino consistency condition in a free theory model.

\paragraph{Ward identity.} Finally, we can see how an extra term in the Ward identity \eqref{wrong_WardQ} appears. The supersymmetric variations of the vector multiplet are given in \eqref{text_susy_tran_first} - \eqref{text_susy_tran_last}. We have to include variations of compensating fields even if we consider correlation functions of operators sourced only by $A_\mu, \lambda, \bl$, and $D$. Two relevant terms are
\begin{align}
& \delta_{\bep} \chi^\alpha = - \I \bep_{\da} \bsigma^{\mu \da \a} A_\mu + O(C), && \delta_{\bep} \mathcal{M} = 2 \bep \bl + O(\chi).
\end{align}
These terms mix the compensating fields $\chi, \mathcal{M}$ with the physical fields $A_\mu$ and $\bl$. This is again the consequence of Wess-Zumino gauge breaking supersymmetry. With the extra variations the Ward identity reads
\begin{align} \label{ward_1pt}
\partial_{\mu} \< \bJ^{\mu \da}_{\bQ} \> & = \I \bsigma_\mu^{\da \a} \lambda_{\a} \< J_A^\mu \> - \left[ \bsigma^{\kappa \lambda \da}_{\ \ \ \ \db} F_{\kappa\lambda} + \I D \delta_{\db}^{\da} \right] \< \bJ^{\db}_{\bl} \> + \bsigma^{\mu \da \a} \partial_\mu \lambda_\a \, \< J_D \> \nn\\
& \qquad  - \I \bsigma^{\mu \da \a} A_\mu \< J_{\chi \a} \> + 2 \bl^{\da} \< J_{\mathcal{M}} \> + O(C, \mathcal{M}, \chi).
\end{align}
The terms $O(C, \mathcal{M}, \chi)$ can be dropped if one considers insertions of operators sourced only by $A_\mu, \lambda, \bl$ and $D$.

\subsection{\texorpdfstring{What is $J_{\chi}$?}{What is Jx?}} \label{sec:WhatChi}

In the previous section we argued that the SUSY anomaly in \eqref{wrong_WardQ} is eliminated by the addition of the $\< J_{\chi \a} \>$ and $\< J_{\mathcal{M}} \>$ terms in \eqref{ward_1pt}. This, however, implies that these operators must be ultralocal, \textit{i.e.}, as local as the anomaly itself. Indeed, the supergauge symmetry \eqref{gauge_tran_first} - \eqref{gauge_tran_last} is the shift symmetry for their sources $\chi$ and $\mathcal{M}$, which implies that
\begin{align} \label{Jchi0}
\< J_{\chi} \> = \< J_{\mathcal{M}} \> = 0 \text{ (up to anomalies)}.
\end{align}
This is at least consistent, but it is possible that we did not remove any anomaly, but rather disguised it by using compensators. Recall that any anomaly can be artificially removed by introducing a compensator.

For example, consider a theory coupled to the background metric, $g_{\mu\nu}$. Classically, Weyl invariance, $\delta_{\sigma} g_{\mu\nu} = - 2 \sigma g_{\mu\nu}$, implies the vanishing of the trace of stress tensor. In a quantum theory, however, conformal anomalies may be present. The generating functional, $W$, is not invariant under the Weyl transformation, $\delta_{\sigma} W[g_{\mu\nu}] \neq 0$, leading to the anomalous trace Ward identity, $g_{\mu\nu} \< T^{\mu\nu} \> = \mathcal{A}_{\sigma}$. The anomaly can be hidden by introducing a compensator. By coupling the dilaton, $\tau$, to the trace of stress tensor $T = g_{\mu\nu}T^{\mu\nu}$ and assigning the shift transformation $\delta_{\sigma} \tau = \sigma$, the generating function is made invariant, $\delta_{\sigma} W[g_{\mu\nu}, \tau] = 0$, and the Ward identity becomes $g_{\mu\nu} \< T^{\mu\nu} \> = \< T \>$, where the 1-point function on the right hand side is obtained by varying $W$ with respect to $\tau$. The theory, however, remains anomalous, because $\< T \> \neq 0$ in the quantum theory, while $T = 0$ classically. Another way to say this is that only $\mathcal{A}_{\sigma}$ is the genuine anomaly satisfying the Wess-Zumino consistency conditions.

In the case of supersymmetry, the source for the supercurrents, $j^{\mu}_{Q \a}, \bj^{\mu \da}_{\bQ}$, is the gravitino, $\psi_{\mu}^{\a}, \bpsi_{\mu \da}$. The compensators for supersymmetry, $\ep, \bep$, can be introduced by the substitution $\psi_{\mu}^{\a} \mapsto \psi_{\mu}^{\a} - \partial_\mu \ep^{\a}$ and its conjugate. Now $\ep^{\a}$ couples to the divergence of the supercurrent and the Ward identity becomes $\partial_\mu \< j^{\mu}_{Q \a} \> = \< \partial_\mu j^{\mu}_{Q \a} \>$.

Is any of the fields $C, \chi, \mathcal{M}$ the compensator for the supercurrent? No: $C$ and $\mathcal{M}$ are scalars, so only $\chi$ is suspect. However, the conformal dimension of the supercurrent equals ${7\over 2}$, its divergence has dimension ${9\over 2}$ and therefore the compensator for supersymmerty has conformal dimension equal to $-{1\over 2}$. On the other hand the conformal dimension of $\chi$ equals ${1\over 2}$ as it couples to the operator $J_{\chi}$ of dimension ${7\over 2}$. Hence, none of the compensating fields is the compensator for supersymmetry. While $\chi$ is not a compensator for supersymmetry, it turns out to be the compensator for $S$-supersymmetry in a specific example considered in this paper (see section \ref{sec:WZsupercurrents}). In a generic case one should think about it as a fermionic (shift-)symmetry.

\subsection{Generating functional and scheme-dependence}

While the Wess-Zumino consistency condition \eqref{fullWZ} is fixed purely by the superalgebra, the form of the counterterm \eqref{ct} is not. For example, any local term $S_{\text{fin}}$ such that $[\delta_\theta, \delta_{\ep,\bep}] S_{\text{fin}} = 0$ can be added to the counterterm \eqref{ct}. Such a term will contribute to the anomalies in \eqref{true_WZ_cond},
\begin{align} \label{true_WZ_condNS}
	& \delta_{\bep} \mathcal{A}_{\theta} = - \I \bep_{\da} \bsigma^{\mu \da \a} \partial_\mu \< J_{\chi \a} \> + \delta_{\bep} \frac{\delta S_{\text{fin}}}{\delta \theta}, && \delta_{\theta} \mathcal{A}_{\bep} = \frac{\delta}{\delta \theta} \delta_{\bep} S_{\text{fin}},
\end{align}
while keeping the full Wess-Zumino consistency condition \eqref{fullWZ} intact. It is only in a supersymmetric scheme, where $\An_{\ep} = \bAn_{\bep} = 0$, that \eqref{true_WZ_cond} is satisfied. As was already remarked in \cite{Kuzenko:2019vvi}, in such a scheme the flavor anomaly $\An_{\theta}$ generally includes extra terms on top of the ``bare" ABJ anomaly, \eqref{ABJ}. In particular, in section \ref{sec:chiral_an} we will calculate relevant parts of the flavor anomaly in the Pauli-Villars renormalization scheme in the free Wess-Zumino theory. As the scheme is manifestly supersymmetric, $\bAn_{\bep}^{PV} = 0$, but the part of the flavor anomaly quadratic in sources takes the form
\begin{align} \label{eq:PVchiralAnomaly}
	\An_{\theta}^{PV} & = - \frac{1}{192 \pi^2} \partial_\mu \left[ \epsilon^{\mu\nu\rho\tau} A_\nu \partial_\rho A_\tau + 3 \lambda \sigma^\mu \bl + 2 \I \lambda \sigma^\mu \bsigma^\nu \partial_\nu \chi - 2 \I \partial^\mu ( \lambda \chi ) \right.\nn\\
	& \qquad\qquad\qquad\qquad \left. + O(C, \mathcal{M}, \chi^2, \bl) \right].
\end{align}

Furthermore, even in a supersymmetric scheme, the counterterm is not uniquely fixed. Indeed, one can add another, explicitly supersymmetric, but not gauge-invariant, counterterm $S_{\text{loc}}$, which modifies both the left and right hand side of \eqref{true_WZ_cond} accordingly. An example of such a manifestly supersymmetric counterterm can be written in terms of superfields as $\int \D^8 z V \bar{D}^2 V D^2 V$, where $V$ is the vector multiplet. The resulting counterterm is non-invariant under the flavor symmetry. Such local terms are expected to appear since PV regularization breaks the chiral invariance explicitly, by virtue of the mass term (we will discuss this in more detail in section \ref{sec:anomaly_PV}).

On the other hand, the Ward identity \eqref{Jchi0} shows that the operators $J_{\chi}$ and $J_{\mathcal{M}}$ are ultralocal in the sense that all their correlation functions are ultralocal. This means that the generating functional $W = W_{\text{loc}}[\chi, \mathcal{M}]$ as the function of the sources $\chi$ and $\mathcal{M}$ is a local expression. We conclude that, in any renormalization scheme, both sides of \eqref{true_WZ_condNS} are determined by a local generating functional $W_{\text{loc}}$ and the ABJ anomaly. In particular, in a supersymmetric scheme, such as the PV scheme, one has
\begin{align} \label{eq:CountAnomRel}
\delta_{\bep} \left(\mathcal{A}_{\theta}^{\text{ABJ}} + \frac{\delta W_{\text{loc}}}{\delta \theta} \right) = - \I \bep_{\da} \bsigma^{\mu \da \a} \partial_{\mu} \frac{\delta W_{\text{loc}}}{\delta \chi^{\a}}.
\end{align}

\section{Supergravity} \label{sec:sugra}

In \cite{Papadimitriou:2019gel}, see also \cite{An:2019zok,Katsianis:2019hhg}, it was argued that supergravity cannot be consistently coupled to a $\mathcal{N} = 1$ superconformal field theory due to an anomaly in the conservation of the supercurrent. The argument follows the same route as the one introduced in section \ref{sec:flavor_form}, with the  $R$-current in place of the flavor current. We will review it in the following section.

Anomalies in supersymmetry were discussed in \cite{Gates:1981yc}, where it was shown how the theory can be made consistent and anomaly-free by introducing a suitable compensator field. Among a variety of different choices, only a single choice, leading to old minimal supergravity, yields the theory consistent. This approach is very similar to the one we adopted for the flavor anomalies in that we will introduce a compensator fermion field which has similar properties to the $\chi$ field described in the previous section.

\subsection{Formulation}

When a $\mathcal{N} = 1$ superconformal theory is coupled to supergravity, the gauge freedom allows one to reduce the degrees of freedom to the vielbein, $e_\mu^a$, sourcing the stress tensor, the gravitino, $\psi_\mu$, sourcing the supercurrent, and the gauge field, $A^R_\mu$, sourcing the $R$-current.

In order to keep the notation consistent with the remainder of this paper, we will follow the conventions of \cite{Wess:1992cp} and use 2-component, Weyl fermions. The original calculations, however, were carried out using conventions of \cite{Freedman:2012zz} with 4-component, Majorana fermions. We keep those results in appendix \ref{app:4comp}.

Among the superconformal transformations we will be interested in $Q$-supersymmetry transformations, parameterized by fermions $\ep, \bep$, $S$-supersymmetry transformations, parameterized by fermions $\eta, \bar{\eta}$ and $R$-transformations parameterized by a real scalar $\theta^R$. For completeness we also include Weyl transformations parameterized by $\sigma$ and local Lorentz parameterized by ${\lambda^a}_b$. The transformations of the sources are
\begin{eqnarray} \label{eq:vielbeinTransf}
\delta e_\mu^a &=& \I ( \bep \bsigma^{a} \psi_{\mu} + \ep \sigma^a \bpsi_\mu ) - \sigma e_\mu^a - {\lambda^a}_{b} e_\mu^b, \\ \label{eq:ARmuTransf}
\delta A_\mu^R &=& \partial_\mu \theta^R - \I ( \ep \phi_\mu - \bep \bphi_\mu) + \I (\eta \psi_\mu - \bar{\eta} \bpsi_\mu), \\ \label{eq:PsimuTransf}
\delta \psi_{\mu\a} &=& \frac{3 \I}{2} \theta^R \psi_{\mu\a} + 2 \mathcal{D}_\mu \ep_{\a} - 2 \I \sigma_{\mu\a\da} \bar{\eta}^{\da} - \frac{1}{2} \sigma \psi_{\mu\a} + \frac{1}{2} \lambda^{ab} \sigma_{ab\a}^{\ \ \ \ \b} \psi_{\mu\b},
\end{eqnarray}
where
\begin{align} \label{eq:PhiSugra}
\phi_{\mu \alpha} & = \frac{2 \I}{3} \sigma^\nu_{\a \da} \left( \mathcal{D}_{[\mu} \bpsi_{\nu]} + \frac{\I}{4} {\epsilon_{\mu\nu}}^{\rho\sigma} \mathcal{D}_{\rho} \bpsi_{\sigma}^{\da} \right)
\end{align}
and
\begin{align}
\mathcal{D}_\mu \ep = \left( D^{\omega}_\mu - \frac{3 \I}{2} A_\mu^R \right) \ep, && \mathcal{D}_\mu \psi_\nu = \left(D_\mu^\omega - \frac{3 \I}{2} A_\mu^R \right) \psi_\nu \; .
\end{align}
By $D_\mu^\omega$ we denote the covariant derivative with connection $\omega$ given by
\begin{align}
\omega_{\mu a b}(e, \psi) & = \omega_{\mu a b}(e) + \frac{\I}{4} \left( \bpsi_a \bsigma_\mu \psi_b + \psi_a \sigma_\mu \bpsi_b + \bpsi_\mu \bsigma_a \psi_b + \psi_\mu \sigma_a \bpsi_b - \bpsi_\mu \bsigma_b \psi_a - \psi_\mu \sigma_b \bpsi_a \right)
\end{align}
In particular,
\begin{align}
\delta \phi_{\mu \a} & = \left( \I P_{\mu\nu} - F_{\mu\nu}^R - \tfrac{\I}{4} \epsilon_{\mu\nu}^{\ \ \: \rho\tau} F_{\rho \tau}^R \right) \sigma^\nu_{\a \da} \bep^{\da} + 2 \mathcal{D}_\mu \eta_{\a} - \frac{3 \I}{2} \theta^R \phi_{\mu \a} \nn\\
& \qquad\qquad + \frac{1}{2} \sigma \phi_{\mu \a} + \frac{1}{2} \lambda^{ab} \sigma_{ab} \phi_{\mu \a},
\end{align}
where 
\begin{align} \label{Pmunu}
\mathcal{D}_\mu \eta = \left( D^{\omega}_\mu + \frac{3 \I}{2} A_\mu^R \right) \eta, && P_{\mu\nu} = \frac{1}{2} \left( R_{\mu\nu} - \frac{1}{6} g_{\mu\nu} R \right).
\end{align}
Note that the transformations of conformal SUGRA have the algebraic property that
\begin{equation}
[\delta_{\ep}, \delta_{\bep}] \ni \delta_{\theta^R} \; ,
\end{equation}
similarly to the SUSY transformations of the vector multiplet in Wess-Zumino gauge.

The conventions of \cite{Papadimitriou:2019gel} (denoted by superscript $P$) are recovered by rescaling the fermionic parameters, $\ep = \ep^P/2$ and $\eta = \eta^P/2$, as well as a rescaling of the gauge field and the gauge parameter, $A^R_\mu = -2 A_\mu^{P}/3$ and $\theta^R = -2 \theta^P/3$. Also, note the change of the sign in the definition of $\phi_{\mu}$, ($\phi_{\mu} = -\phi_{\mu}^P$) as well as in the definition of the $R$- and $Q$-SUSY-anomalies, $\An_{\theta^R}$ and $\An_{\ep}, \bAn_{\bep}$ respectively,
\begin{align}
\delta W = \int \D^4 \bs{x} \sqrt{-g} \left[ \theta^R \mathcal{A}_{\theta^R} + \sigma \An_{\sigma} + \ep \An_{\ep} + \bep \bAn_{\bep} + \eta \An_{\eta} + \bar{\eta} \bAn_{\bar{\eta}} \right],
\end{align}
where $\An_{\sigma}$ is the Weyl anomaly, while $\An_{\eta}, \bAn_{\bar{\eta}}$ are $S$-SUSY anomalies.

In our conventions the ABJ form of the chiral anomaly for the $R$-current reads
\begin{align} \label{eq:ABJSugra}
\mathcal{A}_{\theta^R}^{\text{ABJ}} = \frac{5a - 3c}{16 \pi^2} \epsilon^{\mu\nu\rho\tau} F_{\mu\nu}^R F_{\rho\tau}^R + \frac{c - a}{32 \pi^2}  \epsilon^{\mu\nu\rho\tau} R_{\mu\nu\kappa\lambda} R_{\rho\tau}^{\ \ \: \kappa\lambda}.
\end{align}
The first term is the ABJ anomaly with the coefficient determined by the $a$ and $c$ anomalies and the second term contains a mixed gravitational anomaly proportional to the Pontrjagin density. The Wess-Zumino consistency condition has the form identical  to \eqref{WZ_cond_a}.  As before, the commutator on the left hand side of the equation vanishes. Hence, together with a non-zero variation $\delta_{\ep} \mathcal{A}_{\theta^R}$, the Wess-Zumino condition implies that $\delta_{\theta^R} \mathcal{A}_{\ep}$ does not vanish, leading to an anomaly in supersymmetry. In \cite{Katsianis:2019hhg,Papadimitriou:2019gel} the anomaly is evaluated to be
\begin{align} \label{eq:SUGRAanomaly}
\mathcal{A}_{\ep \a}^{(?)} & =  \frac{3c - 5a}{4 \pi^2} \I \epsilon^{\mu\nu\rho\tau} F_{\rho\tau}^R A^R_{\mu} \phi_{\nu \a} + \frac{a-c}{4 \pi^2} \I \epsilon^{\lambda\kappa\rho\tau} \nabla_\mu ( A_\rho^R R_{\lambda \kappa}^{\ \ \: \mu \nu} ) \sigma_{(\nu \a \da} \bpsi_{\tau)}^{\da} \nn\\
& \qquad\qquad + \frac{c-a}{16 \pi^2} \I \epsilon^{\mu\nu \kappa\lambda} F_{\mu\nu}^R R_{\kappa\lambda}^{\ \ \: \rho\tau} \sigma_{\rho \a \da} \bpsi_\tau^{\da} + O(\psi^3).
\end{align}
Notice the similarity between the first term and \eqref{wrong_An} with $\phi_\nu$ in place of $\sigma_\nu \lambda$.

\subsection{Compensators}

In \cite{Gates:1981yc} it was argued that in the presence of anomalous superconformal symmetry one has to introduce  suitable compensator fields. If the anomalies break all of the main global symmetries ($R$, Weyl and $S$-SUSY) we need a compensator that includes a chiral multiplet.\footnote{At a bare minimum we need a complex scalar to compensate $R$ and Weyl anomalies as well as a Majorana spinor to compensate the $S$-SUSY anomaly. Then we have to add other degrees of freedom to close the supersymmetry algebra off-shell which corresponds to different choices of compensators.} Hence we introduce a chiral multiplet $(Z, \chi^R, \mathcal{F})$ of Weyl weight $w=1$ transforming as
\begin{eqnarray}
\delta Z &=& (\sigma + \I \theta^R) Z + \sqrt{2} \ep \chi^R  \;  , \label{eq:ChiralGravTransf_first} \\
\delta \chi_{\a}^R &=& \frac{1}{2} (3 \sigma - \I \theta^R) \chi_{\a}^R + \I \sqrt{2} \mathcal{D}_\mu Z \sigma^\mu_{\a \da} \bep^{\da} + \sqrt{2} \mathcal{F} \ep_{\a} + 2 \sqrt{2} Z \eta_{\a} \; , \label{eq:ChiralGravTransf_chi} \\
\delta \mathcal{F} &=&  2(\sigma - \I \theta^R) \mathcal{F} + \I \sqrt{2} \bar{\ep} \bsigma^{\mu} \mathcal{D}_\mu \chi^R, \label{eq:ChiralGravTransf_last}
\end{eqnarray}
where the local SUSY-covariant derivatives read
\begin{align}
\mathcal{D}_\mu Z & = \partial_\mu Z - \I A_\mu^R Z - \frac{1}{\sqrt{2}} \psi_\mu \chi^R \; , \\
\mathcal{D}_\mu \chi_{\a}^R & = \left( D_\mu^\omega + \frac{\I}{2} A_\mu^R \right) \chi_{\a}^R - \frac{\I}{\sqrt{2}} \mathcal{D}_\nu Z \sigma_{\a \da}^\nu \bpsi^{\da}_\mu - \frac{1}{\sqrt{2}} \mathcal{F} \psi_{\mu \a} - \sqrt{2} Z \phi_{\mu \a}. \label{Dchi}
\end{align}
Terms proportional to the SUSY parameters $\ep, \bep$ reproduce the transformations of the chiral multiplet \eqref{susy_chiral_first} - \eqref{susy_chiral_last}. 

At  linear level the compensators couple to the relevant anomalies through a term in the effective action
\begin{equation} \label{eq:CompCoupling}
W_{\text{comp}}= \int \D^4 \bs{x} \sqrt{-g} \left( \pi \mathcal{A}_{\theta^R} + \tau \mathcal{A}_\sigma + \frac{1}{2 \sqrt{2}}\chi^R \mathcal{A}_\eta + \frac{1}{2 \sqrt{2}}\bchi^R \bAn_{\bar{\eta}} \right) \; ,
\end{equation}
The Ward identity associated with the $S$-symmetry reads
\begin{align} \label{Ward_eta}
\I \psi_{\mu \a} \< J_R^\mu \> + 2 \I \sigma_{\mu \a \da} \< \bJ_{\bQ}^{\mu \da} \> = \An_{\eta \a}, && - \I \bpsi_\mu^{\da} \< J_R^\mu \> + 2 \I \bsigma_\mu^{\da \a} \< J_{Q \a}^\mu \> = \bAn_{\bar{\eta}}^{\da}.
\end{align}
The analogous expression in  4-component notation is given in \eqref{eq:AchiCurrents4c}.

The compensator fields in equation \eqref{eq:CompCoupling} transform according to \eqref{eq:ChiralGravTransf_first} - \eqref{eq:ChiralGravTransf_last} where we use $(Z=e^{\tau + \I \pi}, \chi^R, \mathcal{F})$ and always set $\pi= \tau= \chi^R= 0$ at the end. From \eqref{eq:ChiralGravTransf_first} we have
\begin{align}
& \delta_{\theta^R} \pi = \theta^R, && \delta_{\sigma} \tau = \sigma, && \delta_\eta \chi_{\a}^R= 2 \sqrt{2} \eta_{\a}.
\end{align}
Here we only include the terms linear in the compensators. In general one also has higher order corrections from expanding $e^{\sigma}$, \textit{etc}. As we can see the compensating fields $\chi^R, \bchi^R$ are the compensators for the $S$-supersymmetry, while $\pi$, $\tau$ correspond to the pion and dilaton respectively. From \eqref{eq:CompCoupling} and \eqref{Ward_eta} we see that the compensators $\chi^R$ and $\bchi^R$ couple to the gamma-contracted supercurrents,
\begin{align} \label{JchiToJQ}
J_{\chi^R \a} & =  \frac{\I}{\sqrt{2}} \sigma_{\mu \a \da} \bJ^{\mu \da}_{\bQ} + \frac{\I}{2 \sqrt{2}} \psi_{\mu \a} J_R^\mu \nn\\
J^{\da}_{\bchi^R} & = \frac{\I}{\sqrt{2}} \bsigma_{\mu}^{\da \a} J^{\mu}_{Q \a} - \frac{\I}{2 \sqrt{2}} \bpsi_\mu^{\da} J_R^\mu .
\end{align}
However, since the $S$-supersymmetry anomaly belongs to same multiplet as the chiral anomaly, we are not worried about introducing a compensator for the symmetry that we know is anomalous. Most importantly, we do not introduce a compensator for $Q$-supersymmetry.

Finally, we are left with the complex field $\mathcal{F}$, which does not represent any gauge degrees of freedom. This is slightly different from the real vector multiplet, where the analogous $F$-terms are gauge degrees of freedom w.r.t.\! the flavor transformations. Including nonzero $\mathcal{F}$ is important for the closure of the SUSY algebra when coupling to old-minimal or 16+16 SUGRA. At the linear level we have
\begin{equation} \label{eq:Fcoupling}
\int \D^4 \bs{x} \sqrt{-g} \left( \mathcal{F} \< J_\mathcal{F} \> + h.c. \right) \; ,
\end{equation}
where $J_{\mathcal{F}}$ is related to the so-called brane current, see \cite{Dvali:1996xe,Dumitrescu:2011iu}. Just like $J_{\chi^R}$ and $J_Z$ the operator $J_{\mathcal{F}}$ is ultralocal. This, in particular, implies that the generating functional, $W = W[Z, \chi_R, \mathcal{F}]$, is a local functional of the three sources.

\subsection{Consequences}

As in the flavor case, the compensating field $\chi^R$ transforms into the physical field $A_\mu^R$ and other compensators. There is a difference in the overall normalization between the two cases, but the physics remains the same.

In order to derive the correct Wess-Zumino consistency condition as well as the Ward identity for the supersymmetric current, we have to include the SUSY transformations of the compensators in \eqref{eq:ChiralGravTransf_first} - \eqref{eq:ChiralGravTransf_last}. As long as we are interested in the correlation functions of the physical operators $T_a^\mu$, $J^{\mu \a}_Q$, and $J_R^\mu$ sourced by $e^a_\mu$, $\psi_{\mu \a}$, and $A^R_\mu$ respectively, we may substitute $Z = 1$ and $\chi^R = \mathcal{F} = 0$ only after the variations are taken. This gives
\begin{align} \label{eq:0variations}
\delta_{\ep,\bep} Z|_0 & =  0, \\ \label{eq:0Chivariation}
\delta_{\ep,\bep} \chi^{R \a}|_0 & =  - \sqrt{2} A_\mu^R \bep_{\da} \bsigma^{\mu \da \a}, \\ \label{eq:0Fvariation}
\delta_{\ep,\bep} \mathcal{F}|_0 & =  - \I \bep_{\da} \left[ ( \bsigma^\mu \sigma^\nu)^{\da}_{\ \db} \bpsi^{\db}_{\mu} A_\nu^R + 2 \bsigma^{\mu \da \a} \phi_{\mu \a} \right].
\end{align}

\paragraph{Wess-Zumino consistency condition.}
Using \eqref{eq:0Chivariation} we can now derive the correct Wess-Zumino consistency condition. In particular, the commutator of  supersymmetry and $R$-symmetry, when acting on the relevant compensators, reads,
\begin{align}
[\delta_{\bep}, \delta_{\theta^R}] \chi^{R \alpha} = \sqrt{2} \bep_{\da} \bsigma^{\mu \da \a} \partial_\mu \theta^R.
\end{align}
Up to a constant, this is the same relation as in the flavor case, \eqref{chi_comm}.
In general \eqref{eq:0Fvariation} implies also a non-vanishing contribution from $\mathcal{F}$. However, in the generating functional $W = W[\mathcal{F}]$ the source $\mathcal{F}$ necessarily couples to at least a pair gravitinos. To see that this should be the case in general we invoke the reality of generating functional, which implies that $\mathcal{F}$ has to couple to a complex scalar. At the linear level in compensators this scalar has to be formed by the fields of conformal SUGRA. This leaves us with the bilinears formed from $\psi$ and $\bar{\psi}$ (and their derivatives). This is analogous to the $\mathcal{M}$ field in the case of the flavor anomaly, where the generating functional would have the form $W[\mathcal{M}] \sim \mathcal{M} \lambda \lambda$, as can be seen from \eqref{ct}. Thus, $\< J_{\mathcal{F}} \> = O(\psi^2)$ and $\< J_{\mathcal{F}} \> \delta_{\ep, \bep} \mathcal{F} |_0 = O(\psi^3)$. This means that we can consistently drop all terms depending on $\< J_{\mathcal{F}} \>$, since they only contribute to $O(\psi^3)$ terms to the SUSY anomaly. We will therefore not consider the effects of including $\mathcal{F}$ here.

As in the flavor case, the Wess-Zumino consistency condition \eqref{derive_WZ} is now consistent with the absence of the $Q$-SUSY anomaly, $\An_{\ep} = 0$ and $\bAn_{\bep} = 0$. In such  case the Wess-Zumino consistency condition becomes
\begin{align} \label{eq:SugraCons}
\delta_{\bep} \mathcal{A}_{\theta^R} & = - \I \sqrt{2} \bep_{\da} \bsigma^{\mu \da \a} \partial_\mu \< J_{\chi^R \alpha} \> && \mathcal{A}_{\bep} = 0 \nn\\
& = - \frac{\I}{2} \bep \bsigma^{\mu} \partial_{\mu}  \mathcal{A}_{\eta} + O(\psi^3),
\end{align}
where in the last line we used \eqref{JchiToJQ}. This equation is identical to \eqref{true_WZ_cond}, up to the multiplicative constant. The constant is fixed here by the relation \eqref{JchiToJQ} between the supercurrent and $J_{\chi^R}$. Furthermore, our conventions are such that the $U(1)$-flavor current matches the $R$-current and hence the two Wess-Zumino consistency conditions \eqref{eq:SugraCons} and \eqref{true_WZ_cond} become equal. In particular, all of this is consistent with the absence of the $Q$-anomaly, $\An_\ep = \An_{\bep} = 0$, and represents the correct Wess-Zumino condition.

\paragraph{Ward identity.} The variation of the compensators \eqref{eq:ChiralGravTransf_first} - \eqref{eq:ChiralGravTransf_last} produces extra terms in the Ward identity. By setting $Z = 1$ and $\chi^R_\alpha = \mathcal{F} = 0$ only after taking the variations we find
\begin{align} \label{Qsusy_ward}
\partial_\mu \< \bJ_{\bQ}^{\mu \da} \> & = \I \bsigma^{a \da \a} \psi_{\mu \a} \< T^\mu_a \> + \I \bphi_{\mu \da} \< J^\mu_R \> - \sqrt{2} A^R_\mu \bsigma^{\mu\da\a} \< J_{\chi^R \a} \> + O(\psi^3, \mathcal{F}).
\end{align}
The last term introduces a piece proportional to the $S$-anomaly in the Ward identity, but the $Q$-anomaly is absent.

\paragraph{Counterterm.} Once again we can define a counterterm by adding terms linear in $\chi^R$, $\bchi^R$, and $\mathcal{F}$ as well as some $R$-symmetry covariant terms quadratic in the gravitino. The vielbein-independent part of the counterterm must resemble that of \eqref{ct} with $\phi_\mu$ playing the role of $\lambda$. Indeed, the counterterm reads
\begin{align} \label{eq:CounterGrav}
S_{\text{ct}}^R = \frac{1}{8 \pi^2}\left[3 (a-c) (P_1 + P_1^{\dagger}) - 2 a (P_2 + P_2^{\dagger}) \right]   \; ,
\end{align}
where
\begin{align} \label{eq:Counter1}
P_1 & = -\sqrt{2} \chi^{R \a} \left( \frac{1}{3} W_{\mu\nu}^{\ \ \: \rho\sigma} \sigma_{\rho\sigma\a}^{\ \ \ \ \b} + \I F_{\mu\nu}^R \delta_{\a}^{\b} \right) \times \left( \I \sigma^{\mu}_{\b\da} \bphi^{\nu\da}  + \mathcal{D}^{\mu} \psi^{\nu}_{\b} \right) \nn\\
& \qquad\qquad\qquad - \frac{2}{3} \epsilon^{\mu\nu\rho\sigma} A_\mu^R \phi_\nu^{\a} \left( 2 \mathcal{D}_{\rho} \psi_{\sigma \a} + \I \sigma_{\rho \a \da} \bphi^{\da}_{\sigma} \right)
\end{align}
and
\begin{align}
P_2 &=  - \sqrt{2} \I \chi^{R \a} \left[ 2 \I F_{\mu\nu}^R - \frac{1}{4} \epsilon_{\mu\nu}^{\ \ \: \rho\sigma} F_{\rho\sigma}^R - P_{\mu\nu} + \frac{1}{6} R g_{\mu\nu} \right] \sigma^{\mu}_{\a\da} \bphi^{\nu\da} \nn\\
& \qquad + A_\nu^R \left( \sqrt{2} \mathcal{D}_\mu \chi^{R \a} + \phi^{\a}_\mu \right) \times \left( \sigma^\mu_{\a \da} \bphi^{\nu \da} + g^{\mu\nu} \sigma^{\rho}_{\a \da} \bphi_\rho^{\da} - \I \epsilon^{\mu\nu\rho\tau} \sigma_{\tau \a \da} \bphi_\rho^{\da} \right) \nn\\
& \qquad + A_\mu^R A_\nu^R P^{\mu\nu}, \label{eq:Counter2}
\end{align}
where $P_{\mu \nu}$ is the Schouten tensor defined in \eqref{Pmunu} and the Weyl tensor reads
\begin{equation}
W^{\rho \sigma \mu \nu}= R^{\rho \sigma \mu \nu}- \frac{1}{2}(R^{\rho \nu} g^{\sigma \mu}-R^{\rho \mu} g^{\sigma \nu}+ R^{\sigma \mu} g^{\rho \nu}-R^{\sigma \nu} g^{\rho \mu})+  \frac{1}{6} R ( g^{\rho \mu} g^{\sigma \nu} - g^{\rho \nu} g^{\sigma \mu}).
\end{equation}
In the expression for the counterterm we have omitted the terms proportional to $\mathcal{F}$ \footnote{These terms are analogous to the $\mathcal{M} \lambda \lambda$ term in \eqref{ct}.}. We have also dropped all terms of higher than quadratic order in $A_\mu^{R}, \psi_\mu$, which are necessary for covariantization w.r.t.\! local SUSY. Furthermore, \eqref{eq:CounterGrav} is explicitly gauge-covariant w.r.t.\! $R$-symmetry, so an equation analogous to \eqref{true_WZ_condNS} holds.

One can explicitly check that the SUSY-variation of \eqref{eq:CounterGrav} cancels the anomaly up to higher order terms as explained above. This means that in the $\chi^R$-dependent terms it is enough to vary $\chi^R$ only. Similarly, in terms containing two fermions and the background gauge field $A_\mu^R$, the variation of $A_\mu^R$ results in 3-fermion terms, which we can drop. Finally, while, after taking the variation, the parameters $\ep$ and $\bep$ may appear under derivatives, one can always combine the terms into commutators, for which the relevant terms read
\begin{align}
[ \mathcal{D}_\mu, \mathcal{D}_\nu ] \ep = - \frac{3}{2} \I F_{\mu\nu} \ep - \frac{1}{2} \sigma^{ab} R_{\mu\nu a b} \ep + \ldots
\end{align}

There exists, however, a simpler method to derive the counterterm \eqref{eq:CounterGrav}. In any supersymmetric renormalization scheme, $\An_{\ep} = \bAn_{\bep} = 0$, but the $R$-symmetry anomaly, $\An_{\theta^R}$, contains additional terms on top of the ABJ anomaly \eqref{eq:ABJSugra}, \textit{i.e.},
\begin{align}
\An_{\theta^R} = \An_{\theta^R}^{\text{ABJ}} + \frac{\delta W_{\text{loc}}}{\delta \theta^R},
\end{align}
where $W_{\text{loc}}$ is the local part of the generating functional, which can be equated to a local, finite counterterm. Then,
\begin{align}
\delta_{\bep} \left(\mathcal{A}_{\theta^R}^{\text{ABJ}} +  \frac{\delta W_{\text{loc}}}{\delta \theta^R} \right)
& = - \I \sqrt{2} \bep_{\da} \bsigma^{\mu \da \a} \, \partial_{\mu} \frac{\delta W_{\text{loc}}}{\delta \chi^{R \a}} \nn\\
& = - \frac{\I}{2} \theta^R \, \bep \bsigma^{\mu} \partial_{\mu} \mathcal{A}_\eta ,
\end{align}
where we used \eqref{eq:CompCoupling}.
This is a gravity analogue of \eqref{eq:CountAnomRel}, with the r.h.s.\! being related to the $S$-SUSY anomaly as expected from \eqref{eq:SugraCons}. Note that these expressions do not match the ones obtained in \cite{Papadimitriou:2019gel}, as our counterterm includes additional terms that are non-invariant under $S$-, Weyl and $R$-transformations analogous to the $A_\mu \lambda \sigma^\mu \bar{\lambda}$ term in \eqref{ct}. As before, the precise form of the $R$-symmetry anomaly can be changed by an additional SUSY-invariant counterterm. Thus, different choices of counterterms will in general lead to different expressions for $\mathcal{A}_\eta$ and $\bAn_{\bar{\eta}}$. Nevertheless, we verified that the $\chi^R$-dependent part of \eqref{eq:CounterGrav} corresponds to the component expansion of the counterterm given in \cite{Kuzenko:2019vvi}. Just as in the flavor case, in order to define a supersymmetric scheme where the anomaly \eqref{eq:SUGRAanomaly} vanishes, we have to supplement the $\chi^R$ term with a gauge-dependent term bi-linear in fermions. This extra term depends only on the fields of conformal SUGRA. In this sense conformal SUGRA plays the role of Wess-Zumino gauge here, as explained in \cite{Kuzenko:2019vvi}. Finally, note that the counterterm analogous to \eqref{eq:CounterGrav} can be defined in any formulation of SUGRA which contains a chiral supermultiplet as a compensator. For example the compensator relevant for 16-16 SUGRA is the vector multiplet (see appendix \ref{App:VecMult}), which contains a chiral multiplet.\footnote{Albeit one scalar degree of freedom (the pion) is ``hidden" inside the gauge field.}

\subsection{Old minimal SUGRA} \label{sec:OMsugra}

One can combine the fields $e^a_\mu, A_\mu^R$ and $\psi_{\mu \a}$ of conformal SUGRA with the compensators $\tau, \pi, \chi^R$ and $\mathcal{F}$ to construct fields invariant under Weyl-, $S$- and $R$-transformations. The resulting gravitino, $\tilde{\psi}_\mu$, vielbein, $\tilde{e}^{a}_{\mu}$, the $R$-current source, $\tilde{A}_\mu^R$, and the scalar $\tilde{\mathcal{F}}$ are defined as follows,
\begin{align}
\tilde{e}^{a}_{\mu} & = e^{a}_{\mu} + \tau e^{a}_{\mu} , \label{eq:eOMsugra} \\
\tilde{A}^R_\mu & = A_\mu^R - \frac{\I}{2 \sqrt{2}} ( \chi^R \psi_\mu - \bchi^R \bpsi_\mu ) - \partial_{\mu} \pi \; , \label{eq:AOMsugra} \\
\tilde{\psi}_{\mu\a} & = \psi_{\mu\a} + \frac{\I}{\sqrt{2}} \sigma_{\mu\a\da} \bchi^{R\da}  -\frac{1}{2} (3\I \pi - \tau) \psi_{\mu\a} \;, \label{eq:psiOMsugra} \\
\tilde{\mathcal{F}} & = \mathcal{F} - 2 (\tau - \I \pi) \mathcal{F}.
\end{align}
We can now apply the variations \eqref{eq:vielbeinTransf} - \eqref{eq:PsimuTransf} and \eqref{eq:ChiralGravTransf_first} - \eqref{eq:ChiralGravTransf_last} to \eqref{eq:eOMsugra} - \eqref{eq:psiOMsugra}.
For example, the relevant SUSY transformations are derived by using the transformation rule of $\chi^R$ derived from \eqref{eq:ChiralGravTransf_chi}
\begin{equation} \label{eq:chiSUGRAtransf}
\delta_{\ep, \bep} \chi^R_{\a} = \sqrt{2} \left[ (A_\mu^R - \I \chi^R \psi_\mu) \sigma^\mu_{\a \da} \bep^{\da} + \mathcal{F} \ep_{\a} \right].
\end{equation}
 After the variations are taken, we substitute $Z = 1$ and $\chi^R = 0$, but we keep $\mathcal{F}$ nonzero as it has now become a part of the multiplet. It is precisely the role of the compensators to cancel $S$-SUSY variations parametrized by $\eta, \bar{\eta}$ as well as Weyl and $R$-symmetry transformations parameterized by $\sigma$ and $\theta^R$. We end up with the SUSY transformations of old minimal supergravity,
\begin{align} \label{eq:vielbeinTransfNM}
\delta \tilde{e}_\mu^a &= \I ( \bep \bsigma^a \tilde{\psi}_\mu + \ep \sigma^a \bar{\tilde{\psi}}_\mu ) - \tilde{\lambda}^a_{b} \tilde{e}_\mu^b, \\ \label{eq:ARmuTransfNM}
\delta \tilde{A}_\mu^R & = - \I \ep^{\a} \left( \tilde{\phi}_{\mu\a} - \tfrac{1}{2} \tilde{A}_\nu^R \sigma^\nu_{\a \da} \bar{\tilde{\psi}}_{\mu}^{\da} + \tfrac{1}{2} \tilde{\mathcal{F}} \tilde{\psi}_{\mu \a} \right) + \I \bep_{\da} \left( \bar{\tilde{\phi}}_{\mu}^{\da} + \tfrac{1}{2} \tilde{A}_\nu^R \bsigma^{\nu\da\a} \tilde{\psi}_{\mu\a} + \tfrac{1}{2} \tilde{\mathcal{F}}^{\ast} \bar{\tilde{\psi}}_{\mu}^{\da} \right), \\
\delta \tilde{\psi}_{\mu \a} & = 2 \mathcal{D}_\mu \ep_{\a} - \I \tilde{A}_\nu^R (\sigma_\mu \bsigma^\nu)_{\a}^{\ \beta} \ep_{\beta} + \I \tilde{\mathcal{F}}^\ast \sigma_{\mu \a \da} \bep^{\da} + \frac{1}{2} \tilde{\lambda}^{ab} \sigma_{ab \a}^{\ \ \ \ \beta} \tilde{\psi}_{\mu\beta}, \\
\delta \tilde{\mathcal{F}} & = - \I \bep_{\da} \left( \tilde{A}_\nu^R (\bsigma^\mu \sigma^\nu)^{\da}_{\ \db} \bar{\tilde{\psi}}^{\db}_{\mu} + \tilde{\mathcal{F}} \bsigma^{\mu\da\a} \psi_{\mu\a} + 2 \bsigma^{\mu\da\a} \phi_\mu \right).
\end{align}
where \eqref{eq:PhiSugra} results in
\begin{equation}
\tilde{\phi}_\mu = \phi_\mu + \frac{1}{\sqrt{2}}\mathcal{D}_\mu \chi^R
\end{equation}
and
\begin{eqnarray}
\tilde{\lambda}^{ab}&=& \lambda^{ab} - \frac{1}{2 \sqrt{2}} \left( \ep \sigma^{ab} \chi^R + \bep \bsigma^{ab} \bchi^R \right).
\end{eqnarray}
The terms involving $\tilde{\lambda}^{ab}$ can be removed by a Lorenz transformation. 

Equations \eqref{eqG:psiOMsugra} - \eqref{eqG:AOMsugra} can be used to deduce the linear couplings of the compensators to the respective currents, \cite{Kaku:1978ea},
\begin{equation} \label{eq:ScurrentsOM}
S_{\text{coupling}} = \int \D^4 \bs{x} \sqrt{-g} \left(\tilde{A}_\mu \< \tilde{J}_R^\mu \>+ \tilde{e}_\mu^a \<\tilde{T}_a^\mu \> + \bar{\tilde{\psi}}_\mu   \<\tilde{J}_Q^\mu \> \right)+ \mathcal{O}(\chi^2, \pi^2, \tau^2, \mathcal{F})  \; .
\end{equation}
Classically, the sources couple to the same currents as the ones of conformal SUGRA, however here we included a tilde superscript to take into account the fact the relevant expressions might differ by local and seagull terms. Using these equations we find the first order ``correction" to the action of conformal SUGRA reads
\begin{equation} \label{eq:Saux}
\delta S_{\text{aux}} = \int \D^4 \bs{x} \sqrt{-g} \left(-\mathcal{D}_\mu \pi \< \tilde{J}_R^\mu \>- \tau \<\tilde{T}_\mu^\mu \> - \frac{1}{\sqrt{2}}\bar{\chi}^R \gamma_\mu \<\tilde{J}_Q^\mu \> \right)+ \mathcal{O}(\chi^2, \pi^2, \tau^2, \mathcal{F})  \; ,
\end{equation}
where we used a SUSY-covariant derivative $\mathcal{D}_\mu \pi= \partial_\mu \pi + \frac{\I}{2\sqrt{2}}\bar{\psi}_\mu \gamma_5 \chi^R$.  The theory with compensators turned on is SUSY invariant, which is equivalent to saying that it is invariant w.r.t. old minimal SUGRA. Thus, we should be able to rewrite the counterterm \eqref{eq:CounterGrav} in terms of the variables of old minimal SUGRA
\begin{equation} \label{eq:Sredef}
S_{\text{ct}}^R = - \int \D^4 \bs{x} \sqrt{-g} \, \bar{\chi}^R \tilde{\mathcal{A}}_\eta + \tilde{S}_{\text{ct}}^R \; ,
\end{equation}
where $\tilde{S}_{\text{ct}}^R$ is a local functional of tilde fields only. The anomaly $\tilde{\mathcal{A}}_\eta$ is absorbed in the definition of the supersymmetry current and in general it will look different from the one implied by \eqref{eq:CounterGrav}. In this case $\tilde{S}_{\text{ct}}^R$ will be manifestly invariant under the gauge symmetries but not invariant w.r.t. SUSY of old minimal SUGRA. At a basic level $\tilde{S}_{\text{ct}}^R$ can be obtained by replacing un-tilded variables with tilded ones in the $\chi^R-$independent part of \eqref{eq:CounterGrav}. This process clearly changes the $\chi^R$ term, which follows directly from \eqref{eqG:psiOMsugra} - \eqref{eqG:AOMsugra}. Furthermore different choices of the counterterm \eqref{eq:CounterGrav} will lead to different expressions for $\tilde{\mathcal{A}}_\eta $ and $\tilde{S}_{\text{ct}}^R$.

\section{Pauli-Villars renormalization} \label{sec:PV_ren}

In section \ref{sec:WZ_model} we will carry out calculations in a free Wess-Zumino model. We will confirm the Wess-Zumino consistency condition \eqref{true_WZ_cond} by evaluating suitable correlation functions and we will show the absence of the SUSY anomaly by evaluating both sides of the Ward identity \eqref{ward_1pt} in this model. To do it, however, certain momentum integrals require renormalization. While any regularization scheme can be used in principle, we want to use a scheme, where two conditions are satisfied: \textit{i}) the scheme is explicitly supersymmetric, and \textit{ii}) the scheme can deal with chiral fields with ease. For this reason we will use Pauli-Villars renormalization scheme with a cut-off regularization.

In this section we define and analyze the Pauli-Villars (PV) renormalization scheme with cut-off regularization. In the context of anomalies, Pauli-Villars regularization is used to evaluate the ABJ anomaly, \cite{Bertlmann:1996xk}, but it can be used in a wider context. In such cases, however, a more detailed description is needed, in order to deal with non-logarithmic divergences.

In this section we present the analysis of the trace (conformal) anomaly in a 2-dimensional free scalar field. We describe how the choice of the extension of the stress tensor from the massless to the massive theory affects the structure of the anomaly. We conclude that if the massive theory preserves a given symmetry (supersymmetry, general covariance, \textit{etc}.) the anomaly is absent in the massless theory.

\subsection{Pauli-Villars renormalization} \label{sec:PV}

All correlators we will consider in the next section have the form of a 1-loop momentum integral of the form
\begin{align} \label{toi_k}
\int \frac{\D^4 \bs{k}}{(2 \pi)^4} \frac{\text{numerator}(\bs{k}; \bs{p}_j; m)}{(k^2 + m^2 + \I \epsilon) ( |\bs{k} - \bs{p}_1|^2 + m^2 + \I \epsilon) \ldots ( |\bs{k} - \bs{p}_1 - \ldots - \bs{p}_{n-1}|^2 + m^2 + \I \epsilon)}.
\end{align}
By using Wick's rotation and Feynman parameters the integral is brought to the sum of terms of the form
\begin{align} \label{l_int}
A_{dnr}(\Delta) = \int_{Eu} \frac{\D^d \bs{l}}{(2 \pi)^d} \frac{l^{2r}}{(l^2 + \Delta)^n} = \frac{\Gamma(n - r - \tfrac{1}{2} d) \Gamma(r + \tfrac{1}{2} d)}{(4 \pi)^{d/2} \Gamma(n) \Gamma(\tfrac{1}{2}d)} \Delta^{r - n + \tfrac{1}{2} d},
\end{align}
where $Eu$ denotes the fact that we consider a Euclidean integral.  $\Delta$ contains Feynman parameters over which another integral will have to be carried out. The mass-dependence enters through $\Delta$ and $\Delta = \Delta_0 + m^2$. Here $\Delta_0$ depends on the Feynman parameters $x_j$ as well as quadratically on external momenta $\bs{p}_i$, but is mass-independent.

If $r \geq n - \tfrac{1}{2} d$, the integral is UV divergent and requires regularization. The most popular regularization is dimensional regularization. While dimensional regularization can deal with chiral terms, \cite{tHooft:1972tcz}, here we will use a cut-off regularization, which is simpler in execution in the presence of such terms. We cut the integral off at some $\Lambda$, so that the integral is taken over momenta $l < \Lambda$. The integral reads
\begin{align} \label{l_int_Lambda}
A_{dnr}^{\Lambda}(\Delta) & = \int_{Eu}^{\Lambda} \frac{\D^d \bs{l}}{(2 \pi)^d} \frac{l^{2r}}{(l^2 + \Delta)^n} = \frac{1}{(4 \pi)^{d/2} \Gamma( \tfrac{1}{2} d)} \times \frac{2}{d+2r} \times \nn\\
& \qquad\qquad \times \frac{\Lambda^{d+2r}}{\Delta^n}  {}_2 F_1 \left(n, \tfrac{1}{2} d + r, \tfrac{1}{2} d + r + 1; -\frac{\Lambda^2}{\Delta} \right).
\end{align}
If $r < n - \tfrac{1}{2} d$, the $\Lambda \rightarrow \infty$ limit can be taken and the expression simplifies to \eqref{l_int}. Otherwise, the expression diverges and we can expand it in powers of $1/\Lambda$. The exact form of the expansion depends on whether
\begin{align} \label{div_k}
k = r + \tfrac{1}{2} d - n
\end{align}
is a non-negative integer or not. If $k$ is not a non-negative integer, then the terms diverging in the $\Lambda \rightarrow \infty$ limit are $\Lambda^{d + 2 r - 2 n - 2 j}$ for $j = 0,1,\ldots,\lfloor k \rfloor$. If $k$ is a non-negative integer, then in addition to those terms a logarithmic divergence containing $\log \Lambda$ appears.

To deal with the divergences in the $\Lambda \rightarrow \infty$ limit two routes can be taken. In a more operator-oriented approach the divergences can be removed by the introduction of the counterterms built up with  sources and operators under consideration. These counterterms are divergent in the $\Lambda \rightarrow \infty$ limit precisely in such a fashion that the divergences cancel. The terms with power divergences $\Lambda^\alpha$ are uniquely fixed and so are the corresponding counterterms. The logarithmic terms, on the other hand, require logarithmic counterterms of the form $\log (\Lambda/m)$, where $m$ is the arbitrary renormalization scale. By introducing such counterterms, correlation functions become scale-dependent and the scaling anomaly emerges. In this approach anomalies in the conserved currents emerge from shifts in the integration variable in loop integrals. However, since integrals with a cut-off are difficult to evaluate, one usually chooses a different approach.

A different approach is the Pauli-Villars (PV) renormalization method. The idea lies in the observation that the expansion of \eqref{l_int_Lambda} around $\Lambda = \infty$ is equivalent to the expansion around $\Delta = 0$. Physically, the $\Delta \rightarrow 0$ limit can be arranged either as the zero-momentum limit or the infinite mass limit. In standard textbooks \cite{Itzykson:1980rh,Bertlmann:1996xk} the method is usually applied to expressions where only a single logarithmic divergence appears. A given amplitude $T(p; m, \Lambda)$ can then be rendered finite by defining
\begin{align} \label{PVdef}
T^{PV}(p; m) = \lim_{\Lambda \rightarrow \infty} \left[ T(p; m, \Lambda) - \lim_{M \rightarrow \infty} T(p; M, \Lambda) \right].
\end{align}
This is sufficient for the analysis of the ABJ anomaly as well as the SUSY anomaly in $\< \bj^{\rho \da}_{\bQ} j_A^{\mu} j_A^\nu j_{\lambda} \>$. If higher order divergences appear, more terms must be subtracted. 

In \cite{Bertlmann:2000da} the PV renormalization scheme is defined by successive subtractions of zero-momentum terms. Equivalently, one can define the PV renormalization by subtracting terms diverging in the infinite mass limit. If we abuse the mathematical notation a little and denote
\begin{align}
\lim_{M \rightarrow \infty} T(p; M, \Lambda) = \text{terms non-vanishing in the } M \rightarrow \infty \text{ limit of } T(p; M, \Lambda),
\end{align}
then the amplitude regulated in the PV scheme is defined by \eqref{PVdef}. If one is interested in the massless theory, then the expansion around $m = 0$ can be calculated with $m$ interpreted as the renormalization scale. If the amplitude is IR-divergent, then $m$ can be regarded as the IR regulator.

This prescription has an advantage of having a more physical implementation. For every physical field $\j$ one introduces a ghost field $\tilde{\j}$ with the wrong commutation relations and very large mass $M$. The two theories do not interact, which means that for each loop diagram in the original theory, one has the identical diagram in the ghost theory, but with the opposite overall sign. Clearly, if we send the mass $M$ of the ghost fields to infinity, the physical amplitude takes form \eqref{PVdef}. In practice, however, this picture is little bit more complicated. This is because in general one needs to introduce multiple massive fields to regulate all divergences even in the free-field theory. For example the power divergences of the type $M^n \Lambda^m$ clearly do not cancel between the massive and massless diagrams in \eqref{PVdef} so we have to introduce more massive PV fields to cancel those. If we denote the statistics of the $i$-th regulator by $s_i=\pm 1$ (with $s_i = +1$ matching the statistics of the original field and $s_i = -1$ standing for the opposite statistics) we have a general condition
\begin{equation} \label{PVsumCond}
\sum_i s_i + s_0= 0 \; ,
\end{equation}
where $s_0$ is the parity of the original field. This condition removes the divergent cosmological constant, whereas further conditions are needed to cancel quadratic divergences etc. Nevertheless we will show in the next section that as far the correlators relevant for this paper go, it is enough to work with just one ``effective" PV regulator.

If any of the original fields is charged under a global symmetry, so are the corresponding PV fields. Therefore the PV fields should be coupled to the corresponding background fields in the path integral. Integrating out these massive fields will leave us with finite, source-dependent counterterms in the action. These terms can be thought of as a choice of  scheme forced on us by the regularization, see the paragraph under \eqref{eq:PVchiralAnomaly}. The mass terms of the ghosts will break some of these global symmetries classically, \textit{e.g.}, the conformal and chiral symmetry. This is the source of the anomaly in the Pauli-Villars regularization scheme. Nevertheless, we can keep diffeomorphisms or supersymmetry unbroken  by coupling the ghosts covariantly to the respective sources. This will lead to some mass-dependent terms in the respective currents. As far as diffeomorphisms go, we will illustrate the consequences thereof on the simple example of a $2D$ free scalar in the section \ref{sec:trace_an}. To preserve supersymmetry classically we have to couple the ghosts to a formulation of SUGRA consistent with the presence of a mass terms.

\subsection{ABJ anomaly} \label{sec:ABJ} 

The main advantage of the Pauli-Villars method for us is the fact that the anomalies can be easily derived in the massless theory by taking the infinite mass limit. As an example, we will consider the ABJ anomaly, \cite{Adler:1969gk,Bell:1969ts}.

Consider the free Wess-Zumino model with the Lagrangian given by \eqref{lag}. The only relevant part of the Lagrangian here is its fermionic part,
\begin{align} \label{lag0}
& L_0 = - \I \bpsi \bsigma^\mu D_\mu \psi, && D_\mu \psi = \left( \partial_\mu + \frac{\I}{2} A_\mu \right) \psi,
\end{align}
from which the axial current is
\begin{align}
J_A^\mu = j^\mu_A = \frac{1}{2} \bpsi \bsigma^\mu \psi.
\end{align}
In particular, seagull terms are absent. The naive Ward identity for the 3-point function of the axial current is simply
\begin{align}
p_{1 \mu} \lla J_A^\mu(\bs{p}_1) J_A^\nu(\bs{p}_2) J_A^{\rho}(\bs{p}_3) \rra = p_{1 \mu} \lla j_A^\mu(\bs{p}_1) j_A^\nu(\bs{p}_2) j_A^{\rho}(\bs{p}_3) \rra = 0 \ \text{(up to anomaly)}.
\end{align}

The 3-point function of the axial current $j_A^\mu$ can be written as
\begin{align} \label{jAjAjA}
& \lla j_A^\mu(\bs{p}_1) j_A^\nu(\bs{p}_2) j_A^{\rho}(\bs{p}_3) \rra = \frac{1}{8} T^{\mu\nu\rho, \kappa\lambda\tau} \int_{Eu} \frac{\D^4 \bs{k}}{(2 \pi)^4} \frac{ k_\lambda (k - p_1)_{\kappa} (k + p_2)_{\tau}}{k^2 (\bs{k} - \bs{p}_1)^2 (\bs{k} + \bs{p}_2)^2},
\end{align}
where the tensor structure $T^{\mu\nu\rho, \kappa\lambda\tau}$ is
\begin{align}
T^{\mu\nu\rho, \kappa\lambda\tau} & = \Tr \left[ \sigma^\kappa \bsigma^\mu \sigma^\lambda \bsigma^\nu \sigma^\tau \bsigma^\rho - \sigma^\lambda \bsigma^\mu \sigma^\kappa \bsigma^\rho \sigma^\tau \bsigma^\nu \right] \nn\\
& = 4 \I \left[ \eta^{\kappa \lambda} \epsilon^{\mu\nu\rho\tau} - \eta^{\mu\kappa} \epsilon^{\nu\rho\lambda\tau} - \eta^{\mu\lambda} \epsilon^{\nu\rho\kappa\tau} - \eta^{\nu\tau} \epsilon^{\mu\rho\kappa\lambda} - \eta^{\rho\tau} \epsilon^{\mu\nu\kappa\lambda} + \eta^{\nu\rho} \epsilon^{\mu\kappa\lambda\tau} \right]
\end{align}
and the subscript \textit{Eu} indicates that the integral should be carried out in the Euclidean signature. This integral can be evaluated explicitly in dimensional regularization, \cite{Bzowski:2020lip}, and its divergence reads
\begin{align} \label{ex_div}
p_{1 \mu} \lla j_A^\mu (\bs{p}_1) j_A^\nu(\bs{p}_2) j_A^\rho(\bs{p}_3) \rra & = \frac{\I}{96 \pi^2} \epsilon^{\nu \rho\kappa\lambda} p_{2 \kappa} p_{3 \lambda}.
\end{align}
On dimensional grounds, there are no counterterms contributing to the 3-point function, at least to the parity-odd sector\footnote{Counterterms of the form $A_\mu A^\mu \partial_\nu A^\nu$, while theoretically possible, do not produce the Levi-Civita tensor.}. Thus, the result must be a physical, scheme-independent anomaly. Indeed, it is easy to check that this corresponds to $\kappa = -1/(192 \pi^2)$ in the ABJ anomaly \eqref{ABJ}.

We will now review the textbook derivation of the anomaly \eqref{ex_div}, which does not require the explicit evaluation of the entire 3-point function. When the momentum $p_{1 \mu}$ is applied to \eqref{jAjAjA}, many terms cancel and the only relevant term becomes
\begin{align} \label{pjA_int1}
p_{1 \mu} \lla j_A^\mu(\bs{p}_1) j_A^\nu(\bs{p}_2) j_A^{\rho}(\bs{p}_3) \rra & = - \frac{\I}{2} \epsilon^{\nu \rho \kappa \lambda} \int_{Eu} \frac{\D^4 \bs{k}}{(2 \pi)^4} \frac{ (k - p_1)_{\kappa} (k + p_2)_{\lambda}}{(\bs{k} - \bs{p}_1)^2 (\bs{k} + \bs{p}_2)^2}.
\end{align}
If one is allowed to shift the momenta according to $\bs{k}' = \bs{k} - \bs{p}_1$, then the resulting integral over $\bs{k}'$ is
\begin{align} \label{pjA_int2}
\int_{Eu} \frac{\D^4 \bs{k}'}{(2 \pi)^4} \frac{ k'_{\kappa} (k' - p_3)_{\lambda}}{k'^2 (\bs{k}' - \bs{p}_3)^2}
\end{align}
and vanishes when contracted with the Levi-Civita tensor. When the cut-off is introduced, the shift is certainly allowed. Assume the original integral in \eqref{pjA_int1} is cut off at about $k \sim \Lambda$. The integral in \eqref{pjA_int2} must be cut off at $| \bs{k}' + \bs{p}_1 | \sim \Lambda'$ instead of $k' \sim \Lambda$. Since the integral is linearly divergent, the shift in the integration limits leads to a finite contribution, as $\Lambda' - \Lambda \sim p_1$.

This is how the anomaly emerges in cut-off regularization followed by the renormalization method using explicit counterterms. However, in case of the ABJ anomaly there are no counterterms available, so the anomaly must always emerge from the regularization. In the case of cut-off regularization, the anomaly appears due to the shift in the integration variable. The emergence of the anomaly in dimensional regularization is also well-understood, \cite{tHooft:1972tcz}.

The issue of momentum shifting can be entirely avoided in Pauli-Villars renormalization. Indeed, any constant or divergent term emerging from the momentum shifts is canceled by the identical ghost term present in the large mass term in \eqref{PVdef}. Hence the two integrals in \eqref{pjA_int1} and \eqref{pjA_int2} are equal in the Pauli-Villars regularization. This, however, does not imply that the anomaly vanishes. Instead, some additional, mass-dependent terms will now be present. One can think that the emergence of the anomaly was moved from the regularization phase to the renormalization phase. This happens because the ability to freely shift the momenta in the integrand comes at the cost of the need to consider correlation functions in the massive theory. The anomaly emerges from the infinite mass terms in \eqref{PVdef}.

In the case of the ABJ anomaly, we have to analyze the massive theory where additional, mass-dependent terms arise. Indeed, the Lagrangian of the massive theory is
\begin{align}
& L_m = L_0 + \Delta L_m, && \Delta L_m = - \frac{m}{2} ( \psi \psi + \bpsi \bpsi),
\end{align} 
where $L_0$ is the massless Lagrangian \eqref{lag0}. The axial current is no longer conserved in the massive theory, but instead we find
\begin{align} \label{pjA_m}
& \partial_\mu j_A^\mu = \I P, && P = \frac{m}{2} ( \bpsi \bpsi - \psi \psi),
\end{align}
up to anomaly. While the contribution from the insertion of $P$ vanishes in the $m \rightarrow 0$ limit, the contribution from $M \rightarrow \infty$ limit in \eqref{PVdef} may be non-vanishing. Indeed, in the massive theory we find
\begin{align} \label{Pintegral}
& p_{1 \mu} \lla j_A^\mu(\bs{p}_1) j_A^\nu(\bs{p}_2) j_A^\rho(\bs{p}_3) \rra_m = \lla P(\bs{p}_1) j_A^\nu(\bs{p}_2) j_A^\rho(\bs{p}_3) \rra_m = \nn\\
& \qquad = - \frac{\I m^2}{16 \pi^2} \epsilon^{\nu\rho\kappa\lambda} p_1^\lambda \int_0^1 \D x_1 \int_0^{1-x_1} \D x_2 \frac{2 x_2 p_1^\kappa + p_2^\kappa - 2 x_1 p_2^\kappa}{\Delta_0(x_1, x_2) + m^2},
\end{align}
where
\begin{align}
& \Delta_0 = x_1 x_2 p_3^2 + x_2 x_3 p_1^2 + x_3 x_1 p_2^2, && x_3 = 1 - x_1 - x_2.
\end{align}
This is a finite integral, but we still have to apply the Pauli-Villars regularization scheme,~\textit{i.e.},
\begin{align} \label{ABJ_P}
\lla P(\bs{p}_1) j_A^\nu(\bs{p}_2) j_A^\rho(\bs{p}_3) \rra_{PV} & = \lim_{m\rightarrow 0} \lla P(\bs{p}_1) j_A^\nu(\bs{p}_2) j_A^\rho(\bs{p}_3) \rra_{m} \nn\\
& \qquad - \lim_{M \rightarrow \infty} \lla P(\bs{p}_1) j_A^\nu(\bs{p}_2) j_A^\rho(\bs{p}_3) \rra_{M}.
\end{align}
The first term vanishes, but the second does not, producing the anomaly. The result matches \eqref{ex_div}.

Clearly, if the operator $P$ in \eqref{pjA_m} were vanishing in the massive theory, there would be no anomaly. In other words if the axial current $j_A^\mu$ had an extension $j_{Am}^\mu$ to the massive theory in such a way that $\partial_\mu j_{Am}^\mu = 0$, there would be no anomaly. This is the main advantage of the Pauli-Villars regularization method: the anomaly emerges if the massive theory breaks the symmetry generating the conserved current. In the next section we will discuss this statement in more detail. 

Before we move on let us comment on the implementation of the PV regulator described above. The attentive reader will have noticed that formally \eqref{pjA_m} vanishes for commuting field variables. To fix this issue one can instead introduce two commuting Weyl fermions with a Dirac mass term and one anticommuting massive Majorana fermion as regulators.\footnote{A related issue arises when using PV fields to regularize supersymmetric Wess-Zumino models~\cite{Katsianis:2020hzd}.}  Their respective statistics was chosen to fulfill \eqref{PVsumCond}. Furthermore in accordance with the global symmetries all of these massive fermions will have the same charges under the flavor $U(1)$ and so will contribute to $P$.
 In the large mass limit the contribution of the Dirac fermion will be twice that of Weyl fermion given by the second line of \eqref{Pintegral}. Therefore we can readily verify that adding up the contributions of the regulators (weighted by their corresponding statistics factors) we get $-2+1 \times$\eqref{Pintegral}, which is exactly the second line of \eqref{ABJ_P}. This justifies the use of \eqref{pjA_m} at the diagrammatic level. A similar argument holds for computation of any dimensionless quantity (eg. anomalies) so we will stick to a single chiral PV regulator for the purposes of section \ref{sec:WZ_model}.

\subsection{Trace anomaly} \label{sec:trace_an}

The axial current discussed in the previous section is anomalous since the axial symmetry cannot be extended to the massive theory. As the opposite case we will consider trace and transverse anomalies in 2-dimensional free scalar theory. 

We start with a free, massless, real scalar $\j$ with Lagrangian $L_0 = - \frac{1}{2} \partial_\mu \j \partial^\mu \j$. The stress tensor $T^{(0)}_{\mu\nu}$ reads
\begin{align}
& T_{\mu\nu}^{(0)} = \hat{\pi}_{\mu\nu}^{\alpha\beta} \partial_\alpha \j \partial_\beta \j, && \hat{\pi}_{\mu\nu}^{\alpha\beta} = \frac{1}{2} ( \delta_\mu^\alpha \delta_\nu^\beta + \delta_\mu^\beta \delta_\nu^\alpha - \eta_{\mu\nu} \eta^{\alpha\beta} ). \label{pihat}
\end{align}
Classicaly, the stress tensor is both conserved and traceless,
\begin{align}
& T_{\mu}^{(0)\mu} = 0, && \partial^\mu T^{(0)}_{\mu\nu} = \Box \j \, \partial_\nu \j.
\end{align}
Here we want to calculate its 2-point function, $\< T_{\mu\nu}^{(0)} T_{\rho\sigma}^{(0)} \>$ and show the textbook result: the 2-point function is conserved but exhibits trace (conformal) anomaly.

In order to calculate the 2-point function, we regulate the theory by adding the usual mass term to the Lagrangian,
\begin{align}
& L_m = L_0 + \Delta L_m, && \Delta L_m = - \frac{1}{2} m^2 \j^2.
\end{align}
In the massive theory we have two natural choices for the stress tensor: the original $T_{\mu\nu}^{(0)}$ and the actual stress tensor $T_{\mu\nu}^{(m)}$ of the massive theory,
\begin{align}
& T_{\mu\nu}^{(m)} = T_{\mu\nu}^{(0)} + \Delta T_{\mu\nu}^{(m)}, && \Delta T_{\mu\nu}^{(m)} = - \frac{1}{2} m^2 \eta_{\mu\nu} \j^2. \label{correction}
\end{align}
Both become $T_{\mu\nu}^{(0)}$ in the massless limit.

We will show now that the choice of extension of the massless stress tensor to the massive theory determines which anomalies appear. For example, notice that in 2 dimensions $\eta^{\mu\nu} \hat{\pi}_{\mu\nu}^{\alpha\beta} = 0$ and therefore $\lla T_{\mu\nu}^{(0)} (\bs{p}) T_{\rho\sigma}^{(0)}(-\bs{p}) \rra_m$ must remain traceless both in the massive theory and when the $m \rightarrow 0$ limit is taken. We conclude that the massless theory has no trace (conformal) anomaly! This seems to go against everything we know about conformal field theories.

To see what is going on here let us compute the 2-point function $\lla T_{\mu\nu}^{(0)} (\bs{p}) T_{\rho\sigma}^{(0)}(-\bs{p}) \rra$ exactly. This correlation function, evaluated in the Pauli-Villars regularization scheme in 2 dimensions for the theory of a free Weyl fermion was studied in the context of gravitational anomalies, \cite{AlvarezGaume:1983ig,Bardeen:1984pm,Tomiya:1985br}. Gravitational anomalies, which we will also call transverse anomalies, result in the violation of the conservation of the stress tensor. Genuine gravitational anomalies, which cannot be removed by local counterterms, appear in chiral theories only and are given by topological terms such as the Pontrjagin density term in \eqref{eq:ABJSugra}. Here, however, we consider a free scalar field, which is non-chiral. Thus, we expect that the only gravitational anomalies present are removable by counterterms. Such counterterms, however, will break conformal invariance and result in conformal anomalies.

Our subsequent discussion follows that of \cite{Bertlmann:2000da}, where the 2-point function is calculated for the Weyl fermions in the Pauli-Villars scheme within dimensional regularization. Unlike there, however, we will study the scalar field and use  cut-off regularization.

\subsubsection{2-point function}

In the massive theory we have
\begin{align} \label{2pt0}
& \lla T_{\mu\nu}^{(0)} (\bs{p}) T_{\rho\sigma}^{(0)}(-\bs{p}) \rra_m = - 2 \I \hat{\pi}_{\mu\nu}^{\alpha\beta} \hat{\pi}_{\rho\sigma}^{\gamma\delta} \int_{Eu} \frac{\D^2 \bs{k}}{(2 \pi)^2} \frac{k_\alpha (k - p)_\beta (k - p)_\gamma k_\delta}{(k^2 + m^2) \left( ( \bs{p}-\bs{k})^2 + m^2 \right)}.
\end{align}
We use the standard Feynman parameterization with $x$ denoting the Feynman parameter,
\begin{align}
& \lla T_{\mu\nu}^{(0)}(\bs{p}) T_{\rho\sigma}^{(0)}(-\bs{p}) \rra_m = - 2 \I \hat{\pi}_{\mu\nu}^{\alpha\beta} \hat{\pi}_{\rho\sigma}^{\gamma\delta} \int_0^1 \D x \int_{Eu} \frac{\D^2 \bs{l}}{(2 \pi)^2} \frac{\text{num}_{\alpha\beta\gamma\delta}}{(l^2 + \Delta)^2},
\end{align}
where the relevant terms containing an even number of $l$'s in the numerator read,
\begin{align}
\text{num}_{\alpha\beta\gamma\delta} & = l_{\alpha} l_{\beta} l_{\gamma} l_{\delta} + x^2 l_{\beta} l_{\gamma} p_{\alpha} p_{\delta} + (x - 1)^2 l_{\alpha} l_{\delta} p_{\beta} p_{\gamma} \nn\\
& \qquad + ( l_{\alpha} l_{\beta} p_{\gamma} p_{\delta} + l_{\alpha} l_{\gamma} p_{\beta} p_{\delta} + l_{\beta} l_{\delta} p_{\alpha} p_{\gamma} + l_{\gamma} l_{\delta} p_{\alpha} p_{\beta} ) x (x - 1) \nn\\
& \qquad + x^2 (x - 1)^2 p_{\alpha} p_{\beta} p_{\gamma} p_{\delta}.
\end{align}
Furthermore, $\Delta = x(1-x) p^2 + m^2$.

Integrals with two and four $\bs{l}$'s in the numerator are UV divergent and require a regulator. By cutting the integrals off at $l = \Lambda$ we find
\begin{align}
\int_{Eu}^{\Lambda} \frac{\D^2 \bs{l}}{(2 \pi)^2} \frac{1}{(l^2 + \Delta)^2} & = \frac{1}{4 \pi \Delta} + O(\Lambda^{-1}), \label{lint1} \\
\int_{Eu}^{\Lambda} \frac{\D^2 \bs{l}}{(2 \pi)^2} \frac{l^2}{(l^2 + \Delta)^2} & = \frac{\log \Lambda}{2 \pi} - \frac{1}{4\pi} (1 + \log \Delta) + O(\Lambda^{-1}), \\
\int_{Eu}^{\Lambda} \frac{\D^2 \bs{l}}{(2 \pi)^2} \frac{l^4}{(l^2 + \Delta)^2} & = \frac{\Lambda^2}{4 \pi} - \frac{\Delta}{\pi} \log \Lambda + \frac{\Delta}{4\pi} (1 + 2 \log \Delta) + O(\Lambda^{-1}). \label{lint3}
\end{align}
Next, integrals over Feynman parameters can be carried out explicitly. This results in a regulated 2-point function, which is not particularly interesting, but depends both on $m$ and $\Lambda$.

Now the 2-point function can be renormalized in the Pauli-Villars scheme defined in equation \eqref{PVdef}. Effectively, the renormalization amounts to the subtraction of all divergences higher than logarithmic in \eqref{lint1} - \eqref{lint3} while simultaneously replacing $\log \Lambda$ with $\log m$. For example, the renormalized value of of the integral \eqref{lint3} is
\begin{align}
\int_{PV} \frac{\D^2 \bs{l}}{(2 \pi)^2} \frac{l^4}{(l^2 + \Delta)^2} & = \frac{\Delta}{4\pi} \left[ 1 + 2 \log \left( \frac{\Delta}{m^2} \right) \right].
\end{align}
The quadratic and logarithmic divergences are cured and the massless limit can be taken. The 2-point function reads
\begin{align} \label{PV0}
& \lla T_{\mu\nu}^{(0)}(\bs{p}) T_{\rho\sigma}^{(0)}(-\bs{p}) \rra_{PV} = - \frac{\I p^2}{24 \pi} \left[ \log \left( \frac{p^2}{m^2} \right) - \tfrac{8}{3} \right] \Pi_{\mu\nu\rho\sigma}(\bs{p}) \nn\\
& \qquad\qquad - \frac{\I}{48 \pi p^2} \left( p^2 \eta_{\mu\nu} - 2 p_\mu p_\nu \right) \left( p^2 \eta_{\rho\sigma} - 2 p_\rho p_\sigma \right),
\end{align}
where $\hat{\pi}_{\mu\nu\rho\sigma} = \eta_{\rho\alpha} \eta_{\sigma\beta} \hat{\pi}_{\mu\nu}^{\alpha\beta}$ and $\Pi_{\mu\nu\rho\sigma}$ is the transverse-traceless projector, which in $d$ dimensions reads
\begin{align}
\Pi_{\mu\nu\rho\sigma}(\bs{p}) & = \pi_{\mu(\rho}(\bs{p}) \pi_{\sigma)\nu}(\bs{p}) - \frac{1}{d-1} \pi_{\mu\nu}(\bs{p}) \pi_{\rho\sigma}(\bs{p}), \\
\pi_{\mu\nu}(\bs{p}) & = \eta_{\mu\nu} - \frac{p_\mu p_\nu}{p^2}. \label{pi}
\end{align}
It seems that the 2-point function contains a logarithmic term. This confusing fact can be explained by the degeneracy of tensor structures. Using the argument of \cite{Bzowski:2017poo} we will argue that the degeneracy of tensor structures in $d=2$ implies $\Pi_{\mu\nu\rho\sigma} = 0$ identically.

In general, having $d$ linearly independent vectors in $d$ (Euclidean) dimensions the metric $\delta^{\mu\nu}$ is not an independent tensor. Having $d-1$ linearly independent vectors $\bs{p}_1, \ldots, \bs{p}_{d-1}$ in $d$ (Euclidean) dimensions, one can find the unique (up to normalization) orthogonal vector, $n^\mu = \epsilon^{\mu \mu_1 \ldots \mu_{d-1}} p_{\mu_1} \ldots p_{\mu_{d-1}}$. In 2 dimensions this means that we can define $n^\mu$ as orthogonal to $p^\mu$ and then back in Lorentzian signature we have
\begin{equation}
n^\mu = \I \epsilon^{\mu\alpha} p_{\alpha}, \qquad\qquad p^2 \eta^{\mu\nu} = n^\mu n^\nu + p^\mu p^\nu.
\end{equation}
When substituting the expression for $\eta^{\mu\nu}$ back to $\Pi_{\mu\nu\rho\sigma}$ we see that the projector vanishes. Thus, the logarithmic term disappears and the 2-point function becomes
\begin{align} \label{PV0d2}
\lla T_{\mu\nu}^{(0)}(\bs{p}) T_{\rho\sigma}^{(0)}(-\bs{p}) \rra_{PV} & = - \frac{\I}{48 \pi p^2} \left( p^2 \eta_{\mu\nu} - 2 p_\mu p_\nu \right) \left( p^2 \eta_{\rho\sigma} - 2 p_\rho p_\sigma \right) \nn\\
& = - \frac{\I}{48 \pi p^2} \left( n_\mu n_\nu - p_\mu p_\nu \right) \left( n_\rho n_\sigma - p_\rho p_\sigma \right).
\end{align}

\subsubsection{Ward identities, compensators, and counterterms}

Since in two dimensions $\hat{\pi}_{\mu\nu}^{\alpha\beta}$ defined in \eqref{pihat} is a projector on traceless tensors, the 2-point function \eqref{PV0d2} is manifestly traceless. On the other hand, since $\bs{p} \cdot \bs{n} = 0$ we get
\begin{align}
\lla T_{\mu}^{(0) \mu}(\bs{p}) T_{\rho\sigma}^{(0)}(-\bs{p}) \rra_{PV} & = 0, \\
p^\mu \lla T_{\mu\nu}^{(0)}(\bs{p}) T_{\rho\sigma}^{(0)}(-\bs{p}) \rra_{PV} & = \frac{\I}{48 \pi} p_\nu \left( p^2 \eta_{\rho\sigma} - 2 p_\rho p_\sigma \right). \label{trans_am}
\end{align}
The trace (conformal) anomaly is absent. Instead we find a transverse (gravitational) anomaly: the stress tensor is not conserved! The anomaly is local, but non-vanishing\footnote{Note that this is not the chiral gravitational anomaly as discussed in \cite{AlvarezGaume:1983ig,Bardeen:1984pm,Bertlmann:2000da}, since the theory is not chiral.}.

While \eqref{PV0d2} violates conservation of the stress tensor, we can shuffle the anomaly around by adding a suitable finite counterterm. This counterterm necessarily violates general covariance exactly in such a way that it restores conservation.

First notice that, among all tensors of rank 4 and of mass dimension 2 built up with the vector $\bs{p}$ and the metric $\eta_{\mu\nu}$, only a single term, $p_\mu p_\nu p_\rho p_\sigma / p^2$, is non-local. Terms such as $p_\mu p_\nu \eta_{\rho \sigma}$ and $p^2 \eta_{\mu\nu} \eta_{\rho \sigma}$ are local and hence can be adjusted at will by a counterterm.

We want to add the following contribution to \eqref{PV0d2} to make it transverse,
\begin{align} \label{Deltafin}
\Delta_{\text{fin}} & = \frac{\I}{24 \pi} \left( \eta_{\mu\nu} p_\rho p_\sigma + p_\mu p_\nu \eta_{\rho\sigma} \right) - \frac{\I}{16 \pi} p^2 \eta_{\mu\nu} \eta_{\rho\sigma}.
\end{align}
Let $h^{\mu\nu}$ denote the source for $T_{\mu\nu}^{(0)}$. In the massive theory $h^{\mu\nu}$ is not exactly equal to the metric $g^{\mu\nu}$, since $T_{\mu\nu}^{(0)}$ is not the true stress tensor. For this reason, we treat $h_{\mu\nu}$ as a linear coupling and $T^{(0)}_{\mu\nu} = 2 \, \delta S/\delta h^{\mu\nu}$, where we set $h_{\mu\nu} = 0$ after the derivatives are taken. The finite counterterm producing this contribution to the 2-point function reads
\begin{align} \label{Sfin_conf}
S_{\text{fin}} = \int \D^2 \bs{x} \left[ \frac{1}{96 \pi} h^{\rho}_{\rho} \partial_\mu \partial_\nu h^{\mu\nu} - \frac{1}{128 \pi} h_{\mu\nu} \partial^2 h^{\mu\nu} \right],
\end{align}
where raising and lowering is carried out with $\eta_{\mu\nu}$. This counterterm cannot be obtained from a covariant term containing the actual metric in place of $h^{\mu\nu}$. Indeed, the only such term would be the ``improvement term" proportional to $\sqrt{g} R$, where $R$ is the Ricci scalar. The variation of $\sqrt{g} R$, however, is transverse and hence $\sqrt{g} R$ cannot reproduce $S_{\text{fin}}$.

With the contribution from this counterterm added to \eqref{PV0d2}, the 2-point function becomes
\begin{align} \label{PV0d2ct}
\lla T_{\mu\nu}^{(0)}(\bs{p}) T_{\rho\sigma}^{(0)}(-\bs{p}) \rra'_{PV} & = - \frac{\I}{12 \pi p^2} \left( p^2 \eta_{\mu\nu} - p_\mu p_\nu \right) \left( p^2 \eta_{\rho\sigma} - p_\rho p_\sigma \right) \nn\\
& = - \frac{\I}{12 \pi p^2} n_\mu n_\nu n_\rho n_\sigma.
\end{align}
We put a prime on the 2-point function to indicate the inclusion of the finite counterterm contribution. The 2-point function is now manifestly conserved but
\begin{align}
\lla T_{\mu}^{(0) \mu}(\bs{p}) T_{\rho\sigma}^{(0)}(-\bs{p}) \rra'_{PV} & = - \frac{\I}{12 \pi} \left( p^2 \eta_{\rho\sigma} - p_\rho p_\sigma \right), \\
p^\mu \lla T_{\mu\nu}^{(0)}(\bs{p}) T_{\rho\sigma}^{(0)}(-\bs{p}) \rra'_{PV} & = 0.
\end{align}
We obtain the standard form of the trace (conformal) anomaly, while the transverse (gravitational) anomaly is absent. By adding the counterterm \eqref{Sfin_conf} we killed the transverse anomaly, but generated the trace anomaly. We have chosen ``a lesser evil" by keeping the stress tensor conserved on quantum level, but with non-vanishing trace.

Equivalently, one can achieve the same effect by introducing a suitable compensator. This realizes the idea described in appendix B of \cite{Kuzenko:2019vvi}. First, we assign transformation properties of the source $h_{\mu\nu}$ under diffeomorphisms and Weyl transformations,
\begin{align}
	& \delta_{\xi} h_{\mu\nu} = - 2 \eta_{\alpha (\mu} \partial_{\nu)} \xi^\alpha, && \delta_{\sigma} h_{\mu\nu} = - 2 \sigma h_{\mu\nu}.
\end{align}
The associated Ward identities read
\begin{align}
	& \partial_{\nu} \< T^{(0)\mu\nu} \> = \mathcal{A}^{\mu}_{\xi}, && \eta_{\mu\nu} \< T^{(0)\mu\nu} \> = 0.
\end{align}
The trace Ward identity is exact and anomaly-free, while the transverse Ward identity is anomalous. From \eqref{trans_am} the anomaly reads
\begin{align}
	\mathcal{A}^{\mu}_{\xi} = \frac{1}{96 \pi} \partial^{\mu} \left( \partial^2 h_{\alpha}^{\alpha} - 2 \partial_\alpha \partial_\beta h^{\alpha \beta} \right) + O(h^2).
\end{align}

In order to move the anomaly from the transverse to the trace Ward identity, one can use the counterterm \eqref{Sfin_conf}. Here, however, we want to consider introducing compensators. Since we want to break conformal symmetry, while keeping the stress tensor conserved, we introduce a compensator $\tau$ for the Weyl symmetry. We assign the following transformations to $\tau$,
\begin{align}
	& \delta_{\sigma} \tau = \sigma, && \delta_{\xi} \tau = - \frac{1}{2} \partial_\mu \xi^\mu.
\end{align}
Clearly, $\tau$ is the compensator for the Weyl symmetry. The transformation under $\xi^\mu$ is such that $\gamma_{\mu\nu} = h_{\mu\nu} + 2 \tau \eta_{\mu\nu}$ satisfies $\eta^{\mu\nu} \delta_{\xi} \gamma_{\mu\nu} = 0$. In other words, $\gamma_{\mu\nu}$ is the first order term in the expansion of the metric, $\hat{g}_{\mu\nu} = \eta_{\mu\nu} + \gamma_{\mu\nu} + O(\gamma^2)$ defined as $\hat{g}_{\mu\nu} = e^{2 \tau} g_{\mu\nu}$.  The metric $\hat{g}_{\mu\nu}$, and thus $\gamma_{\mu\nu}$, are inert under Weyl transformations, $\delta_{\sigma} \gamma_{\mu\nu} = 0$. This corresponds to old minimal SUGRA fields being inert under $R$-symmetry transformations.

The compensator $\tau$ should couple to the trace of the stress tensor. This is easy to achieve in the massive theory, where $\tau$ couples (linearly) to $T = - m^2 \j^2$. In the massless theory this operator becomes null, in the same sense as $j_{\chi}$ becomes null. Using equations of motion, we can define $T$ without involvement of the mass as $T = - \j \, \partial^2 \j$. Just like $j_{\chi}$, it is proportional to the equations of motion.

Now, varying both $h_{\mu\nu}$ and the compensator, the Ward identities read
\begin{align}
	& \partial_{\nu} \< T^{(0)\mu\nu} \> + \frac{1}{2} \partial^\mu \< T \> = 0, && \eta_{\mu\nu} \< T^{(0)\mu\nu} \> = - \< T \>.
\end{align}
The transverse Ward identity becomes anomaly-free. Indeed, by direct calculations one finds
\begin{align}
	\lla T(\bs{p}) T_{\rho\sigma}^{(0)}(-\bs{p}) \rra_{PV} & = - \lim_{M \rightarrow \infty} \lla - M^2 \j^2 (\bs{p}) \: T_{\rho\sigma}^{(0)} (-\bs{p}) \rra_M \nn\\
	& = - \frac{\I}{48 \pi} (p^2 \eta_{\rho\sigma} - 2 p_{\rho} p_{\sigma}).
\end{align}
The compensating term took place of the anomaly, $\partial^\mu \< T \> = - 2 \mathcal{A}_{\xi}^\mu$. On the other hand, the trace Ward identity is now anomalous. Since $\tau$ is the compensator for Weyl transformations, the expectation value of the operator it sources, $T$, represents the anomaly. Notice also that while $\gamma_{\mu\nu}$ is inert under Weyl transformations, it couples (linearly) to $T^{(m)}_{\mu\nu}$. The trace anomaly of this stress tensor is therefore expressible entirely in terms of the operator which couples to the compensator~$\tau$.

As we can see, the role of the compensator is not to remove the anomaly entirely, but to move it from one Ward identity to another one. The anomaly cannot be removed. It can, however, be removed from either one Ward identity and moved to the other one. Since the conservation of the stress tensor is more fundamental and holds in massive theories as well, it is natural to keep the anomalous terms in the trace.

\subsubsection{Massive stress tensor} \label{sec:massive_tensor}

Instead of using $T_{\mu\nu}^{(0)}$ in the regulated theory, we could have been using the actual stress tensor of the massive theory, $T_{\mu\nu}^{(m)}$. This tensor is conserved and hence using it as a regulator of the massless theory should yield the conserved 2-point function, free of the transverse (gravitational) anomaly. Let us check this claim.

Since $T_{\mu\nu}^{(m)}$ is the sum of the previously analyzed $T_{\mu\nu}^{(0)}$ and the ``correction" in \eqref{correction}, we have to calculate two more correlators,
\begin{align}
& \lla T_{\mu\nu}^{(0)}(\bs{p}) \Delta T_{\rho\sigma}^{(m)}(-\bs{p}) \rra_m = \I m^2 \hat{\pi}_{\mu\nu}^{\alpha\beta} \eta_{\rho\sigma} \int_{Eu} \frac{\D^2 \bs{k}}{(2 \pi)^2} \frac{k_\alpha (k - p)_\beta}{(k^2 + m^2) \left( (\bs{p}-\bs{k})^2 + m^2 \right)}, \\
& \lla \Delta T_{\mu\nu}^{(m)}(\bs{p}) \Delta T_{\rho\sigma}^{(m)}(-\bs{p}) \rra_m = - \frac{\I}{2} m^4 \eta_{\mu\nu} \eta_{\rho\sigma} \int_{Eu} \frac{\D^2 \bs{k}}{(2 \pi)^2} \frac{1}{(k^2 + m^2) \left( (\bs{p}-\bs{k})^2 + m^2 \right)}.
\end{align}

As before, we compute these correlators in the massive theory with a cut-off $\Lambda$ and sum everything together to get the regulated $\lla T_{\mu\nu}^{(m)}(\bs{p}) T_{\rho\sigma}^{(m)}(-\bs{p}) \rra$ in the massive theory. We then apply \eqref{PVdef}. After subtracting the divergences according to the Pauli-Villars regularization procedure the additional contribution from the ``correction term" is
\begin{align} \label{PVcorr}
\lla T_{\mu\nu}^{(0)}\Delta T_{\rho\sigma}^{(m)} \rra_{PV} + \lla \Delta T_{\mu\nu}^{(m)} T_{\rho\sigma}^{(0)} \rra_{PV} + \lla \Delta T_{\mu\nu}^{(m)} \Delta T_{\rho\sigma}^{(m)} \rra_{PV} = \Delta_{\text{fin}},
\end{align}
where $\Delta_{\text{fin}}$ is exactly equal to the correction term \eqref{Deltafin}. We end up with the 2-point function of the conserved stress tensor in the massless theory, \eqref{PV0d2ct},
\begin{align} \label{PVmd2}
\lla T_{\mu\nu}^{(m)}(\bs{p}) T_{\rho\sigma}^{(m)}(-\bs{p}) \rra_{PV} & = \lla T_{\mu\nu}^{(0)}(\bs{p}) T_{\rho\sigma}^{(0)}(-\bs{p}) \rra'_{PV}
\end{align}
In deriving this expression we used the actual stress tensor $T_{\mu\nu}^{(m)}$ in the regulated, massive theory. Since the stress tensor is conserved in the regulated theory, the resulting 2-point function is conserved as well. Instead, the 2-point function exhibits the scaling anomaly,
\begin{align}
\lla T_{\mu}^{(m) \mu}(\bs{p}) T_{\rho\sigma}^{(m)}(-\bs{p}) \rra_{PV} & = - \frac{\I}{12 \pi} \left( p^2 \eta_{\rho\sigma} - p_\rho p_\sigma \right), \\
p^\mu \lla T_{\mu\nu}^{(m)}(\bs{p}) T_{\rho\sigma}^{(m)}(-\bs{p}) \rra_{PV} & = 0.
\end{align}
As we can see, the choice of the regularization of the operator dictates which symmetries become anomalous.

\subsubsection{Summary}

Consider a massless theory with a conserved current $j_0^\mu$. If the current can be extended to a conserved current $j_m^\mu$ in the massive theory in such a way that $j_m^\mu = j_0^\mu + O(m)$, then the corresponding symmetry is non-anomalous. This argument is usually invoked for the vector current, \cite{Bertlmann:1996xk}.

In general, consider any Ward identity of the form $\partial_\mu \< j_0^\mu \ldots \> = \text{rhs}$, where $\ldots$ denotes the insertion of any operators. If the current $j_0^\mu$ can be extended to the massive theory and the Ward identity holds in the massive theory with the same right hand side, then the anomaly is absent. If, on the other hand, the right hand side in the massive theory gets modified by mass-dependent terms, the anomaly may emerge.

In particular, this argument holds for the conservation of the stress tensor and the supersymmetry currents. In the previous section we have shown that by choosing $T^{(m)}_{\mu\nu}$ -- the true stress tensor of the massive theory -- as the massive extension of the stress tensor of the massless theory, the gravitational anomaly is eliminated. Analogously, if the massive theory is supersymmetric, there is no anomaly in the conservation of the supercurrent $j^{\mu\alpha}_{Q}, \bj^{\mu \da}_{\bQ}$ in the massless theory. It is, however, necessary to pick the actual conserved supercurrent of the massive theory as the massive extension of the massless supercurrent. Furthermore, if other operators related to each other by supersymmetry are involved, their extensions to the massive theory must also obey the SUSY algebra. This argument is sufficient to argue that there is no SUSY anomaly in the massless Wess-Zumino model. We will present detailed checks on this statement in the following section.

\section{Wess-Zumino model} \label{sec:WZ_model}

In this section we present various consistency checks in a free Wess-Zumino model. We are interested in anomalies in the massless, superconformal theory. However, as we will be employing the Pauli-Villars renormalization method, we are forced to consider the massive Wess-Zumino model as well. The action of the massive model reads
\begin{align}
\label{Wzfreem}
S = \int \D^4 \bs{x} \left[ - \partial_\mu \j^\ast \partial^\mu \j - \I \bpsi \bsigma^\mu \partial_\mu \psi + F F^\ast + m \left( \j F + \j^\ast F^\ast - \frac{1}{2} ( \psi \psi + \bar{\psi} \bar{\psi} ) \right) \right].
\end{align}
We couple this model to an external vector multiplet. The coupled action is
\begin{align} \label{Sm}
S_V = \int \D^4 \bs{x} \D^4 \theta \left[ \Phi^+ e^V \Phi + \frac{1}{2} m ( \Phi \Phi + \Phi^+ \Phi^+) \right],
\end{align}
where $\Phi = (\j, \psi, F)$ is the dynamical chiral multiplet \eqref{chiral_multi} and $V$ is the vector multiplet of sources, \eqref{vector_multi}. The component fields of the vector multiplet source corresponding operators. These are as follows:
\begin{align} \label{ops}
\begin{array}{ccccccccccc}
C && \chi^\alpha && \mathcal{M} && A_\mu && \lambda^\beta && D \\
j_C && j_{\chi \alpha} && j_{\mathcal{M}} && j^\mu_A && j_{\lambda \beta} && j_D
\end{array}
\end{align}
Note that the vector multiplet of sources couples to the mass-independent part only. This means that the operators sourced by the component fields are identical both in the massive and massless theory. The relevant operators for us are
\begin{align} \label{text_ops_first}
j_A^\mu & = \frac{\I}{2} \left( \j^\ast \partial^\mu \j - \j \partial^\mu \j^\ast \right) + \frac{1}{2} \bar{\psi} \bar{\sigma}^\mu \psi, \\
j_{\lambda \beta} & = \frac{\I}{\sqrt{2}} \j^\ast \psi_{\beta}, \\
j_{D} & = \frac{1}{2} \j \j^\ast, \\
j_{\chi \alpha} & = \frac{1}{\sqrt{2}} \left( \j \sigma^\mu_{\a \da} \partial_\mu \bar{\psi}^{\da} - \I F^\ast \psi_{\alpha} \right), \\
j_{\mathcal{M}} & = \frac{\I}{2} F^\ast \j. \label{text_ops_last}
\end{align}
The operators listed come from the linear coupling of the component fields of the vector multiplet to the dynamical chiral multiplet: the Lagrangian in \eqref{Sm} contains also seagull terms. The fully expanded Lagrangian is given in \eqref{lag}.

\subsection{Supercurrents} \label{sec:WZsupercurrents}

The supercurrents $j_{Q \alpha}^\mu$ and $\bar{j}_{\bQ}^{\mu \dot{\alpha}}$ can be defined by the Noether theorem as
\begin{align}
\delta_{\ep, \bep} S = - \int \D^4 \bs{x} \left[ \partial_\mu \ep^{\a} \: j_{Q \a}^\mu + \partial_\mu \bep_{\da} \: \bj_{\bQ}^{\mu \da} \right].
\end{align}
The form of the supercurrents depends on the mass term and it will be useful to split the massless and massive contributions. We define the currents $j_{Q \alpha}^\mu$ and $\bar{j}_{\bQ}^{\mu \dot{\alpha}}$ as SUSY currents in the massless theory, while $j_{Q m \alpha}^\mu$ and $\bar{j}_{\bQ m}^{\mu \dot{\alpha}}$ denote supercurrents in the massive theory. With the variations in \eqref{susy_chiral_first} - \eqref{susy_chiral_last} one finds
\begin{align} \label{jQ}
j_{Q m \alpha}^\mu & = j_{Q \alpha}^\mu + \Delta j_{Q \alpha}^{\mu}, & \bar{j}_{\bQ m}^{\mu \dot{\alpha}} & = \bar{j}_{\bQ}^{\mu \dot{\alpha}} + \Delta \bar{j}_{\bQ}^{\mu \dot{\alpha}},
\end{align}
where
\begin{align}
j_{Q \alpha}^\mu & = - \sqrt{2} \, \partial_\kappa \j^\ast \, ( \sigma^\kappa \bar{\sigma}^\mu)_{\alpha}^{\ \beta} \psi_\beta, & \bar{j}_{\bQ}^{\mu \dot{\alpha}} & = - \sqrt{2} \, \partial_\kappa \j \, ( \bar{\sigma}^{\kappa} \sigma^{\mu})^{\da}_{\ \db} \bpsi^{\db}, \label{supercurrents} \\
\Delta j_{Q \alpha}^{\mu} & = m \I \sqrt{2} \j^\ast \sigma^\mu_{\alpha \da} \bpsi^{\da}, &
\Delta \bar{j}_{\bQ}^{\mu \dot{\alpha}} & = m \I \sqrt{2} \, \j \bsigma^{\mu \da \a} \psi_{\a}. \label{corrections}
\end{align}
 In general, however, the supercurrents obtained from coupling to supergravity contain additional improvement terms, analogous -- and related via supersymmetry -- to improvement terms for the stress tensor. In $d=4$ the improved supercurrent is
\begin{align}
\bar{j}_{\bQ \text{imp}}^{\mu \dot{\alpha}} = - \sqrt{2} \, \partial_\kappa \j \, ( \bar{\sigma}^{\kappa} \sigma^{\mu})^{\da}_{\ \db} \bpsi^{\db} + m \I \sqrt{2} \, \j \bsigma^{\mu \da \a} \psi_{\a} + \tfrac{2}{3} \sqrt{2} (2 \bsigma)^{\kappa \mu \da}_{\ \ \ \: \db} \partial_\kappa (\j \psi^{\db}).
\end{align}
The last term is the improvement term. It has a vanishing divergence, so that it does not alter the conserved charge $\bar{Q}^{\da}$. Note that the above supersymmetry current can also be defined by coupling the theory to old-minimal supergravity classically and extracting the term linear in the gravitino\footnote{The theory in~\eqref{Wzfreem} also allows for a {\it R}-supercurrent multiplet. When $m=0$ the {\it R}-charge of the chiral multiplet is arbitrary. If $m\neq0$ there is still a  {\it R}-symmetry under which the fermions are uncharged. This {\it R}-symmetry is not anomalous, hence~\eqref{Wzfreem} can be coupled to new minimal supergravity.}. There will also be a superconformal $R$-current analogous to $j_\mu$ that can be obtained from the classical coupling to $\tilde{A}_\mu^R$ (see section \ref{sec:OMsugra} for the definition of  tilde fields). In the following analysis we will focus on the flavor anomaly so we will not write the operators relevant for section \ref{sec:sugra} explicitly. The corresponding Ward identities in the gravitational sector are completely analogous if we replace the flavor currents with $R$-current insertions. A Pauli Villars regularization scheme is used in~\cite{Katsianis:2020hzd} to analyze~\eqref{Wzfreem} coupled to background old minimal supergravity.

The massless Wess-Zumino model is superconformal and thus can be coupled to conformal supergravity. With the improvement term $S$-supersymmetry implies that the sigma-contraction of the cupercurrent vanishes on-shell. Indeed, one finds
\begin{align} \label{sigmaJQ}
\sigma_{\mu \a \da} \bar{j}_{\bQ \text{imp}}^{\mu \dot{\alpha}}|_{m=0} = 4 j_{\chi \a} + 2 \sqrt{2} \I F^{\ast} \psi_{\a},
\end{align}
where $j_{\chi \alpha}$ is sourced by the compensator field $\chi$ and given in \eqref{text_ops_last}. As promised, the operator sourced by $\chi$ is effectively equal to the gamma-contracted supercurrent. In other words $\chi$ is the compensator for $S$-supersymmetry but not for $Q$-supersymmetry.

\subsection{1- and 2-point functions}

In this section, as a warm-up exercise, let us calculate a pair of 2-point functions, $\< j_A^\mu j_A^\nu \>$ and $\< \bj_{\bQ}^{\mu \da} j_{Q \beta}^\nu \>$ in the PV regularization scheme. Before we do it, however, let us first remark that in the massless theory, in the PV scheme, all 1-point functions (of operators of positive conformal dimension) vanish. This follows from the fact that in the massive theory such 1-point functions may be non-vanishing, but must contain a positive power of the mass. All such terms therefore are subtracted in \eqref{PVdef}. Thus, in the massless theory, $\< \O \>_{PV} = 0$ for all operators other than identity. This agrees with conformality of the massless Wess-Zumino model.

Let us now consider the 2-point function. Since 1-point functions vanish in the PV scheme,
\begin{align}
& \lla J_A^\mu(\bs{p}) J_A^\nu(-\bs{p}) \rra_{PV} = \lla j_A^\mu(\bs{p}) j_A^\nu(-\bs{p}) \rra_{PV} = \nn\\
& \qquad = \lla J_{\j}^\mu(\bs{p}) J_{\j}^\nu(-\bs{p}) \rra_{PV} + \lla j_{\psi}^\mu(\bs{p}) j_{\psi}^\nu(-\bs{p}) \rra_{PV}.
\end{align}
The 2-point functions become
\begin{align}
\lla j_{\j}^\mu(\bs{p}) j_{\j}^\nu(-\bs{p}) \rra_{PV} & = - \frac{\I}{192 \pi^2} p^2 \pi^{\mu\nu} \left[ \log \left( \frac{p^2}{m^2} \right) - \tfrac{8}{3} \right], \\
\lla j_{\psi}^\mu(\bs{p}) j_{\psi}^\nu(-\bs{p}) \rra_{PV} & = - \frac{\I}{96 \pi^2} p^2 \pi^{\mu\nu} \left[ \log \left( \frac{p^2}{m^2} \right) - \tfrac{5}{3} \right] + \frac{\I}{192 \pi^2} p^2 \eta^{\mu\nu},
\end{align}
where $\pi^{\mu\nu}$ is the transverse projector defined in \eqref{pi}. While the scalar current, $j_{\j}^\mu$, is conserved, the fermionic part, $j_{\psi}^\mu$, is not. This is not unexpected: as we already now there is no conserved extension of the axial current to the massive theory. We find a removable flavor anomaly in the PV scheme. To write it down, we sum the 2-point functions above and find
\begin{align} \label{jAjA}
\lla j_A^\mu(\bs{p}) j_A^\nu(-\bs{p}) \rra_{PV} = - \frac{\I}{64 \pi^2} p^2 \left\{ \pi^{\mu\nu} \left[ \log \left( \frac{p^2}{m^2} \right) - 2 \right] - \frac{1}{3} \eta^{\mu\nu} \right\}.
\end{align}
Its divergence equals
\begin{align}
p_\mu \lla j_A^\mu(\bs{p}) j_A^\nu(-\bs{p}) \rra_{PV} = \frac{\I}{192 \pi^2} p^2 p^\nu.
\end{align}
Using \eqref{JA_ward} we read off the linear term of the flavor anomaly in the PV scheme,
\begin{align}
\mathcal{A}^{PV}_{\theta} = - \frac{1}{192 \pi^2} \partial^2 \partial_\mu A^\mu + \ldots
\end{align}
In principle, this anomaly could be removed by a finite counterterm
\begin{align}
S_{\text{fin}} = - \frac{1}{96 \pi^2} \int \D^4 \bs{x} \, A_\mu \partial^2 A^{\mu}.
\end{align}
We will not do it here, as we want to consistently apply the PV regularization scheme. It is also worth mentioning that such a form of the flavor anomaly was found in \cite{Guadagnini:1985ea} in an explicitly supersymmetric scheme.

We end this section by checking that the 2-point function of the supercurrents is conserved. This is once again consistent with the absence of the SUSY anomaly. In the massive theory the SUSY currents are given by \eqref{jQ}. As we will see now, it is essential to include the correction terms. Indeed, the full 2-point function reads
\begin{align}
& \lla \bj^{\mu \da}_{\bQ m}(\bs{p}) j^{\nu}_{Q m \beta}(-\bs{p}) \rra_{PV} = \frac{\I}{24 \pi^2} p^2 p_{\tau} \bsigma^{\tau \da \a} \epsilon_{\a \b} \pi^{\mu\nu} \left[ \log \left( \frac{p^2}{m^2} \right) - \tfrac{5}{3} \right] \nn\\
& \qquad\qquad - \frac{1}{48 \pi^2} p^2 p_{\kappa} \bsigma_{\tau}^{\da \a} \epsilon_{\a \b} \epsilon^{\kappa\tau\mu\nu} \left[ \log \left( \frac{p^2}{m^2} \right) - \tfrac{2}{3} \right]
\end{align}
and is conserved. However, the 2-point function involving the massless current in place of the massive one has a non-vanishing divergence. With the correction terms given in \eqref{corrections} one finds,
\begin{align}
& \lla \bj^{\mu \da}_{\bQ m}(\bs{p}) j^{\nu}_{Q m \beta}(-\bs{p}) \rra_{PV} = \lla \bj^{\mu \da}_{\bQ}(\bs{p}) j^{\nu}_{Q \beta}(-\bs{p}) \rra_{PV} + \left[ \lla \bj^{\mu \da}_{\bQ}(\bs{p}) \Delta j^{\nu}_{Q \beta}(-\bs{p}) \rra_{PV} \right.\nn\\
& \qquad \qquad \left. + \lla \Delta \bj^{\mu \da}_{\bQ}(\bs{p}) j^{\nu}_{Q \beta}(-\bs{p}) \rra_{PV} + \lla \Delta \bj^{\mu \da}_{\bQ}(\bs{p}) \Delta j^{\nu}_{Q \beta}(-\bs{p}) \rra_{PV} \right],
\end{align}
where the correction terms read
\begin{align}
& \lla \bj^{\mu \da}_{\bQ}(\bs{p}) \Delta j^{\nu}_{Q \beta}(-\bs{p}) \rra_{PV} = - \frac{\I}{96 \pi^2} p^2 \bsigma^{\da \a}_{\tau} \epsilon_{\alpha\beta} \left[ p^\mu \eta^{\nu\tau} - p^\nu \eta^{\mu\tau} - p^\tau \eta^{\mu\nu} - \I \epsilon^{\mu\nu\kappa\tau} p_{\kappa} \right], \\
& \lla \Delta \bj^{\mu \da}_{\bQ}(\bs{p}) \Delta j^{\nu}_{Q \beta}(-\bs{p}) \rra_{PV} = - \frac{\I}{96 \pi^2} p^2 \bsigma^{\da \a}_{\tau} \epsilon_{\alpha\beta} \left[ p^\mu \eta^{\nu\tau} + p^\nu \eta^{\mu\tau} - p^\tau \eta^{\mu\nu} - \I \epsilon^{\mu\nu\kappa\tau} p_{\kappa} \right].
\end{align}
Thus, the divergences read
\begin{align}
& p_{\mu} \lla \bj^{\mu \da}_{\bQ m}(\bs{p}) j^{\nu}_{Q m \beta}(-\bs{p}) \rra_{PV} = 0, \\
& p_{\mu} \lla \bj^{\mu \da}_{\bQ}(\bs{p}) j^{\nu}_{Q \beta}(-\bs{p}) \rra_{PV} = \frac{\I}{96 \pi^2} \bsigma_{\tau}^{\da \a} \epsilon_{\a \b} p^2 \left( 2 p^\nu p^\tau - p^2 \eta^{\nu \tau} \right).
\end{align}	
Once again, we see how important it is to regulate the supercurrent in the SUSY-invariant fashion.

\subsection{Wess-Zumino consistency condition} \label{sec:WZ_cond}

Here we want to check explicitly the Wess-Zumino condition \eqref{true_WZ_cond}, which we rewrite here for completeness,
\begin{align} \label{true_WZ_conda}
\delta_{\bep} \mathcal{A}_{\theta} = - \I \bep_{\da} \bsigma^{\mu \da \a} \partial_\mu \< J_{\chi \a} \>.
\end{align}
This identity is consistent with the absence of the SUSY-anomaly, $\bAn_{\bep} = 0$. We want to check it by taking some functional derivatives of both its sides and compare the resulting correlation functions. In the course of the evaluation we have to stick to a SUSY-preserving renormalization scheme, in order to keep $\bAn_{\bep} = 0$. Furthermore, since terms we are looking for are local, it is crucial to include all scheme-dependent terms in our calculations.

Calculations of the flavor anomaly in an explicitly supersymmetry-preserving scheme were carried out in \cite{Guadagnini:1985ar} and explicit expressions were worked out in \cite{Guadagnini:1985ea,Harada:1985wa,Krivoshchekov:1985ep}. However, in order to check \eqref{true_WZ_conda} we would also need to calculate its right hand side in the same scheme. To be consistent, we will use the Pauli-Villars regularization scheme throughout the calculations. The flavor anomaly will then take form
\begin{align} \label{flavor_an}
\mathcal{A}_{\theta}^{\text{PV}} = \mathcal{A}^{\text{ABJ}}_{\theta} + \delta_{\theta} S_{\text{loc}}^{\text{PV}},
\end{align}
where $\mathcal{A}^{\text{ABJ}}_{\theta}$ is the ABJ anomaly \eqref{ABJ} while $\delta_{\theta} S_{\text{loc}}^{\text{PV}}$ is a scheme-dependent part. Its form, however, is uniquely fixed by the choice of the Pauli-Villars regularization scheme. 

Having calculated the flavor anomaly, we can take functional derivatives of the Wess-Zumino condition \eqref{true_WZ_conda} and compare the two sides. Since we are guaranteed to have the ABJ anomaly present, we will obtain a non-trivial check by taking functional derivatives with respect to $A_\mu$ and $\lambda_\beta$. Hence, in order to test the Wess-Zumino consistency condition we test the following identity,
\begin{align} \label{to_check}
\lla J_A^\mu(\bs{p}_1) J_{\lambda \beta}(\bs{p}_2) \frac{\partial \An_{\theta}}{\partial \bep_{\da}}(\bs{p}_3) \rra_{PV} = \bsigma^{\rho \da \a} p_{3 \rho} \lla J_A^\mu(\bs{p}_1) J_{\lambda \beta}(\bs{p}_2) J_{\chi \alpha} (\bs{p}_3) \rra_{PV}
\end{align}
in the Pauli-Villars regularization scheme.

\subsubsection{Correlator $\< J^\mu_A J_{\chi \alpha} J_{\lambda \beta} \>$} \label{sec:3ptJchi}

We start with the right hand side of \eqref{to_check}. First, the seagull terms in the Lagrangian \eqref{lag} result in the relation,
\begin{align} \label{seagull:text_Achilambda}
\lla J^\mu_A(\bs{p}_1) J_{\chi \alpha}(\bs{p}_2) J_{\lambda \beta}(\bs{p}_3) \rra_m & = \lla j^\mu_A(\bs{p}_1) j_{\chi \alpha}(\bs{p}_2) j_{\lambda \beta}(\bs{p}_3) \rra_m - \frac{\I}{2} \sigma^\mu_{\a \da} \lla \bj^{\da}_{\bl}(\bs{p}_{12}) j_{\lambda \b}(\bs{p}_3) \rra_m,
\end{align}
where $\bs{p}_{12} = \bs{p}_1 + \bs{p}_2$. Next, since $j_{\chi \alpha}$ is sourced by a compensator, it is proportional to the equations of motion. In particular, if we define
\begin{align}
& j^{\text{(eq)}}_{\chi \alpha} = \frac{\I}{\sqrt{2}} \left( \j \frac{\delta}{\delta \psi^\alpha} - \psi_\alpha \frac{\delta}{\delta F} \right) S
\end{align}
then, in the massless theory, $j_{\chi \alpha} = j^{\text{(eq)}}_{\chi \alpha}|_{m=0}$. We can use the Schwinger-Dyson equations to write
\begin{align} \label{ward:jAjCjL}
\lla j_A^\mu(\bs{p}_1) j_{\chi \alpha}(\bs{p}_2) j_{\lambda \beta}(\bs{p}_3) \rra = \frac{\I}{2} \sigma^{\mu}_{\a \da} \lla \bar{j}_{\bl}^{\dot{\alpha}} (\bs{p}_{12}) \: j_{\lambda \beta}(\bs{p}_3) \rra
\end{align}
in the massless theory, up to an anomaly. This, together with \eqref{seagull:text_Achilambda}, shows that, up to an anomaly, $\lla J^\mu_A J_{\chi \alpha} J_{\lambda \beta} \rra = 0$. This is consistent with \eqref{Jchi0}.

In order to find the anomaly in \eqref{ward:jAjCjL} we consider the massive theory. In the massive theory, however, $j_{\chi \alpha}$ is no longer proportional to the equations of motion. Instead
\begin{align}
& j_{\chi \alpha} = j^{\text{(eq)}}_{\chi \alpha} + \Delta j_{\chi \alpha}, && \Delta j_{\chi \alpha} = \I m \sqrt{2} \j \psi_{\alpha}.
\end{align}
This gives us the pseudo-Schwinger-Dyson identity\footnote{We call it a pseudo Schwinger-Dyson identity, since it relates its left hand side to correlation functions that are not necessarily local.},
\begin{align} \label{schwinger_dyson_m}
& \lla j_A^\mu(\bs{p}_1) j_{\chi \alpha}(\bs{p}_2) j_{\lambda \beta}(\bs{p}_3) \rra_m = \frac{\I}{2} \sigma^{\mu}_{\a \da} \lla \bar{j}_{\bl}^{\dot{\alpha}} (\bs{p}_{12}) \: j_{\lambda \beta}(\bs{p}_3) \rra_m - \I \epsilon_{\beta\alpha} \lla j_A^\mu(\bs{p}_1) j_D(\bs{p}_{23}) \rra_m \nn\\
& \qquad\qquad + \I m \sqrt{2} \lla j_A^\mu(\bs{p}_1) \lwick \j \psi_{\a} \rwick(\bs{p}_2) \, j_{\lambda \beta}(\bs{p}_3) \rra_m.
\end{align}
The second term on the right hand side vanishes, $\< j_A^\mu j_D \>_m = 0$. When \eqref{schwinger_dyson_m} is substituted into \eqref{seagull:text_Achilambda} we find
\begin{align}
& \lla J_A^\mu(\bs{p}_1) J_{\chi \alpha}(\bs{p}_2) J_{\lambda \beta}(\bs{p}_3) \rra_m = \lla j_A^\mu(\bs{p}_1) \, \Delta j_{\chi \alpha}(\bs{p}_2) \, j_{\lambda \beta}(\bs{p}_3) \rra_m.
\end{align}
This is the same mechanism for anomaly emergence as in the cases discussed in section \ref{sec:PV_ren}. Here, however, we discover the possibility of a new kind of anomaly: an anomaly in the equations of motion. Notice also that, in principle, we could have declared $j^{\text{(eq)}}_{\chi \alpha}$ as the regulated version of the massless $j_{\chi \alpha}$. This, however, would violate supersymmetry, which requires that the form of the operator $j_{\chi \alpha}$ remains the same regardless of the mass.

Now we proceed with the standard Feynman parameters,
\begin{align}
& \lla J_A^\mu(\bs{p}_1) J_{\chi \alpha}(\bs{p}_2) J_{\lambda \beta}(\bs{p}_3) \rra_{PV} = \nn\\
& \qquad = - \lim_{M \rightarrow \infty} \lla j_A^\mu(\bs{p}_1) \, \Delta j_{\chi \alpha}(\bs{p}_2) \, j_{\lambda \beta}(\bs{p}_3) \rra_M \nn\\
& \qquad = - \lim_{M \rightarrow \infty} M^2 \int_{Eu} \frac{\D^4 \bs{k}}{(2 \pi)^4}\frac{k^\mu \epsilon_{\b \a} + (2 \sigma)^{\kappa \mu \ \gamma}_{\ \ \: \beta} \epsilon_{\gamma \a} \left( k_{\kappa} - \frac{p_{1 \kappa}}{2} \right)}{(k^2 + M^2) (|\bs{k} - \bs{p}_1|^2 + M^2) (|\bs{k} + \bs{p}_2|^2 + M^2)} \nn\\
& \qquad = \frac{1}{32 \pi^2} \int_0^1 \D x_1 \int_0^{1 - x_1} \D x_2 \left[ (2 \sigma)^{\kappa \mu \ \gamma}_{\ \ \: \beta} \epsilon_{\gamma \alpha} \left( p_{1 \kappa} (1 - 2 x_2) + 2 x_1 p_{2 \kappa} \right) + 2 \epsilon_{\beta \alpha} (x_1 p_2^\mu - x_2 p_1^\mu) \right] \nn\\
& \qquad = \frac{1}{192 \pi^2} \left[ (2 \sigma)^{\kappa \mu \ \gamma}_{\ \ \: \beta} \epsilon_{\gamma \alpha} ( p_{1 \kappa} + 2 p_{2 \kappa} ) + 2 \epsilon_{\beta \alpha} (p_2^\mu - p_1^\mu) \right], \label{jAjcjl}
\end{align}
Note that the 3-point function in \eqref{to_check} requires swapping $J_{\chi \alpha}$ and $J_{\lambda \beta}$ which swaps $\bs{p}_2$ with $\bs{p}_3$ and produces an additional minus sign by anticommuting the fermions. After some algebra one finds
\begin{align} \label{to_get}
& \bsigma^{\rho \da \a} p_{3 \rho} \lla J_A^\mu(\bs{p}_1) J_{\lambda \beta}(\bs{p}_2) J_{\chi \alpha} (\bs{p}_3) \rra_{PV} = \frac{\epsilon_{\beta \alpha}}{192 \pi^2} \left[ \I \epsilon^{\mu\kappa\lambda\tau} p_{1\kappa} p_{3\lambda} \bsigma_{\tau}^{\da \a}  \right. \nn\\
& \qquad\qquad\qquad \left. + \bsigma^{\mu \da \a} \bs{p}_3 \cdot (\bs{p}_2 - \bs{p}_3) + \bsigma_{\kappa}^{\da \a} (p_1^\kappa p_3^\mu + 2 p_1^\mu p_3^\kappa) \right].
\end{align}

\subsubsection{Flavor anomaly} \label{sec:chiral_an}

Now we calculate the left hand side of \eqref{to_check}. To do it, we first must calculate the relevant terms in the flavor anomaly \eqref{flavor_an}. These are terms that contain exactly two sources and under the supersymmetry transformation $\delta_{\bep}$ transform into expressions containing a single $A_{\mu}$ and a single $\lambda^\beta$. It is easy to see that there are four types of terms we have to worry about, as listed in the table below. In order to recover the exact form of such terms, we can evaluate various 3-point functions, as indicated in the table.
\begin{center}
\begin{tabular}{|l|c|c|c|c|}
\hline Terms: & $A_\mu A_\nu$ & $A_\mu D$ & $\bl_{\db} \lambda_{\beta}$ & $\chi_{\alpha} \lambda_{\beta}$ \\ \hline
Correlators: & $\< \partial j_A^\mu j_A^\nu j_A^\rho \>$ & $\< \partial j_A^\mu j_A^\nu j_D \>$ & $\< \partial j_A^\mu j_{\bl \db} j_{\lambda \beta} \>$ & $\< \partial j_A^\mu j_{\chi \alpha} \lambda_{\lambda \beta} \>$ \\ \hline
\end{tabular}
\end{center}
The anomalies can be easily extracted from the 3-point functions obtained by the functional differentiation of the generating functional, \textit{i.e.}, the correlators involving ``capital" operators $J_k$. Feynman calculus, on the other hand, provides means to calculate correlation functions of the ``small" operators $j_k$. To relate the two, we have to resolve the seagull terms, which are given in \eqref{seagulls_3pt1} - \eqref{seagulls_3pt2}. As we can see, the 3-point functions of interest have vanishing seagull terms.

The ABJ anomaly was calculated in section \ref{sec:ABJ}. Here we will calculate anomalous terms in other relevant correlation functions. As discussed in section \ref{sec:PV}, these anomalous terms -- unlike the ABJ anomaly -- can be removed by a finite counterterm. However, since we use the single, consistent, supersymmetry-preserving, Pauli-Villars regularization scheme, the specific form of these anomalous terms is fixed. While supersymmetric, the scheme does not preserve the conservation of the axial current $j_A^\mu$. Hence, additional, anomaly-like terms may appear in various correlation functions involving the divergence of the current. For example, with the operator $P$ defined in \eqref{pjA_m} we find
\begin{align}
& p_{1 \mu} \lla j^\mu_A(\bs{p}_1) \bj_{\bl \db}(\bs{p}_2) j_{\lambda \b}(\bs{p}_3) \rra_{PV} =  \nn\\
& = - \lim_{M \rightarrow \infty} \lla P (\bs{p}_1) \, \bj_{\bl}^{\da}(\bs{p}_2) j_{\lambda \a}(\bs{p}_3) \rra_M \nn\\
& = \lim_{M \rightarrow \infty} \frac{M^2}{2} p_{1 \kappa} \sigma^{\kappa}_{\beta \db} \int_{Eu} \frac{\D^4 \bs{k}}{(2 \pi)^4} \frac{1}{(k^2 + M^2) ( |\bs{k} - \bs{p}_1|^2 + M^2) ( |\bs{k} + \bs{p}_2|^2 + M^2)} \nn\\
& = \frac{\I}{32 \pi^2} p_{1 \kappa} \sigma^{\kappa}_{\beta \db} \int_0^1 \D x_1 \int_0^{1-x_1} \D x_2 \lim_{M \rightarrow \infty} \frac{M^2}{\Delta_0 + M^2} \nn\\
& = \frac{\I}{64 \pi^2} p_{1 \kappa} \sigma^{\kappa}_{\beta \db}.
\end{align}
For completeness we also list
\begin{align}
p_{1 \mu} \lla j_A^\mu (\bs{p}_1) j_A^\nu(\bs{p}_2) j_A^\rho(\bs{p}_3) \rra_{PV} & = \frac{\I}{96 \pi^2} \epsilon^{\kappa\lambda \nu \rho} p_{2 \kappa} p_{3 \lambda}, \\
p_{1 \mu} \lla j_A^\mu (\bs{p}_1) j_A^\nu(\bs{p}_2) j_D(\bs{p}_3) \rra_{PV} & = 0.
\end{align}
Finally, by contracting \eqref{jAjcjl}, which is the exact result in the PV scheme, we find
\begin{align}
& p_{1 \mu} \lla J_A^\mu(\bs{p}_1) J_{\chi \alpha}(\bs{p}_2) J_{\lambda \beta}(\bs{p}_3) \rra_{PV} = - \frac{1}{96 \pi^2} \left[ (2 \sigma)^{\mu \nu \ \gamma}_{\ \ \: \beta} \epsilon_{\gamma \alpha} p_{1 \mu} p_{2 \nu} + \epsilon_{\beta \alpha} \bs{p}_1 \cdot (\bs{p}_1 - \bs{p}_2) \right].
\end{align}

All together we find the relevant terms in the flavor anomaly in the PV regularization scheme,
\begin{align} \label{an_PV}
\An_{\theta}^{PV} & = - \frac{1}{192 \pi^2} \partial_\mu \left[ \epsilon^{\mu\nu\rho\tau} A_\nu \partial_\rho A_\tau + 3 \lambda \sigma^\mu \bl + 2 \I \lambda \sigma^\mu \bsigma^\nu \partial_\nu \chi - 2 \I \partial^\mu ( \lambda \chi ) \right.\nn\\
& \qquad\qquad\qquad\qquad \left. + O(C, \mathcal{M}, \chi^2, \bl) \right]
\end{align}
and its supersymmetric variation,
\begin{align}
\delta_{\bep} \An_{\theta}^{PV} & = \frac{1}{192 \pi^2} \bep_{\da} \partial_{\mu} \left[ -\frac{\I}{2} \epsilon^{\mu \kappa \lambda \tau} F_{\kappa \lambda} \bsigma_{\tau}^{\da \a} + \bsigma_\nu^{\da \a} ( \partial^\mu A^\nu + \partial^\nu A^\mu ) \right.\nn\\
& \qquad\qquad\qquad\qquad \left. +\,2 ( \bsigma^{\mu \da \a} \partial_\nu A^\nu + \bsigma^{\nu \da \a} A_{\nu} \partial^\mu ) - 3 \I D \bsigma^{\mu \da \a} \right] \lambda_{\alpha}.
\end{align}
By taking functional derivatives with respect to $A_\mu$ and $\lambda^\beta$ we arrive at the left hand side of \eqref{to_check},
\begin{align}
& \lla J_A^\mu(\bs{p}_1) J_{\lambda \beta}(\bs{p}_2) \frac{\partial \An_{\theta}}{\partial \bep_{\da}}(\bs{p}_3) \rra_{PV} = \frac{\epsilon_{\beta \alpha}}{192 \pi^2} \left[ \I \epsilon^{\mu\kappa\lambda\tau} p_{1\kappa} p_{3\lambda} \bsigma_{\tau}^{\da \a} \right. \nn\\
& \qquad\qquad\qquad \left. + \bsigma^{\mu \da \a} \bs{p}_3 \cdot (\bs{p}_2 - \bs{p}_3) + \bsigma_{\kappa} (p_1^\kappa p_3^\mu + 2 p_1^\mu p_3^\kappa) \right].
\end{align}
This matches \eqref{to_get} exactly.

In this section we have shown that the relation \eqref{to_check} holds in the PV-regulated Wess-Zumino model. This relation is a direct consequence of the correct Wess-Zumino consistency condition \eqref{true_WZ_conda} with no SUSY-anomaly, $\bAn_{\bep} = 0$.

\subsection{Ward identity} \label{sec:ward}

We want to independently check the Ward identity \eqref{ward_1pt}, which reads,
\begin{align} \label{ward_1pta}
\partial_{\mu} \< \bJ^{\mu \da}_{\bQ} \> & = \I \bsigma_\mu^{\da \a} \lambda_{\a} \< J_A^\mu \> - \left[ \bsigma^{\kappa \lambda \da}_{\ \ \ \ \db} F_{\kappa\lambda} + \I D \delta_{\db}^{\da} \right] \< \bJ^{\db}_{\bl} \> + \bsigma^{\mu \da \a} \partial_\mu \lambda_\a \, \< J_D \> \nn\\
& \qquad  - \I \bsigma^{\mu \da \a} A_\mu \< J_{\chi \a} \> + 2 \bl^{\da} \< J_{\mathcal{M}} \> + O(C, \mathcal{M}, \chi)
\end{align}
and show that it is non-anomalous. In particular, we want to show that the inclusion of the operators $J_{\chi}$ and $J_{\mathcal{M}}$ removes the suspected anomaly \eqref{wrong_An}. The simplest correlator which contains a non-vanishing contribution from the insertion of $J_{\chi \alpha}$ is $\< J_{\chi \alpha} J_A^{\mu} J_{\lambda \beta} \>$. Hence, we take 3 functional derivatives of \eqref{ward_1pta} with respect to $\lambda^\beta$ as well as $A_\mu$ and $A_\nu$ to obtain
\begin{align} \label{SUSY_WARD}
& p_{Q \rho} \lla \bar{J}^{\rho \dot{\alpha}}_{\bar{Q}}(\bs{p}_Q) J_A^\mu(\bs{p}_A) J_A^\nu(\bs{p}_B) J_{\lambda \beta}(\bs{p}_L) \rra = \nn\\
& \qquad + \I (2 \bsigma)^{\mu\rho\da}_{\ \ \ \ \db} \, p_{A \rho} \lla \bar{J}_{\bar{\lambda}}^{\dot{\beta}}(\bs{p}_{QA}) J_A^\nu(\bs{p}_B) J_{\lambda \beta}(\bs{p}_L) \rra + (\bs{p}_A \leftrightarrow \bs{p}_B, \mu \leftrightarrow \nu), \nn\\
& \qquad + \blue{\I \bsigma^{\mu\da\a} \lla J_{\chi \alpha}(\bs{p}_{QA}) J_A^\nu(\bs{p}_B) J_{\lambda \beta}(\bs{p}_L) \rra + (\bs{p}_A \leftrightarrow \bs{p}_B, \mu \leftrightarrow \nu)} \nn\\
& \qquad - \I \bsigma^{\rho \da \a} \epsilon_{\a \b} p_{L \rho} \lla J_A^\mu(\bs{p}_A) J_A^\nu(\bs{p}_B) J_{D}(\bs{p}_{QL}) \rra \nn\\
& \qquad + \I \bsigma^{\rho \da \a} \epsilon_{\a \b} \lla J_{A \rho}(\bs{p}_{QL}) J_A^\mu(\bs{p}_A) J_A^\nu(\bs{p}_B) \rra,
\end{align}
where we denote $\bs{p}_{IJ} = \bs{p}_I + \bs{p}_J$. In the spirit of the SUSY transformations \eqref{text_susy_tran_first} - \eqref{text_susy_tran_last} we highlight the line that would be missing if one used the gauge-breaking SUSY-transformations \eqref{susy_tran_WZ_first} - \eqref{susy_tran_WZ_last}. In this section we want to show that this Ward identity is anomaly-free. We will use both the ancient method of shifting the momentum running in the loop as well as the analysis in the Pauli-Villars regularization scheme.

\subsubsection{Canonical Ward identity}

As we use the standard Feynman diagram methods, we will look for the anomaly in the canonical Ward identity. To derive it, we need to work out SUSY transformations of the operators involved. Since $\Phi = (\j, \psi, F)$ fit into the chiral multiplet, their SUSY transformations are given by \eqref{susy_chiral_first} - \eqref{susy_chiral_last}. From this we find,
\begin{align} \label{djA}
\delta_{\ep, \bep} j_A^\mu & = \ep^{\a} (2 \sigma)^{\mu\nu \ \beta}_{\ \ \alpha} \partial_\nu j_\lambda^{\b} + (\partial^\mu \ep^{\alpha}) j_{\lambda \alpha} + \I \ep \sigma^\mu \bar{j}_{\bar{\chi}} \nn\\
& \qquad\qquad + \bep_{\da} (2 \bar{\sigma})^{\mu\nu \dot{\alpha}}_{\ \ \ \ \dot{\beta}} \partial_\nu \bar{j}_{\bar{\lambda}}^{\dot{\beta}} + (\partial^\mu \bep_{\dot{\alpha}}) \bar{j}_{\bar{\lambda}}^{\dot{\alpha}} + \I \bar{\ep} \bsigma^\mu j_{\chi}, \\
\delta_{\ep, \bep} j_{\lambda \alpha} & = - j_{\mathcal{M}} \ep_{\alpha} + \left( \partial_\mu j_D - \I j_{A \mu} \right) \bep_{\db} \bsigma^{\mu \db \b} \epsilon_{\a \b}, \\
\delta_{\ep, \bep} j_D & = - \I \ep j_\lambda + \I \bar{\ep} \bar{j}_{\bar{\lambda}},
\end{align}
where all the operators are listed in \eqref{text_ops_first} - \eqref{text_ops_last}. This leads to the following Ward identity
\begin{align} \label{susy_ward}
& p_{Q \rho} \lla \bar{j}^{\rho \dot{\alpha}}_{\bar{Q}}(\bs{p}_Q) j_A^\mu(\bs{p}_A) j_A^\nu(\bs{p}_B) j_{\lambda \beta}(\bs{p}_L) \rra = \nn\\
& \qquad \I \left[ - \delta^{\da}_{\db} p_Q^\mu + (2 \bar{\sigma})^{\mu\rho\dot{\alpha}}_{\ \ \ \ \dot{\beta}} p_{QA \rho} \right] \times \lla \bar{j}_{\bar{\lambda}}^{\dot{\beta}}(\bs{p}_{QA}) j_A^\nu(\bs{p}_B) j_{\lambda \beta}(\bs{p}_L) \rra + (\bs{p}_A \leftrightarrow \bs{p}_B, \mu \leftrightarrow \nu) \nn\\
& \qquad + \I \bsigma^{\mu \dot{\alpha} \alpha} \lla j_{\chi \alpha}(\bs{p}_{QA}) j_A^\nu(\bs{p}_B) j_{\lambda \beta}(\bs{p}_L) \rra + (\bs{p}_A \leftrightarrow \bs{p}_B, \mu \leftrightarrow \nu) \nn\\
& \qquad - \I \bsigma^{\rho \da \a} \epsilon_{\a \b} p_{QL \rho} \lla j_A^\mu(\bs{p}_A) j_A^\nu(\bs{p}_B) j_D(\bs{p}_{QL}) \rra \nn\\
& \qquad + \I \bsigma^{\rho \da \a} \epsilon_{\a \b} \lla j_A^\mu(\bs{p}_A) j_A^\nu(\bs{p}_B) j_{A \rho}(\bs{p}_{QL}) \rra,
\end{align}
where we denote $\bs{p}_{IJ} = \bs{p}_I + \bs{p}_J$. Other Ward identities involving the divergence of supercurrent are presented in appendix \ref{app:Ward}.

We will show that the Ward identity \eqref{susy_ward} is non-anomalous. Notice that in the derivation of this statement we have not used the Ward identity \eqref{ward_1pta}. Therefore, if we relate the ``small" operators $j_k$ to the ``capital" ones, $J_k$, we can derive \eqref{SUSY_WARD} from \eqref{susy_ward} and therefore confirm \eqref{ward_1pta}. To do it, we need to work out the seagull terms from the Lagrangian in \eqref{lag}. These are presented in appendix \ref{sec:seagulls}. This also provides an important check on our results. We can check commutativity of the following diagram

\begin{equation}
\xymatrix@R+=40pt@C+=120pt{ \< \bJ^{\rho \da}_{\bQ} J_A^\mu J_A^\nu J_{\lambda \beta} \> \ar[r]^{\text{seagulls \eqref{seagulls_4pt}}} \ar[d]^{\partial_\rho} & \< \bj^{\rho \da}_{\bQ} j_A^\mu j_A^\nu j_{\lambda \beta} \> + \text{local terms} \ar[d]^{\partial_\rho} \\
	\partial_\rho \< \bJ^{\rho \da}_{\bQ} J_A^\mu J_A^\nu J_{\lambda \beta} \>  \ar[d]^{\text{Ward identity \eqref{SUSY_WARD}}}  & \partial_{\rho} \left[ \< \bj^{\rho \da}_{\bQ} j_A^\mu j_A^\nu j_{\lambda \beta} \> + \text{local terms} \right] \ar[d]^{\text{Ward identities (\ref{susy_ward}, \ref{susyward3}, \ref{susyward3a})}} \\
	\text{Rhs of \eqref{SUSY_WARD}} \ar[r]^{\text{seagulls (\ref{seagulls_3pt1} - \ref{seagulls_3pt2} \& \ref{seagulls_3ptq1} - \ref{seagulls_3ptq2})}}  &  \text{Rhs of \eqref{susy_ward}}}
\end{equation}

If one goes first to the right by using \eqref{seagulls_4pt} and then down by means of the Ward identities \eqref{susy_ward} - \eqref{susyward3a} one gets
\begin{align}
& p_{Q \rho} \lla \bar{J}^{\rho \dot{\alpha}}_{\bar{Q}}(\bs{p}_Q) J^\mu(\bs{p}_A) J_A^\nu(\bs{p}_B) J_{\lambda \beta}(\bs{p}_L) \rra = \nn\\
& \qquad \I (2 \bar{\sigma})^{\mu\rho\dot{\alpha}}_{\ \ \ \ \dot{\beta}} p_{A \rho} \lla \bar{j}_{\bar{\lambda}}^{\dot{\beta}}(\bs{p}_{QA}) j_A^\nu(\bs{p}_B) j_{\lambda \beta}(\bs{p}_L) \rra + (\bs{p}_A \leftrightarrow \bs{p}_B, \mu \leftrightarrow \nu) \nn\\
& \qquad + \I \bsigma^{\mu \dot{\alpha} \alpha} \lla j_{\chi \alpha}(\bs{p}_{QA}) j_A^\nu(\bs{p}_B) j_{\lambda \beta}(\bs{p}_L) \rra + (\bs{p}_A \leftrightarrow \bs{p}_B, \mu \leftrightarrow \nu) \nn\\
& \qquad - \I \bsigma^{\rho \da \a} \epsilon_{\a \b} p_{L \rho} \lla j_A^\mu(\bs{p}_A) j_A^\nu(\bs{p}_B) j_D(\bs{p}_{QL}) \rra \nn\\
& \qquad + \I \bsigma^{\rho \da \a} \epsilon_{\a \b} \lla j_A^\mu(\bs{p}_A) j_A^\nu(\bs{p}_B) j_{A \rho}(\bs{p}_{QL}) \rra \nn\\
& \qquad - \eta^{\mu\nu} \lla \bj^{\da}_{\bl}(\bs{p}_{QAB}) j_{\lambda \beta}(\bs{p}_{L}) \rra \nn\\
& \qquad + \eta^{\mu\nu} \bsigma^{\rho \da \a} \epsilon_{\a \b} p_{L \rho} \lla j_{D}(\bs{p}_{AB}) j_{D}(\bs{p}_{QL}) \rra.
\end{align}
Now one can go over the diagram down first and confirm this Ward identity. In particular, if a SUSY-anomaly is present in the Ward identity \eqref{SUSY_WARD}, it is equal to the SUSY-anomaly in \eqref{susy_ward}.

Note that if one replaces the third line of \eqref{susy_ward} containing $j_{\chi \alpha}$ by the naive Schwinger-Dyson identity \eqref{ward:jAjCjL}, the anomaly reported in~\cite{Papadimitriou:2019yug} reappears. However, there is no reason to apply this substitution as the supersymmetric variation of $j_A^\mu$ in \eqref{djA} is uniquely determined by the variations of the component fields of the chiral multiplet.

\subsection{Anomaly by momentum shifting} \label{sec:anom_shift}

In this section we explicitly check that the anomaly in absent in the Ward identity \eqref{susy_ward} using the method of shifting the momentum running in the loop integrals.

\subsubsection{The 4-point function}

The full 4-point function is the sum of the 4 terms,
\begin{align} \label{4pt}
& \< \bar{j}^{\rho \dot{\alpha}}_{\bar{Q}}(\bs{p}_Q) j_A^\mu(\bs{p}_A) j_A^\nu(\bs{p}_B) j_{\lambda \beta}(\bs{p}_L) \> = \nn\\
& \qquad = \< \bar{j}^{\rho \dot{\alpha}}_{\bar{Q}}(\bs{p}_Q) j_{\j}^\mu(\bs{p}_A) j_{\j}^\nu(\bs{p}_B) j_{\lambda \beta}(\bs{p}_L) \> \nn\\
& \qquad\qquad + \< \bar{j}^{\rho \dot{\alpha}}_{\bar{Q}}(\bs{p}_Q) j_{\j}^\mu(\bs{p}_A) j_{\psi}^\nu(\bs{p}_B) j_{\lambda \beta}(\bs{p}_L) \> + ( \bs{p}_A \ \leftrightarrow \ \bs{p}_B, \mu \leftrightarrow \nu ) \nn\\
& \qquad\qquad + \< \bar{j}^{\rho \dot{\alpha}}_{\bar{Q}}(\bs{p}_Q) j_{\psi}^\mu(\bs{p}_A) j_{\psi}^\nu(\bs{p}_B) j_{\lambda \beta}(\bs{p}_L) \>.
\end{align}
We reverted here to correlation functions without the momentum conserving delta function pulled out. It will be convenient to keep all Dirac deltas explicitly. We have
\begin{align} \label{jQjfjfjl}
& \< \bar{j}^{\rho \dot{\alpha}}_{\bar{Q}}(\bs{p}_Q) j_{\j}^\mu(\bs{p}_A) j_{\j}^\nu(\bs{p}_B) j_{\lambda \beta}(\bs{p}_L) \> = \I ( \bsigma^\kappa \sigma^\rho \bsigma^\lambda)^{\da \a} \epsilon_{\a \b} \times \nn\\
& \qquad\qquad \times \prod_{I} \int_{Eu} \frac{\D^4 \bs{k}_I}{(2 \pi)^4} \frac{k_{Q \kappa} \left( k_A - \frac{p_A}{2} \right)^\mu \left( k_B - \frac{p_B}{2} \right)^\nu k_{L \lambda}}{k_Q^2 k_A^2 k_B^2 k_L^2} \times \nn\\
& \qquad\qquad \times (2 \pi)^{16} \delta(\bs{k}_L + \bs{p}_Q - \bs{k}_Q) \delta(\bs{k}_Q + \bs{p}_A - \bs{k}_A) \delta(\bs{k}_A + \bs{p}_B - \bs{k}_B) \delta(\bs{k}_B + \bs{p}_L - \bs{k}_L) \nn\\
& \qquad + ( \bs{p}_A \ \leftrightarrow \ \bs{p}_B, \mu \leftrightarrow \nu ).
\end{align}
\begin{align} \label{jQjfjpsijl}
& \< \bar{j}^{\rho \dot{\alpha}}_{\bar{Q}}(\bs{p}_Q) j_{\j}^\mu(\bs{p}_A) j_{\psi}^\nu(\bs{p}_B) j_{\lambda \beta}(\bs{p}_L) \> = - \frac{\I}{2} ( \bsigma^\kappa \sigma^\rho \bsigma^\xi \sigma^\nu \bsigma^\lambda)^{\da \a} \epsilon_{\a \b} \times \nn\\
& \qquad\qquad \times \prod_{I} \int_{Eu} \frac{\D^4 \bs{k}_I}{(2 \pi)^4} \frac{k_{Q \kappa} \left( k_A - \frac{p_A}{2} \right)^\mu k_{B \xi} k_{L \lambda}}{k_Q^2 k_A^2 k_B^2 k_L^2} \times \nn\\
& \qquad\qquad \times (2 \pi)^{16} \delta(\bs{k}_L + \bs{p}_B - \bs{k}_B) \delta(\bs{k}_B + \bs{p}_Q - \bs{k}_Q) \delta(\bs{k}_Q + \bs{p}_A - \bs{k}_A) \delta(\bs{k}_A + \bs{p}_L - \bs{k}_L).
\end{align}
\begin{align} \label{jQjpsijpsijl}
& \< \bar{j}^{\rho \dot{\alpha}}_{\bar{Q}}(\bs{p}_Q) j_{\psi}^\mu(\bs{p}_A) j_{\psi}^\nu(\bs{p}_B) j_{\lambda \beta}(\bs{p}_L) \> = \frac{\I}{4} ( \bsigma^\kappa \sigma^\rho \bsigma^\xi \sigma^\mu \bsigma^\eta \sigma^\nu \bsigma^\lambda)^{\da \a} \epsilon_{\a \b} \times \nn\\
& \qquad\qquad \times \prod_{I} \int_{Eu} \frac{\D^4 \bs{k}_I}{(2 \pi)^4} \frac{k_{Q \kappa} k_{A \xi} k_{B \eta} k_{L \lambda}}{k_Q^2 k_A^2 k_B^2 k_L^2} \times \nn\\
& \qquad\qquad \times (2 \pi)^{16} \delta(\bs{k}_L + \bs{p}_B - \bs{k}_B) \delta(\bs{k}_B + \bs{p}_A - \bs{k}_A) \delta(\bs{k}_A + \bs{p}_Q - \bs{k}_Q) \delta(\bs{k}_Q + \bs{p}_L - \bs{k}_L) \nn\\
& \qquad + ( \bs{p}_A \ \leftrightarrow \ \bs{p}_B, \mu \leftrightarrow \nu ),
\end{align}
where the products are taken over $I = Q, A, B, L$. All subsequent calculations and identities will be checked term-by-term.

To evaluate the anomaly in \eqref{susy_ward} we move all terms to the left hand side and denote the difference by $\mathcal{A}^{\mu\nu\dot{\alpha}}_{\ \ \ \ \beta}$. We will now check the absence of the anomaly by employing the ancient method of shifting the integration momentum in the loop integrals. Such a scheme almost certainly violates supersymmetry, but this is not a problem. If the anomaly is absent, but the scheme is not supersymmetric, then the resulting difference, $\mathcal{A}^{\mu\nu\dot{\alpha}}_{\ \ \ \ \beta}$, will be removable by a counterterm. On dimensional grounds only a single counterterm exists,
\begin{equation}
S_{\text{fin}} \sim \int \D^4 \bs{x} A^\mu A_\mu \, \lambda \sigma^{\kappa} \bpsi_{\kappa}.
\end{equation}
Hence, if $\mathcal{A}^{\mu\nu\dot{\alpha}}_{\ \ \ \ \beta}$ is proportional to $p_{Q \kappa} \bsigma^{\kappa \da \a} \epsilon_{\a \beta}$, there is no SUSY anomaly. 

\subsubsection{Calculations} \label{sec:AnomShiftCalc}

Now we act with $p_{Q \rho}$ on each term in the 4-point function \eqref{4pt}. First, denote $\bs{p}_{IJ} = \bs{p}_I + \bs{p}_J$ and look for the delta function of the form $\delta(\bs{k}_I + \bs{p}_Q - \bs{k}_Q)$ in each term. This means that we can write $\bs{p}_Q = \bs{k}_Q - \bs{k}_I$ underneath the integral and use the fact that $\sigma^\mu \bsigma^\nu k_{\mu} k_{\nu} = - k^2$. For example, in the term written explicitly in \eqref{jQjfjfjl} we find $\delta(\bs{k}_L + \bs{p}_Q - \bs{k}_Q)$. Hence, $\bs{p}_Q = \bs{k}_Q - \bs{k}_L$ and the important part of the numerator of the integrand reads
\begin{align} \label{sigmas_k}
& \bsigma^\kappa \sigma^\rho \bsigma^\lambda p_{Q \rho} k_{Q \kappa} k_{L \lambda} = \bsigma^\kappa \sigma^\rho \bsigma^\lambda (k_{Q \rho} - k_{L \rho}) k_{Q \kappa} k_{L \lambda} \nn\\
& \qquad = \bsigma^\kappa \left[ k_L^2 k_{Q \kappa} - k_Q^2 k_{L \kappa} \right].
\end{align}
We can cancel factors in the numerator and the denominator now and in total we find
\begin{align} \label{pexp1}
& p_{Q \rho} \< \bar{j}^{\rho \dot{\alpha}}_{\bar{Q}}(\bs{p}_Q) j_{\j}^\mu(\bs{p}_A) j_{\j}^\nu(\bs{p}_B) j_{\lambda \beta}(\bs{p}_L) \> = \I \bsigma^{\kappa \da \a} \epsilon_{\a \beta} \times \nn\\
& \qquad\qquad \times \prod_{I} \int_{Eu} \frac{\D^4 \bs{k}_I}{(2 \pi)^4} \frac{\left( k_A - \frac{p_A}{2} \right)^\mu \left( k_B - \frac{p_B}{2} \right)^\nu}{k_A^2 k_B^2} \times (2 \pi)^4 \delta(\bs{k}_L + \bs{p}_Q - \bs{k}_Q) \times \left[ \right.\nn\\
& \qquad\qquad\qquad \frac{k_{Q \kappa}}{k_Q^2} (2 \pi)^{12} \delta(\bs{k}_Q + \bs{p}_A - \bs{k}_A) \delta(\bs{k}_A + \bs{p}_B - \bs{k}_B) \delta(\bs{k}_B + \bs{p}_{QL} - \bs{k}_Q) \nn\\
& \qquad\qquad\qquad \left. - \frac{k_{L \kappa}}{k_L^2} (2 \pi)^{12} \delta(\bs{k}_L + \bs{p}_{QA} - \bs{k}_A) \delta(\bs{k}_A + \bs{p}_B - \bs{k}_B) \delta(\bs{k}_B + \bs{p}_L - \bs{k}_L) \right] \nn\\
& \qquad + ( \bs{p}_A \ \leftrightarrow \ \bs{p}_B, \mu \leftrightarrow \nu ).
\end{align}
Analogous expressions hold for \eqref{jQjfjpsijl} and \eqref{jQjpsijpsijl}. We must compare now this expression to the right hand side of the supersymmetric Ward identity \eqref{susy_ward}. First, we rewrite its right hand side in a way that is better suited for the comparison. The combinations of the operators on the right hand side of the Ward identity in momentum space are
\begin{align}
& \I \left[ - \delta^{\da}_{\db} p_Q^\mu + (2 \bar{\sigma})^{\mu\rho\dot{\alpha}}_{\ \ \ \ \dot{\beta}} p_{QA \rho} \right] \times \bar{j}_{\bar{\lambda}}^{\dot{\beta}}(\bs{p}_{QA}) + \I \bsigma^{\mu \dot{\alpha} \alpha} j_{\chi \alpha}(\bs{p}_{QA}) = \nn\\
& \qquad = - \sqrt{2} \int \frac{\D^4 \bs{k}}{(2 \pi)^4} \lwick \j(\bs{k}) \bar{\psi}^{\dot{\beta}}(\bs{p}_{QA} - \bs{k}) \rwick \left[ \frac{1}{2} ( \bar{\sigma}^\kappa \sigma^\mu )^{\dot{\alpha}}_{\ \dot{\beta}} k_{\kappa} + \left( k^\mu - \frac{p_A^\mu}{2} \right) \delta^{\da}_{\db} \right] \nn\\
& \qquad\qquad\qquad + \frac{1}{\sqrt{2}} \bsigma^{\mu \da \a} \int \frac{\D^4 \bs{k}}{(2 \pi)^4} \lwick \psi_{\alpha}(\bs{k}) F^{\ast}(\bs{p}_{QA} - \bs{k}) \rwick \label{manip1} \\
& - \I \bsigma^{\rho \da \a} \epsilon_{\a \b} j_D(\bs{p}_{QL}) + \I \bsigma^{\rho \da \a} \epsilon_{\a \b} j_{A \rho}(\bs{p}_{QL}) = \nn\\
& \qquad = - \I \bsigma^{\rho \da \a} \epsilon_{\a \b} \int \frac{\D^4 \bs{k}}{(2 \pi)^4} k_{\rho} \, \lwick \j(\bs{k}) \j^\ast(\bs{p}_{QL} - \bs{k}) \rwick + \, \I \int \frac{\D^4 \bs{k}}{(2 \pi)^4} \lwick \bar{\psi}^{\da}(\bs{p}_{QL} - \bs{k}) \psi_{\beta}(\bs{k}) \rwick. \label{manip2}
\end{align}
Using these expressions one can write down the momentum integrals representing the right hand side of the Ward identity. This is a little long, but expressions on both sides split into a set of simpler terms that must match each other. This happens because:
\begin{itemize}
\item the terms corresponding to bosonic and fermionic parts $j_{\j}^\mu$ and $j_{\psi}^\mu$ of the current $j_A^\mu$ match separately and
\item the 3-point function-like terms depend on 3 momenta, one of which is $\bs{p}_{QA}$, $\bs{p}_{QB}$, or $\bs{p}_{QL}$. Again, such terms must match each other on both sides of the Ward identity.
\end{itemize}

A further simplification comes from the fact that, in order to calculate the anomaly, we do not have to keep track of all terms. First, in the method of momentum shifting the anomaly emerges from momentum shifting in linearly divergent integrals. Since the 4-point function \eqref{4pt} has dimension zero in momentum space, its divergence is linearly divergent at most. The linearly divergent integrals in \eqref{pexp1} are those with exactly three loop momenta $\bs{k}$ in the numerator. This means we drop all external momenta in the numerator. 

Secondly, the integrals that require shifting are exactly those which have $k_Q^2$ in their denominators. Concentrating on \eqref{pexp1} and using \eqref{manip1} we can easily guess which terms in \eqref{susy_ward} correspond to two bosonic currents $j_{\j}^\mu$ and $j_{\j}^\nu$ on the right hand side of the Ward identity. Dropping all explicit external momenta in the numerator these terms are
\begin{align}
&  \I \bsigma^{\kappa \da \a} \epsilon_{\a \beta} \times \prod_{I} \int_{Eu} \frac{\D^4 \bs{k}_I}{(2 \pi)^4} \frac{k_A^\mu k_B^\nu}{k_A^2 k_B^2} \times (2 \pi)^4 \delta(\bs{k}_L + \bs{p}_Q - \bs{k}_Q) \times \left[ \right.\nn\\
& \qquad\qquad\qquad \frac{k_{L \kappa}}{k_L^2} (2 \pi)^{12} \delta(\bs{k}_L + \bs{p}_A - \bs{k}_A) \delta(\bs{k}_A + \bs{p}_B - \bs{k}_B) \delta(\bs{k}_B + \bs{p}_{QL} - \bs{k}_L) \nn\\
& \qquad\qquad\qquad \left. - \frac{k_{L \kappa}}{k_L^2} (2 \pi)^{12} \delta(\bs{k}_L + \bs{p}_{QA} - \bs{k}_A) \delta(\bs{k}_A + \bs{p}_B - \bs{k}_B) \delta(\bs{k}_B + \bs{p}_L - \bs{k}_L) \right] \nn\\
& \qquad + ( \bs{p}_A \ \leftrightarrow \ \bs{p}_B, \mu \leftrightarrow \nu ),
\end{align}
The integrals in the second line above and the second line of \eqref{pexp1} match exactly. However, to match the first lines we have to shift the loop momentum. Hence, the difference between \eqref{pexp1} and the corresponding part of the right hand side of the Ward identity \eqref{susy_ward} is
\begin{align}
& \I \prod_{I \in \{Q,A,B,L\}} \int_{Eu} \frac{\D^4 \bs{k}_I}{(2 \pi)^4} \frac{k_A^\mu k_B^\nu k_Q^\kappa}{k_A^2 k_B^2 k_Q^2} \times \nn\\
& \qquad \times (2 \pi)^{16} \delta(\bs{k}_Q + \bs{p}_A - \bs{k}_A) \delta(\bs{k}_A + \bs{p}_B - \bs{k}_B) \delta(\bs{k}_B + \bs{p}_{QL} - \bs{k}_Q) \delta(\bs{k}_L + \bs{p}_Q - \bs{k}_Q) \nn\\
& - \I \prod_{J \in \{Q,A,B\}} \int_{Eu} \frac{\D^4 \bs{k}_J}{(2 \pi)^4} \frac{k_A^\mu k_B^\nu k_L^\kappa}{k_A^2 k_B^2 k_L^2} \nn\\
& \qquad \times \delta(\bs{k}_L + \bs{p}_A - \bs{k}_A) \delta(\bs{k}_A + \bs{p}_B - \bs{k}_B) \delta(\bs{k}_B + \bs{p}_{QL} - \bs{k}_L) = \nn\\
& = \I (2 \pi)^4 \delta(\bs{p}_{QABL}) \left[ \int_{Eu} \frac{\D^4 \bs{k}_L}{(2 \pi)^4} \frac{(k_L + p_Q)^\kappa (k_L + p_A + p_Q)^\mu (k_L + p_A + p_B + p_Q)^\nu}{(k_L + p_Q)^2 (k_L + p_A + p_Q)^2 (k_L + p_A + p_B + p_Q)^2} \right.\nn\\
& \qquad\qquad\qquad\qquad \left. - \int_{Eu} \frac{\D^4 \bs{k}_L}{(2 \pi)^4} \frac{k_L^\kappa (k_L + p_A)^\mu (k_L + p_A + p_B)^\nu}{k_L^2 (k_L + p_A)^2 (k_L + p_A + p_B)^2} \right] \nn\\
& = (2 \pi)^4 \delta(\bs{p}_{QABL}) A^{\kappa \mu \nu}.
\end{align}
The difference between the two integrals equals
\begin{equation}
A^{\kappa \mu \nu} = \frac{\I}{192 \pi^2} \left( p_Q^\kappa \eta^{\mu\nu} + p_Q^\mu \eta^{\kappa\nu} + p_Q^\nu \eta^{\kappa\mu} \right),
\end{equation}
which follows from the textbook shift identity.

Now we repeat the procedure for the divergence of all terms \eqref{jQjfjfjl} - \eqref{jQjpsijpsijl}. We find the difference between the left and right hand sides of \eqref{susy_ward}, \textit{i.e.}, the anomaly, to be equal
\begin{align} \label{anom_fin_corr}
\mathcal{A}^{\mu\nu\da}_{\ \ \ \ \beta} & = 2 \bsigma^{\kappa \da \a} \epsilon_{\a \b} A^{\mu\nu}_{\ \ \ \kappa} - \frac{1}{2} \left( (\bsigma^\kappa \sigma^\nu \bsigma^\lambda)^{\da \a} \epsilon_{\a \b} A^{\mu}_{\ \lambda \kappa} + (\mu \leftrightarrow \nu) \right) \nn\\
& \qquad\qquad + \frac{1}{4} \left( (\bsigma^\kappa \sigma^\mu \bsigma^\eta \sigma^\nu \bsigma^\lambda)^{\da \a} \epsilon_{\a \b} A^{\eta \lambda \kappa} + (\mu \leftrightarrow \nu) \right) \nn\\
& = \frac{\I}{64 \pi^2} \eta^{\mu\nu} p_{Q \kappa} \bsigma^{\kappa \da \a} \epsilon_{\a \b}.
\end{align}
This anomaly is readily removable by a finite counterterm,
\begin{equation} \label{anom_fin}
S_{\text{fin}} = \frac{1}{64 \pi^2} \int \D^4 \bs{x} A^\mu A_\mu \, \lambda \sigma^{\kappa} \bpsi_{\kappa}.
\end{equation}
Hence, we have shown that both canonical Ward identity \eqref{susy_ward} as well as \eqref{SUSY_WARD} are free of genuine SUSY anomalies. Consequently, we confirm the Ward identity \eqref{ward_1pta}. Notice, however, that the above counterterm is not invariant under the flavor symmetry and hence will generate a mixed gravitational contribution to the ABJ anomaly \eqref{ABJ}.

\subsection{Anomaly in the Pauli-Villars regularization} \label{sec:anomaly_PV}

The fact that the massless Wess-Zumino model has the explicit supersymmetric extension to the massive theory is sufficient for the absence of the SUSY anomalies. Nevertheless, we have checked that the both sides of the Ward identity \eqref{susy_ward} match in the massive theory.

As discussed in section \ref{sec:PV_ren} the mechanism behind the emergence of anomalies in the Pauli-Villars renormalization scheme is the appearance of non-local mass-dependent terms in the massive Ward identity. The left hand side of the Ward identity \eqref{susy_ward} contains two 4-point functions involving the massless supercurrent $\bar{j}^{\rho \dot{\alpha}}_{\bQ}$ and its massive correction term $\Delta \bar{j}^{\rho \dot{\alpha}}_{\bar{Q}}$ given in \eqref{jQ}. Both correlator are evaluated in the massive theory now. We want to show that their sum equals exactly to the right hand side of the Ward identity \eqref{susy_ward} and contains only 3- and lower-point functions. In particular it is enough to show that no non-trivial 4-point functions are present, but we can recover the right hand side exactly. Furthermore, we already know that the leading, mass-independent terms reproduce the 3-point functions on the r.h.s. of \eqref{susy_ward} for the massless theory.

This is a straightforward, although long calculation. As an example, we will show the cancellation of such terms in the correlation function with one scalar and one fermionic flavor current, $\< \bar{j}^{\rho \dot{\alpha}}_{\bar{Q} m} j_{\j}^\mu j_{\psi}^\nu j_{\lambda \beta} \>_m$.

\subsubsection{Left hand side}

We split the supercurrent into its massless form and the massive correction, $\bj^{\mu \da}_{\bQ m} = \bj^{\mu \da}_{\bQ} + \Delta \bj^{\mu \da}_{\bQ}$. First consider the correlation function of the massless supercurrent in the massive theory. We find
\begin{align} \label{jQjfjpsijl_m}
& \< \bar{j}^{\rho \dot{\alpha}}_{\bar{Q}}(\bs{p}_Q) j_{\j}^\mu(\bs{p}_A) j_{\psi}^\nu(\bs{p}_B) j_{\lambda \beta}(\bs{p}_L) \>_m = - \frac{\I}{2} ( \bsigma^\kappa \sigma^\rho \bsigma^\xi \sigma^\nu \bsigma^\lambda)^{\da \a} \epsilon_{\a \b} \times \nn\\
& \qquad\qquad \times \prod_{I} \int_{Eu} \frac{\D^4 \bs{k}_I}{(2 \pi)^4} \frac{k_{Q \kappa} \left( k_A - \frac{p_A}{2} \right)^\mu k_{B \xi} k_{L \lambda}}{(k_Q^2 + m^2) (k_A^2 + m^2) (k_B^2 + m^2) (k_L^2 + m^2)} \times \nn\\
& \qquad\qquad \times (2 \pi)^{16} \delta(\bs{k}_L + \bs{p}_B - \bs{k}_B) \delta(\bs{k}_B + \bs{p}_Q - \bs{k}_Q) \delta(\bs{k}_Q + \bs{p}_A - \bs{k}_A) \delta(\bs{k}_A + \bs{p}_L - \bs{k}_L) \nn\\
& \qquad + \frac{\I}{2} m^2 ( \bsigma^\kappa \sigma^\rho \bsigma^\nu )^{\da \a} \epsilon_{\a \b} \times \nn\\
& \qquad\qquad \times \prod_{I} \int_{Eu} \frac{\D^4 \bs{k}_I}{(2 \pi)^4} \frac{k_{Q \kappa} \left( k_A - \frac{p_A}{2} \right)^\mu}{(k_Q^2 + m^2) (k_A^2 + m^2) (k_B^2 + m^2) ( |\bs{p}_Q - \bs{k}_Q|^2 + m^2)} \times \nn\\
& \qquad\qquad \times (2 \pi)^{16} \delta(\bs{k}_L + \bs{k}_B) \delta(\bs{p}_B - \bs{k}_B + \bs{p}_Q - \bs{k}_Q) \delta(\bs{k}_Q + \bs{p}_A - \bs{k}_A) \delta(\bs{k}_A + \bs{p}_L - \bs{k}_L),
\end{align}
where, as before, the index $I$ runs over $Q,A,B,L$. The first term is equal to the expression \eqref{jQjfjpsijl} but with massive propagators. The second term comes from the fact that in the massive theory contractions of any two spinors are non-vanishing, see propagators \eqref{corr_mid} - \eqref{corr_last}. In this case $\bpsi$ present in $\bar{j}^{\rho \dot{\alpha}}_{\bar{Q}}$ is contracted with $\bpsi$ in $j_{\psi}^{\nu}$, while $\psi$ there can be contracted with $\psi$ in $j_{\lambda \beta}$.

Now we apply $p_{Q \rho}$ to \eqref{jQjfjpsijl_m} and we rework the numerator of the first term analogously to \eqref{sigmas_k}. We apply the split, $\bs{p}_Q = \bs{k}_Q - \bs{k}_B$ but now we have to take the mass into consideration when cancelling factors,
\begin{align} \label{manip_m}
& \bsigma^\kappa \sigma^\rho \bsigma^\xi \sigma^\nu \bsigma^\lambda p_{Q \rho} k_{Q \kappa} k_{B \xi} = \bsigma^\kappa \sigma^\rho \bsigma^\lambda (k_{Q \rho} - k_{B \rho}) k_{Q \kappa} k_{B \xi} \nn\\
& \qquad = \bsigma^\kappa \sigma^\nu \bsigma^\lambda \left[ (k_B^2 + m^2) k_{Q \kappa} - (k_Q^2 + m^2) k_{B \kappa} + m^2 (k_{B \kappa} - k_{Q \kappa}) \right] \nn\\
& \qquad = \bsigma^\kappa \sigma^\nu \bsigma^\lambda \left[ (k_B^2 + m^2) k_{Q \kappa} - (k_Q^2 + m^2) k_{B \kappa} \right] - m^2 \bsigma^\kappa \sigma^\nu \bsigma^\lambda p_{Q \kappa}.
\end{align}
The first term cancels the corresponding factors in the denominator as before. Since all operators retain their form in the massive theory, these terms reproduce the appropriate terms on the right hand side the Ward identity by the same calculations as in the previous sections. Hence, we are left with the second term, which has the form of a mass-dependent 4-point function-like expression. In total, we find the mass-dependent terms in the divergence of \eqref{jQjfjpsijl_m}
\begin{align} \label{p4pt_psi}
& p_{Q \rho} \< \bar{j}^{\rho \dot{\alpha}}_{\bar{Q}}(\bs{p}_Q) j_{\j}^\mu(\bs{p}_A) j_{\psi}^\nu(\bs{p}_B) j_{\lambda \beta}(\bs{p}_L) \>_m = \frac{\I}{2} m^2 ( \bsigma^\rho \sigma^\nu \bsigma^\lambda)^{\da \a} \epsilon_{\a \b} p_{Q \rho} \times \nn\\
& \qquad\qquad \times \prod_{I} \int_{Eu} \frac{\D^4 \bs{k}_I}{(2 \pi)^4} \frac{\left( k_A - \frac{p_A}{2} \right)^\mu k_{L \lambda}}{(k_Q^2 + m^2) (k_A^2 + m^2) (k_B^2 + m^2) (k_L^2 + m^2)} \times \nn\\
& \qquad\qquad \times (2 \pi)^{16} \delta(\bs{k}_L + \bs{p}_B - \bs{k}_B) \delta(\bs{k}_B + \bs{p}_Q - \bs{k}_Q) \delta(\bs{k}_Q + \bs{p}_A - \bs{k}_A) \delta(\bs{k}_A + \bs{p}_L - \bs{k}_L) \nn\\
& \qquad + \frac{\I}{2} m^2 ( \bsigma^\kappa \sigma^\rho \bsigma^\nu)^{\da \a} \epsilon_{\a \b} p_{Q \rho} \times \nn\\
& \qquad\qquad \times \prod_{I} \int_{Eu} \frac{\D^4 \bs{k}_I}{(2 \pi)^4} \frac{\left( k_A - \frac{p_A}{2} \right)^\mu k_{Q \kappa}}{(k_Q^2 + m^2) (k_A^2 + m^2) (k_B^2 + m^2) ( | \bs{p}_Q - \bs{k}_Q|^2 + m^2)} \times \nn\\
& \qquad\qquad \times (2 \pi)^{16} \delta(\bs{k}_Q + \bs{p}_A - \bs{k}_A) \delta(\bs{k}_A + \bs{p}_L - \bs{k}_L) \delta(\bs{p}_Q - \bs{k}_Q + \bs{p}_B - \bs{k}_B) \delta(\bs{k}_B + \bs{k}_L) \nn\\
& \qquad + O(m^0).
\end{align}

Now the question is whether the 4-point functions are canceled by the correction term involving $\Delta \bar{j}_{\bQ}^{\mu \dot{\alpha}}$. If this is the case, the anomaly is absent. We find
\begin{align} \label{p4pt_Delta}
& \< \Delta \bar{j}^{\rho \dot{\alpha}}_{\bar{Q}}(\bs{p}_Q) j_{\j}^\mu(\bs{p}_A) j_{\psi}^\nu(\bs{p}_B) j_{\lambda \beta}(\bs{p}_L) \>_m = - \frac{\I}{2} m^2 ( \bsigma^\rho \sigma^\nu \bsigma^\lambda)^{\da \a} \epsilon_{\a \b} \times \nn\\
& \qquad\qquad \times \prod_{I} \int_{Eu} \frac{\D^4 \bs{k}_I}{(2 \pi)^4} \frac{\left( k_A - \frac{p_A}{2} \right)^\mu k_{L \lambda}}{(k_Q^2 + m^2) (k_A^2 + m^2) (k_B^2 + m^2) (k_L^2 + m^2)} \times \nn\\
& \qquad\qquad \times (2 \pi)^{16} \delta(\bs{k}_L + \bs{p}_B - \bs{k}_B) \delta(\bs{k}_B + \bs{p}_Q - \bs{k}_Q) \delta(\bs{k}_Q + \bs{p}_A - \bs{k}_A) \delta(\bs{k}_A + \bs{p}_L - \bs{k}_L) \nn\\
& \qquad - \frac{\I}{2} m^2 ( \bsigma^\rho \sigma^\kappa \bsigma^\nu)^{\da \a} \epsilon_{\a \b} \times \nn\\
& \qquad\qquad \times \prod_{I} \int_{Eu} \frac{\D^4 \bs{k}_I}{(2 \pi)^4} \frac{\left( k_A - \frac{p_A}{2} \right)^\mu ( p_{Q \kappa} - k_{Q \kappa})}{(k_Q^2 + m^2) (k_A^2 + m^2) (k_B^2 + m^2) ( | \bs{p}_Q - \bs{k}_Q|^2 + m^2)} \times \nn\\
& \qquad\qquad \times (2 \pi)^{16} \delta(\bs{k}_Q + \bs{p}_A - \bs{k}_A) \delta(\bs{k}_A + \bs{p}_L - \bs{k}_L) \delta(\bs{p}_Q - \bs{k}_Q + \bs{p}_B - \bs{k}_B) \delta(\bs{k}_B + \bs{k}_L).
\end{align}
Now we apply $p_{Q \rho}$ to this expression and add it to \eqref{p4pt_psi}. The first lines cancel each other, while the combination of the numerators of the second lines is
\begin{align} \label{add_m2_terms}
\bsigma^\kappa \sigma^\rho \bsigma^\nu k_{Q \kappa} p_{Q \rho} - \bsigma^\rho \sigma^\kappa \bsigma^\nu p_{Q \rho} ( p_{Q \kappa} - k_{Q \kappa}) = \bsigma^\nu \left[ ( |\bs{p}_Q - \bs{k}_Q|^2 + m^2 ) - (k_Q^2 + m^2) \right].
\end{align}
Hence the sum of \eqref{p4pt_psi} and \eqref{p4pt_Delta} produces a 3-point function-like contribution. As we can see, the 4-point function-like mass-dependent contributions are indeed canceled by the mass correction to the supercurrent. There is no SUSY anomaly.

\subsubsection{Exact match}

If we want to, we can also match the remaining 3-point function-like terms to the right hand side of the Ward identity \eqref{susy_ward}. In section \ref{sec:AnomShiftCalc} we have already shown that the terms leading in mass match. Here we have found two additional terms in \eqref{add_m2_terms} proportional to $m^2$, which can be written as
\begin{align}
& p_{Q \rho}\< \bar{j}^{\rho \dot{\alpha}}_{\bar{Q} m}(\bs{p}_Q) j_{\j}^\mu(\bs{p}_A) j_{\psi}^\nu(\bs{p}_B) j_{\lambda \beta}(\bs{p}_L) \>_m = \frac{\I}{2} m^2 \bsigma^{\nu \da \a} \epsilon_{\a \b} \times \nn\\
& \qquad\qquad \times \prod_{I} \int_{Eu} \frac{\D^4 \bs{k}_I}{(2 \pi)^4} \left[ \frac{\left( k_A - \frac{p_A}{2} \right)^\mu}{(k_Q^2 + m^2) (k_A^2 + m^2) (k_B^2 + m^2)} \times \right.\nn\\
& \qquad\qquad \times (2 \pi)^{16} \delta(\bs{k}_Q + \bs{p}_A - \bs{k}_A) \delta(\bs{k}_A + \bs{p}_L - \bs{k}_L) \delta(\bs{p}_Q - \bs{k}_Q + \bs{p}_B - \bs{k}_B) \delta(\bs{k}_B + \bs{k}_L) \nn\\
& \qquad\qquad\qquad\qquad \left. - \: (\bs{p}_A \leftrightarrow \bs{p}_B) \right] + O(m^0).
\end{align}
The two terms can be matched to the right hand side of the Ward identity. One term appears when the contraction of $F^{\ast}$ with $\j^\ast$ is taken into account while evaluating the 3-point function $\< j_{\chi \alpha}(\bs{p}_{QA}) j^{\nu}_{\j}(\bs{p}_B) j_{\lambda \beta}(\bs{p}_L) \>$. The other term appears in the evaluation of $\< j_{\chi \alpha}(\bs{p}_{QA}) j^{\nu}_{\psi}(\bs{p}_B) j_{\lambda \beta}(\bs{p}_L) \>$ and $\< \bj_{\bl}^{\da}(\bs{p}_{QA}) j^{\nu}_{\psi}(\bs{p}_B) j_{\lambda \beta}(\bs{p}_L) \>$. In the massive theory $\psi$ present in $j_{\lambda \beta}$ can be contracted with $\psi$ in $j_{\psi}^\nu$, while $\bpsi$ there can be contracted with $\bpsi$ in the first operator. Notice that all these additional massive terms appear in 3-point functions to which the counterterm \eqref{ct} contributes. This is the manifestation of the fact that when the PV regularization scheme is applied, the large mass limit contributes additional local terms to the generating functional.

Analogous calculations follow for the remaining two correlation functions in \eqref{4pt}. All terms that could contribute in the infinite mass limit match between the two sides of the Ward identity \eqref{susy_ward}. This is simply a consequence of the fact that the Ward identity holds in the massive theory with the conserved SUSY current. We conclude that there is no SUSY anomaly in the canonical Ward identity \eqref{susy_ward} in the Pauli-Villars renormalization scheme.

As a final remark let us underline the connection between the correction term $\Delta \bj_{\bQ}^{\mu \da}$ and scheme-dependence. In section \ref{sec:massive_tensor} we have shown that local terms following from the counterterm \eqref{Sfin_conf} are equivalent to the large mass limit contribution from $\Delta T^{(m)}_{\mu\nu}$. In the same fashion we can show that the inclusion of the correction term $\Delta \bj_{\bQ}^{\mu \da}$ parameterizes scheme-dependence of the SUSY anomalies and is equivalent to the inclusion of the counterterm \eqref{anom_fin}. Indeed, by taking the large mass limit one finds
\begin{align}
\lim_{M \rightarrow \infty} \lla \Delta \bj_{\bQ}^{\mu \da}(\bs{p}_Q) \: j_A^\mu(\bs{p}_A) \: j_A^\nu(\bs{p}_B) \: j_{\lambda \beta}(\bs{p}_L) \rra = \frac{\I}{192 \pi^2} \eta^{\mu\nu} p_{Q \rho} \bsigma^{\rho \da \a} \epsilon_{\a \b},
\end{align}
the expression proportional to the removable SUSY anomaly \eqref{anom_fin_corr} we have found in the non-SUSY invariant scheme. Thus, an extension of the massless supercurrent to the massive theory of the form $\bj^{\mu \da}_{\bQ m} = \bj^{\mu \da}_{\bQ} + \xi \Delta \bj^{\mu \da}_{\bQ}$ for any $\xi \neq 1$ would result in an anomaly in PV regularization.\\
Similarly if we tried to use dimensional regularization we would encounter additional supersymmetry-breaking finite $\frac{\epsilon}{\epsilon}$ terms that are to be removed by local counterterms of the type \eqref{anom_fin}. Indeed this was demonstrated by explicit calculation in a recent work~\cite{Eleftheriou:2020ray}.

\section{Summary}

In this paper we have addressed recent claims~\cite{Papadimitriou:2017kzw,Papadimitriou:2019yug,Katsianis:2019hhg,Papadimitriou:2019gel,An:2019zok} that supersymmetry is anomalous in the presence of a flavor or {\it R}-symmetry anomaly. In order to investigate the claim, we carry out detailed calculations in an explicitly supersymmetric Pauli-Villars renormalization scheme. Our analysis supports the results of \cite{Kuzenko:2019vvi}. Our conclusions are as follows:
\begin{enumerate}
\item In the context of a $\mathcal{N} = 1$ theory with a chiral $U(1)$ flavor symmetry we show that in Wess-Zumino gauge the presence of the chiral 't Hooft anomaly implies an anomaly in supersymmetry. By going away from Wess-Zumino gauge the anomaly can be removed from supersymmetry at the cost of generating specific terms in the flavor anomaly on top of the ABJ anomaly. Coupling the theory to the full vector supermultiplet the compensating fields can be used to transfer the anomaly from supersymmetry to the chiral flavor symmetry. Additionally, we derive the suitable counterterm \eqref{ct} as well as the correct Wess-Zumino consistency condition \eqref{fullWZ} and SUSY Ward identity \eqref{ward_1pt}.
\item In section \ref{sec:sugra} we show a similar mechanism in the context of coupling a $\mathcal{N} = 1$ theory to background conformal supergravity. The anomaly in the SUSY Ward identity can be removed at the cost of modifying other Ward identities. To do it, we introduce an additional chiral multiplet, whose fields couple to the conformal, $R$-, and $S$-SUSY anomalies. In this way we remove the $Q$-SUSY anomaly and generate additional contributions to the anomalies in the aforementioned symmetries. The procedure only affects the symmetries that have already been broken by anomalies, thus only additional constraints, such as a priori choice of current multiplet, can deem it inapplicable. Similarly to the flavor case, we derived the suitable counterterm \eqref{eq:CounterGrav}, as well as the modified Wess-Zumino consistency condition \eqref{eq:SugraCons} and the $Q$-SUSY Ward identity \eqref{Qsusy_ward}. The introduction of the extra chiral multiplet is tantamount to coupling the theory to old minimal supergravity.
\item The simpler model of the interplay between trace and transverse anomalies in the theory of a free massless real boson displays similar features. As reviewed in section \ref{sec:trace_an}, the two anomalies are two faces of the same transverse-trace anomaly. In particular, one can reshuffle the anomalous contribution from the transverse Ward identity to the trace Ward identity and vice versa. Similarly the anomalies considered here can be moved from supersymmetry to other symmetries. However in the SUSY case one cannot restore any other symmetry at the cost of introducing an anomaly in supersymmetry. If some additional constraints, a specific current multiplet, or gauge conditions are imposed, it may be impossible to remove the anomalous contributions from the SUSY Ward identity. 
\item In section \ref{sec:WZ_model} we have presented a detailed analysis of various correlation functions for the free massless Wess-Zumino theory coupled to a background vector multiplet. We analyzed a number of correlators, which are susceptible to the SUSY Ward identities (\ref{ward_1pt}, \ref{susy_ward}) and flavor Ward identity \eqref{JA_ward} as well as the Wess-Zumino condition \eqref{true_WZ_cond}. We show that in the explicitly supersymmetric Pauli-Villars renormalization scheme the SUSY anomaly is absent. Instead, the flavor anomaly picks up a number of specific terms on top of the usual ABJ anomaly, \eqref{an_PV}. 
\end{enumerate}

\begin{acknowledgments}

AB is supported by the Advanced ERC grant SM-grav, No 669288. The work of GF is supported by the ERC under the STG grant 639220 and by Vetenskapsr\aa{}det under grant 2018-05572. VP acknowledges discussions with F.~Ciceri and S.~Theisen while GF thanks M.~Dine. During the work on the project VP was supported by ERC STG grant 639220 and AB was supported by Knut and Alice Wallenberg Foundation under Grant No. 113410212.

\end{acknowledgments}

\appendix

\section{Multiplets and variations}

Conventions follow \cite{Wess:1992cp}. In particular, all spinors are anti-commuting and their contractions are
\begin{align}
\psi \chi & = \psi^{\alpha} \chi_{\alpha}, & \bpsi \bchi & = \bpsi_{\da} \bchi^{\da}, & \psi \sigma^\mu \bchi & = \psi^{\alpha} \sigma^\mu_{\a \da} \bchi^{\da}, & \bpsi \bsigma^\mu \chi & = \bpsi_{\da} \bsigma^{\mu \da \a} \chi_{\a}.
\end{align}
For more details, consult appendices A and B of \cite{Wess:1992cp}. Fermionic functional derivatives act from the left.

\subsection{Chiral multiplet}

The chiral multiplet $\Phi = (\j, \psi, F)$ consist of the physical fields $\j, \psi$ and auxiliary field $F$. It is given by
\begin{align} \label{chiral_multi}
\Phi & = \j + \sqrt{2} \theta \psi + \I \theta \sigma^\mu \btheta \partial_\mu \j + \theta \theta F + \frac{\I}{\sqrt{2}} \theta \theta \btheta \bsigma^\mu \partial_\mu \psi + \frac{1}{4} \theta \theta \btheta \btheta \Box \j, \\
\Phi^+ & = \j^\ast + \sqrt{2} \btheta \bpsi - \I \theta \sigma^\mu \btheta \partial_\mu \j^\ast + \btheta \btheta F^\ast + \frac{\I}{\sqrt{2}} \btheta \btheta \theta \sigma^\mu \partial_\mu \bpsi + \frac{1}{4} \theta \theta \btheta \btheta \Box \j^\ast.
\end{align}
The supersymmetry transformations of the component fields are
\begin{align}
\delta_{\ep, \bep} \j & = \sqrt{2} \ep \psi, & \delta_{\ep, \bep} \j^\ast & = \sqrt{2} \bep \bpsi, \label{susy_chiral_first} \\
\delta_{\ep, \bep} \psi_{\alpha} & = \I \sqrt{2} \sigma^\mu_{\a \da} \bep^{\da} \partial_\mu \j + \sqrt{2} \ep_{\a} F, & \delta_{\ep, \bep} \bpsi_{\da} & = - \I \sqrt{2} \ep^{\a} \sigma_{\a \da}^\mu \partial_\mu \j^\ast + \sqrt{2} \bep_{\da} F^\ast, \\
\delta_{\ep, \bep} F & = \I \sqrt{2} \bep \bsigma^{\mu} \partial_\mu \psi, & \delta_{\ep, \bep} F^{\ast} & = \I \sqrt{2} \ep \sigma^\mu \partial_\mu \bpsi. \label{susy_chiral_last}
\end{align}

\subsection{Vector multiplet} \label{App:VecMult}

The vector multiplet $V = (C, \chi, \mathcal{M}, A_\mu, \lambda, D)$ consist of the fields $A_\mu, \lambda, D$ compensators fields $C, \chi, \mathcal{M}$. It is given by
\begin{align} \label{vector_multi}
V & = C + \I \theta \chi - \I \btheta \bchi + \frac{\I}{2} \theta \theta \mathcal{M} - \frac{\I}{2} \btheta \btheta \mathcal{M}^{\ast} - \theta \sigma^\mu \btheta A_\mu + \I \theta \theta \btheta \left[ \bl + \frac{\I}{2} \bsigma^\mu \partial_\mu \chi \right] \nn\\
& \qquad - \I \btheta \btheta \theta \left[ \lambda + \frac{\I}{2} \sigma^\mu \partial_\mu \bchi \right] + \frac{1}{2} \theta \theta \btheta \btheta \left[ D + \frac{1}{2} \Box C \right].
\end{align}
The supersymmetry transformations of the component fields are
\begin{align}
\delta_{\ep,\bep} A_\mu & = \I \ep (\sigma_\mu \bl - \I \partial_\mu \chi) + \I \bep (\bsigma_\mu \lambda - \I \partial_\mu \bchi), \label{susy_tran_first} \\
\delta_{\ep,\bep} \lambda^{\alpha} & = - \frac{1}{2} \ep^{\b} (2 \sigma)^{\mu \nu \ \a}_{\: \ \ \b} F_{\mu\nu} + \I D \ep^{\a}, \\
\delta_{\ep,\bep} D & = - \ep \sigma^\mu \partial_\mu \bl + \bep \bsigma^\mu \partial_\mu \lambda, \\
\delta_{\ep,\bep} \chi^\alpha & = - \I \bep_{\da} \bsigma^{\mu \da \a} (A_\mu - \I \partial_\mu C) + \mathcal{M} \ep^\alpha, \\
\delta_{\ep,\bep} \mathcal{M} & = 2 \I \bep \bsigma^{\mu} \partial_\mu \chi + 2 \bep \bl, \\
\delta_{\ep,\bep} C & = \I (\ep \chi - \bep \bchi). \label{susy_tran_last}
\end{align}
The supergauge transformations of the vector multiplet are given by
\begin{align}
\delta_{\Lambda} C & = \sigma, \label{gauge_tran_first} \\
\delta_{\Lambda} \chi & = - \I \sqrt{2} \Upsilon, \\
\delta_{\Lambda} \mathcal{M} & = - 2 \I \mathfrak{f}, \\
\delta_{\Lambda} A_\mu & = \partial_\mu \theta, \\
\delta_{\Lambda} \lambda & = 0, \\
\delta_{\Lambda} D & = 0, \label{gauge_tran_last}
\end{align}
where $\Lambda = ( \frac{1}{2} \sigma + \frac{\I}{2} \theta, \Upsilon, \mathfrak{f})$ is the chiral multiplet, $\sigma$ and $\theta$ are real and $\theta$ is identified with the usual gauge transformation parameter.

\section{Wess-Zumino model}

\subsection{Lagrangian}

The Lagrangian of the Wess-Zumino model coupled to the vector multiplet of sources is
\begin{align} \label{lag}
L & = \Phi^+ e^{g V} \Phi|_{g=1} + L_m \nn\\
& = e^{g \sC} \left[ L_0 + g L_1 + g^2 L_2 + g^3 L_3 + g^4 L_4 \right]_{g=1} + L_m,
\end{align}
where
\begin{align}
L_0 & = \frac{1}{4} \left( \j \Box \j^\ast + \j^\ast \Box \j - 2 \partial_\mu \j^\ast \partial^\mu \j \right) - \frac{\I}{2} \left( \bar{\psi} \bar{\sigma}^\mu \partial_\mu \psi - \partial_\mu \bar{\psi} \bar{\sigma}^\mu \psi \right) + F F^\ast,
\end{align}
\begin{align}
L_1 & = \frac{\I}{2} \sA_\mu \left( \j^\ast \partial_\mu \j - \j \partial_\mu \j^\ast \right) + \frac{1}{2} \sA_\mu \, \bar{\psi} \bar{\sigma}^\mu \psi \nn\\
& \qquad + \frac{1}{2 \sqrt{2}} \left( \j \, \schi \sigma^\mu \partial_\mu \bar{\psi} - \j \, \partial_\mu \schi \sigma^\mu \bar{\psi} + \j^\ast \, \partial_\mu \bar{\schi} \bar{\sigma}^\mu \psi - \j^\ast \, \bar{\schi} \bar{\sigma}^\mu \partial_\mu \psi \right.\nn\\
& \qquad\qquad \left. + \partial_\mu \j^\ast \, \bar{\schi} \bar{\sigma}^\mu \psi - \partial_\mu \j \, \schi \sigma^\mu \bpsi + 2 \I F \, \bar{\schi} \bar{\psi} - 2 \I  F^\ast \schi \psi \right) \nn\\
& \qquad + \frac{\I}{\sqrt{2}} \left( \j^\ast \, \sl \psi - \j \, \bar{\sl} \bar{\psi} \right) \nn\\
& \qquad + \frac{\I}{2} F^\ast \j \smcM - \frac{\I}{2} F \j^\ast \smcM^\ast + \frac{1}{2} \j \j^\ast \left( \sD + \frac{1}{2} \Box \sC \right),
\end{align}
\begin{align}
L_2 & = - \frac{1}{4} \sA_\mu \sA^\mu \j \j^\ast - \frac{\I}{2 \sqrt{2}} \sA_\mu \left( \j \schi \sigma^\mu \bpsi + \j^\ast \bar{\schi} \bar{\sigma}^\mu \psi \right) \nn\\
& \qquad - \frac{\I}{4} \j \j^\ast \left( \bar{\schi} \bar{\sigma}^\mu \partial_\mu \schi - \partial_\mu \bar{\schi} \bar{\sigma}^\mu \schi \right) + \frac{\I}{4} \left( \j \partial_\mu \j^\ast - \j^\ast \partial_\mu \j \right) \bar{\schi} \bar{\sigma}^\mu \schi \nn\\
& \qquad + \frac{1}{2} \psi \schi \, \bar{\psi} \bar{\schi} + \frac{1}{4} \left( F \j^\ast \, \bar{\schi} \bar{\schi} + F^\ast \j \, \schi \schi \right) \nn\\
& \qquad - \frac{1}{2} \j \j^\ast \left( \sl \schi + \bar{\sl} \bar{\schi} \right) \nn\\
& \qquad - \frac{1}{2 \sqrt{2}} \j \, \bar{\psi} \bar{\schi} \, \smcM - \frac{1}{2 \sqrt{2}} \j^\ast \, \psi \schi \, \smcM^\ast + \frac{1}{4} \j \j^\ast \smcM \smcM^\ast
\end{align}
\begin{align}
L_3 & = \frac{1}{4} \sA_\mu \j \j^\ast \, \bar{\schi} \bar{\sigma}^\mu \schi + \frac{\I}{4 \sqrt{2}} \left( \j \, \bar{\psi} \bar{\schi} \, \schi \schi - \j^\ast \, \psi \schi \, \bar{\schi} \bar{\schi}\right) \nn\\
& \qquad + \frac{\I}{8} \j \j^\ast \left( \bar{\schi} \bar{\schi} \, \smcM - \schi \schi \, \smcM^\ast \right),
\end{align}
\begin{align}
L_4 & = \frac{1}{16} \j \j^\ast \, \schi \schi \, \bar{\schi} \bar{\schi},
\end{align}
\begin{align}
L_m = m \left( \j F + \j^\ast F^\ast - \frac{1}{2} ( \psi \psi + \bar{\psi} \bar{\psi} ) \right).
\end{align}
Non-dynamical source are in bold. In the Wess-Zumino gauge $\chi = C = \mathcal{M} = 0$ and we find
\begin{align}
L_0^{\text{WZ}} & = L_0, \\
L_1^{\text{WZ}} & = \frac{\I}{2} \sA_\mu \left( \j^\ast \partial_\mu \j - \j \partial_\mu \j^\ast \right) + \frac{1}{2} \sA_\mu \, \bar{\psi} \bar{\sigma}^\mu \psi \nn\\
& \qquad + \frac{\I}{\sqrt{2}} \left( \j^\ast \, \sl \psi - \j \, \bar{\sl} \bar{\psi} \right) + \frac{1}{2} \j \j^\ast \sD, \\
L_2^{\text{WZ}} & = - \frac{1}{4} \sA_\mu \sA^\mu \j \j^\ast, \\
L_3^{\text{WZ}} & = L_4^{\text{WZ}} = 0.
\end{align}

\subsection{Propagators}

Non-vanishing propagators are
\begin{align}
& \< \j(\bs{p}) \j^\ast(\bs{p}') \> = (2 \pi)^4 \delta(\bs{p} + \bs{p}') \frac{-\I}{p^2 + m^2}, \\
& \< \j(\bs{p}) F(\bs{p}') \> = (2 \pi)^4 \delta(\bs{p} + \bs{p}') \frac{\I m}{p^2 + m^2}, \\
& \< F(\bs{p}) F^\ast(\bs{p}') \> = (2 \pi)^4 \delta(\bs{p} + \bs{p}') \frac{\I p^2}{p^2 + m^2}, \\
& \< \psi_\alpha(\bs{p}) \bar{\psi}_{\dot{\beta}}(\bs{p}') \> = (2 \pi)^4 \delta(\bs{p} + \bs{p}') \frac{-\I \sigma^\mu_{\alpha \dot{\beta}} p_\mu}{p^2 + m^2}, \label{corr_mid} \\
& \< \psi_\alpha(\bs{p}) \psi^\beta(\bs{p}') \> = (2 \pi)^4 \delta(\bs{p} + \bs{p}') \frac{-\I m \delta_{\alpha}^{\beta}}{p^2 + m^2}, \\
& \< \bpsi^{\da}(\bs{p}) \bpsi_{\db}(\bs{p}') \> = (2 \pi)^4 \delta(\bs{p} + \bs{p}') \frac{-\I m \delta^{\da}_{\db}}{p^2 + m^2}. \label{corr_last}
\end{align}

\subsection{Seagull terms} \label{sec:seagulls}

Using the Lagrangian \eqref{lag} we can express the correlation functions of the operators $J_{\phi_k}$ through the correlators of the $j_{\phi_k}$. We write $\bs{p}_{IJ} = \bs{p}_I + \bs{p}_J$.

2-point functions are identical,
\begin{equation}
\< J_{\phi_1} J_{\phi_2} \> = \< j_{\phi_1} j_{\phi_2} \>.
\end{equation}

Relations between 3-point functions are
\begin{align}
\lla J^\mu_A(\bs{p}_1) J^\nu_A(\bs{p}_2) J^\rho_A(\bs{p}_3) \rra & = \lla j^\mu_A(\bs{p}_1) j^\nu_A(\bs{p}_2) j^\rho_A(\bs{p}_3) \rra, \label{seagulls_3pt1} \\
\lla J^\mu_A(\bs{p}_1) J^\nu_A(\bs{p}_2) J_D(\bs{p}_3) \rra & = \lla j^\mu_A(\bs{p}_1) j^\nu_A(\bs{p}_2) j_D(\bs{p}_3) \rra \nn\\
& \qquad \qquad + \I \eta^{\mu\nu} \lla j_D(\bs{p}_{12}) j_D(\bs{p}_3) \rra, \\
\lla J^\mu_A(\bs{p}_1) \bar{J}_{\bar{\lambda} \dot{\beta}}(\bs{p}_2) J_{\lambda \beta}(\bs{p}_3) \rra & = \lla j^\mu_A(\bs{p}_1) \bar{j}_{\lambda \dot{\beta}}(\bs{p}_2) j_{\lambda \beta}(\bs{p}_3) \rra, \\
\lla J^\mu_A(\bs{p}_1) J_{\chi \alpha}(\bs{p}_2) J_{\lambda \beta}(\bs{p}_3) \rra & = \lla j^\mu_A(\bs{p}_1) j_{\chi \alpha}(\bs{p}_2) j_{\lambda \beta}(\bs{p}_3) \rra \nn\\
& \qquad\qquad - \frac{\I}{2} \sigma^\mu_{\a \da} \lla \bj^{\da}_{\bl}(\bs{p}_{12}) j_{\lambda \b}(\bs{p}_3) \rra, \label{seagull:Achilambda} \\
\lla J_D(\bs{p}_1) J_{\chi \alpha}(\bs{p}_2) J_{\lambda \beta}(\bs{p}_3) \rra & = \lla j_D(\bs{p}_1) j_{\chi \alpha}(\bs{p}_2) j_{\lambda \beta}(\bs{p}_3) \rra \nn\\
& \qquad\qquad - \I \epsilon_{\alpha \beta} \lla j_D(\bs{p}_{1}) j_D(\bs{p}_{23}) \rra. \label{seagulls_3pt2}
\end{align}

By coupling the theory to gravitino, $\bpsi_{\mu\da}$, \cite{Sohnius:1982fw}, one can also find,
\begin{align} \label{seagulls_3ptq1}
\lla \bJ^{\rho \da}_{\bQ}(\bs{p}_Q) J_D(\bs{p}_D) J_{\lambda \beta}(\bs{p}_L) \rra & = \lla \bj^{\rho \da}_{\bQ}(\bs{p}_Q) j_D(\bs{p}_D) j_{\lambda \beta}(\bs{p}_L) \rra \nn\\
& \qquad\qquad + \I a \bsigma^{\rho \da \a} \epsilon_{\a \b} \lla j_D(\bs{p}_{QL}) j_D(\bs{p}_D) \rra, \\
\lla \bJ^{\rho \da}_{\bQ}(\bs{p}_Q) J_A^\mu(\bs{p}_D) J_{\lambda \beta}(\bs{p}_L) \rra & = \lla \bj^{\rho \da}_{\bQ}(\bs{p}_Q) j_A^\mu(\bs{p}_A) j_{\lambda \beta}(\bs{p}_L) \rra \nn\\
& \qquad\qquad - \I ( \bsigma^{\mu} \sigma^{\rho})^{\da}_{\ \db} \lla \bj_{\bl}^{\db}(\bs{p}_{QA}) j_{\lambda \beta}(\bs{p}_L) \rra. \label{seagulls_3ptq2}
\end{align}

Finally, for the 4-point function of interest,
\begin{align} \label{seagulls_4pt}
& \lla \bJ^{\rho \da}_{\bQ}(\bs{p}_Q) J_A^\mu(\bs{p}_A) J_A^\nu(\bs{p}_B) J_{\lambda \beta}(\bs{p}_L) \rra = \lla \bj^{\rho \da}_{\bQ}(\bs{p}_Q) j_A^\mu(\bs{p}_A) j_A^\nu(\bs{p}_B) j_{\lambda \beta}(\bs{p}_L) \rra \nn\\
& \qquad\qquad - \I ( \bsigma^\mu \sigma^\rho)^{\da}_{\ \db} \lla \bj^{\db}_{\bl}(\bs{p}_{QA}) j_A^\nu(\bs{p}_B) j_{\lambda \beta}(\bs{p}_L) \rra + (\bs{p}_A \leftrightarrow \bs{p}_B, \mu \leftrightarrow \nu) \nn\\
& \qquad\qquad + \I \eta^{\mu\nu} \lla \bj^{\rho \da}_{\bQ}(\bs{p}_{Q}) j_D(\bs{p}_{AB}) j_{\lambda \beta}(\bs{p}_L) \rra \nn\\
& \qquad\qquad + \I \bsigma^{\rho \da \a} \epsilon_{\a \b} \lla j^{\mu}_A(\bs{p}_{A}) j^{\nu}_A(\bs{p}_{B}) j_D(\bs{p}_{QD}) \rra \nn\\
& \qquad\qquad - \eta^{\mu\nu} \bsigma^{\rho \da \a} \epsilon_{\a \b} \lla j_{D}(\bs{p}_{AB}) j_D(\bs{p}_{QD}) \rra.
\end{align}

\subsection{Ward identities} \label{app:Ward}

The following 3-point functions are useful as well,
\begin{align} \label{susyward3}
	& p_{Q \rho} \lla \bar{j}^{\rho \dot{\alpha}}_{\bar{Q}}(\bs{p}_Q) j_D(\bs{p}_D) j_{\lambda \beta}(\bs{p}_L) \rra = \nn\\
	& \qquad + \I \lla \bar{j}_{\bar{\lambda}}^{\dot{\alpha}}(\bs{p}_{QD}) j_{\lambda \beta}(\bs{p}_L) \rra, \nn\\
	& \qquad - \I \bsigma^{\rho \da \a} \epsilon_{\a \b} p_{QL \rho} \lla j_D(\bs{p}_D) j_D(\bs{p}_{QL}) \rra
\end{align}
and
\begin{align} \label{susyward3a}
	& p_{Q \rho} \lla \bar{j}^{\rho \dot{\alpha}}_{\bar{Q}}(\bs{p}_Q) j_A^\mu(\bs{p}_A) j_{\lambda \beta}(\bs{p}_L) \rra = \nn\\
	& \qquad + \I \left[ - p_Q^\mu \delta^{\da}_{\db} + (2 \bsigma)^{\mu \rho \da}_{\ \ \ \ \db} p_{QA \rho} \right] \lla \bar{j}_{\bar{\lambda}}^{\dot{\beta}}(\bs{p}_{QA}) j_{\lambda \beta}(\bs{p}_L) \rra \nn\\
	& \qquad + \I \bsigma^{\rho \da \a} \epsilon_{\a \b} \lla j_A^\mu(\bs{p}_{A}) j_{A \rho}(\bs{p}_{QL}) \rra.
\end{align}

\section{Results in 4-component notation} \label{app:4comp}

We want to convert from 4-component Majorana spinors of \cite{Freedman:2012zz} to 2-component Weyl spinors of \cite{Wess:1992cp}. Both books use the mostly plus convention for the metric $\eta_{\mu\nu} \sim (-1,1,1,1)$ and the standard expressions for the Pauli matrices $\sigma_i = \sigma^i$, $i=1,2,3$. They differ in the definition of conjugation, though, as well as the definition of $\sigma^0$. By FVP we denote the conventions of \cite{Freedman:2012zz} and by WB the conventions of \cite{Wess:1992cp}. One has
\begin{align}
& \sigma^0_{\FVP} = - \sigma^0_{\WB}, && \sigma^i_{\FVP} = \sigma^i_{\WB}, \\
& \bar{\sigma}^0_{\FVP} = \bar{\sigma}^0_{\WB}, && \bar{\sigma}^i_{\FVP} = - \bar{\sigma}^i_{\WB}.
\end{align}
While in both conventions gamma matrices are defined in the same way, they satisfy different Clifford algebra conditions,
\begin{align}
& \gamma^\mu = \left( \begin{array}{cc} 0 & \sigma^\mu \\ \bsigma^\mu & 0 \end{array} \right), && \{ \gamma^\mu, \gamma^\nu \} = \pm 2 \eta^{\mu\nu} \bs{1},
\end{align}
with plus sign for FVP and minus for WB. To convert from one convention to another one could expand everything in components, but that is obviously not a convenient way to proceed. Instead, we want to reproduce the FVP Clifford algebra with the WB sigma matrices. This means that we substitute,
\begin{align}
\gamma^\mu_{\FVP} \: \longmapsto \: \left( \begin{array}{cc} 0 & \I \sigma^\mu_{\WB} \\ \I \bsigma^\mu_{\WB} & 0 \end{array} \right).
\end{align}
Note that the matrix on the right hand side is not equal to $\gamma^\mu_{\FVP}$, but it obeys the same algebra as $\gamma^\mu_{\FVP}$, thus all invariants are identical. With the conjugation in \cite{Freedman:2012zz} defined as $\bar{\Psi} = \I \Psi^{\dagger} \gamma^0_{\FVP}$ we obtain the following conversion rules.
\begin{align}
& \Psi \mapsto \left( \begin{array}{c} \psi \\ \bpsi \end{array} \right), && \bar{\Psi} \mapsto ( \psi, \bpsi), \\
& P_L \Psi \mapsto \left( \begin{array}{c} \psi \\ 0 \end{array} \right), && P_R \Psi \mapsto \left( \begin{array}{c} 0 \\ \bpsi \end{array} \right), \\
& \gamma^\mu \mapsto \left( \begin{array}{cc} 0 & \I \sigma^\mu \\ \I \bsigma^\mu & 0 \end{array} \right), && \gast \mapsto \left( \begin{array}{cc} 1 & 0 \\ 0 & -1 \end{array} \right), \\
& \gamma^{\mu\nu} \mapsto - \left( \begin{array}{cc} (2 \sigma)^{\mu\nu} & 0 \\ 0 & (2\bsigma)^{\mu\nu} \end{array} \right).
\end{align}
We dropped indices $\FVP$ and $\WB$ on the left and right hand sides respectively. In particular
\begin{align}
\bar{\Psi} X \: & \longmapsto \: \psi \chi + \bpsi \bchi, & \I \bar{\Psi} \gast X \: & \longmapsto \: \I \psi \chi - \I \bpsi \bchi, \\
\bar{\Psi} \gamma^\mu X \: & \longmapsto \: \I \bpsi \bsigma^\mu \chi + \I \psi \sigma^\mu \bchi, & \I \bar{\Psi} \gast \gamma^\mu X \: & \longmapsto \: \bpsi \bsigma^\mu \chi - \psi \sigma^\mu \bchi.
\end{align}
All those bi-linears are self-adjoint.

\subsection{Supergravity}

Here we present our results in 4-component notation. Conventions of \cite{Papadimitriou:2019gel} are recovered by rescaling of the gauge field and the gauge parameter, $A^R_\mu \mapsto -2 A_\mu^R/3$ and $\theta^R \mapsto -2 \theta^R/3$. Note the change of the sign in the definition of $\phi_{\mu}$. In this subsection we keep the fermionic parameters divided by two w.r.t. to the remainder of the paper. In order to match the rest of the paper, one must rescale here $\ep \mapsto 2\ep$ and $\eta \mapsto 2\eta$.

The SUSY variations are
\begin{eqnarray} \label{eqG:vielbeinTransf}
\delta e_\mu^a &=& \frac{1}{2} \bar{\ep}\gamma^a \psi_\mu - \sigma e_\mu^a - {\lambda^a}_{b} e_\mu^b, \\ \label{eqG:ARmuTransf}
\delta A_\mu^R &=& \partial_\mu \theta^R + \frac{1}{2} \I \bar{\ep} \gast \phi_\mu + \frac{1}{2} \I \bar{\eta} \gast \psi_\mu, \\ \label{eqG:PsimuTransf}
\delta \psi_\mu &=& \frac{3}{2} \I \theta^R \gast \psi_\mu + \left( D^{\omega}_\mu - \frac{3}{2} \I A_\mu^R \gast \right) \ep - \gamma_a e_\mu^a \eta - \frac{1}{2} \sigma \phi_\mu - \frac{1}{4} \lambda^{ab} \gamma_{ab} \psi_\mu,
\end{eqnarray}
where
\begin{equation} \label{eqG:PhiSugra}
\phi_\mu = \frac{2}{3}\gamma^\nu \left( \mathcal{D}_{[\mu}\psi_{\nu]}-\frac{\I}{4}{\epsilon_{\mu \nu}}^{\rho \sigma} \gamma_5 \mathcal{D}_\rho \psi_\sigma \right)
\end{equation}
and
\begin{align}
\mathcal{D}_\mu \psi_\nu & = \left(D_\mu^\omega - \frac{3}{2} \I \gast A_\mu^R \right) \psi_\nu \; .
\end{align}
By $D^\omega_\mu$ we denote the covariant derivative with connection $\omega$ given by
\begin{align}
\omega_{\mu a b}(e, \psi) & = \omega_{\mu a b}(e) + \frac{1}{4} \left( \bpsi_a \gamma_\mu \psi_b + \bpsi_\mu \gamma_a \psi_b - \bpsi_\mu \gamma_b \psi_a \right).
\end{align}
Finally the SUSY transformation rule for the derived field $\phi$ reads
\begin{equation}
\delta_{\ep} \phi_\mu = \frac{1}{2}\left(\I \gamma_5 F_{\mu \nu}^R - \frac{1}{4} {\epsilon_{\mu \nu}}^{\rho \sigma} F_{\rho \sigma}^R \right)\gamma^\nu \varepsilon+\frac{1}{2} P_{\mu \nu}\gamma^\nu \varepsilon \; .
\end{equation}
Variations of the compensators are
\begin{eqnarray}
\delta Z &=& (\sigma + \I \theta^R) Z +  \frac{1}{\sqrt{2}} \bar{\ep}P_L \chi^R  \;  ; \\
\delta P_L \chi^R &=&  \frac{1}{2} (3\sigma - \I \theta^R)  P_L \chi^R +  \frac{1}{\sqrt{2}}P_L (\gamma^\mu \mathcal{D}_\mu Z + \mathcal{F})  \ep + \sqrt{2}Z P_L \eta \; ; \\ \label{eqG:ChiralGravTransf}
\delta \mathcal{F} &=&  2 (\sigma- \I \theta^R) \mathcal{F} + \frac{1}{\sqrt{2}}\bar{\ep}\gamma^\mu P_L  \left[(D_\mu^\omega + \frac{1}{2} \I A_\mu^R) \chi^R - \frac{1}{\sqrt{2}} (\gamma^\nu \mathcal{D}_\nu Z + \mathcal{F})\psi_\mu - \sqrt{2}Z \phi_\mu \right] \; ,
\end{eqnarray}
where
\begin{align}
\mathcal{D}_\mu Z & = \partial_\mu Z- \I A_\mu^R Z- \frac{1}{\sqrt{2}}\bar{\psi}_\mu P_L \chi^R \;, \\
\mathcal{D}_\mu P_L \chi^R & = P_L \left(D_\mu^\omega+ \frac{\I}{2} A_\mu \right) \chi^R -  \frac{1}{\sqrt{2}}\left( \gamma^\nu \mathcal{D}_\nu Z + \mathcal{F} \right)\psi_\mu - \sqrt{2} Z \phi_\mu
\end{align}

The fields in old minimal SUGRA are defined as
\begin{align}
\tilde{\psi}_\mu & = \psi_\mu + \frac{1}{\sqrt{2}} \gamma_\mu \chi^R - \frac{1}{2} (3 \I\pi \gamma_5 - \tau) \psi_\mu \;, \label{eqG:psiOMsugra} \\
\tilde{e}^{a}_{\mu} & = e^{a}_{\mu} - \tau e^{a}_{\mu} , \\
\tilde{A}^R_\mu & = A_\mu^R - \frac{1}{\sqrt{2}} \I \bar{\chi}^R \gamma_5 \psi_\mu  - \partial_{\mu} \pi \; . \label{eqG:AOMsugra}
\end{align}
and their variations read
\begin{eqnarray} \label{eqG:vielbeinTransfNM}
\delta \tilde{e}_\mu^a &=& - {\tilde{\lambda}_a}^b e_\mu^b + \frac{1}{2} \bar{\ep}\gamma^a \tilde{\psi}_\mu \;  ; \\ \label{eqG:ARmuTransfNM}
\delta \tilde{A}_\mu^R &=&  - \frac{\I}{2} \bar{\ep} \gamma_5 \tilde{\phi}_\mu + \frac{1}{4} \bar{\ep}(\gamma^\nu\tilde{A}^{R}_{\nu} - \I \gamma_5 \tilde{M} - \tilde{N}) \tilde{\psi}_\mu \; ; \\
\delta \tilde{\psi}_\mu &=& - \frac{1}{4} \tilde{\lambda}^{ab} \gamma_{ab} \tilde{\psi}_\mu + \left(D_\mu^\omega - \frac{3 \I}{2} \tilde{A}_\mu^R \gamma_5 + \frac{1}{2} \gamma_\mu(-\I \gamma_5\gamma_\nu\tilde{A}^{R\nu} + \tilde{M}+\I \gamma_5 \tilde{N}) \right) \ep \label{eqG:PsimuTransfNM}
\; ,
\end{eqnarray}
where $\mathcal{F} = \tilde{M}+\I \tilde{N}$ and we defined
\begin{equation} \label{eq:TilPhimu}
\tilde{\phi}_\mu= \phi_\mu + \frac{1}{\sqrt{2}}\mathcal{D}_\mu \bar{\chi}^R \;
\end{equation}
The reported SUSY anomaly reads
\begin{align}
\mathcal{A}_{\ep} & =  \frac{3c - 5a}{8 \pi^2} \I \epsilon^{\mu\nu\rho\tau} F_{\rho\tau}^R A^R_{\mu} \gamma^5 \phi_\nu + \frac{a-c}{8 \pi^2} \epsilon^{\lambda\kappa\rho\tau} \nabla_\mu ( A_\rho^R R_{\lambda \kappa}^{\ \ \: \mu \nu} ) \gamma_{(\nu} \psi_{\tau)} \nn\\
& \qquad\qquad + \frac{c-a}{32 \pi^2} \epsilon^{\mu\nu \kappa\lambda} F_{\mu\nu}^R R_{\kappa\lambda}^{\ \ \: \rho\tau} \gamma_\rho \psi_\tau + O(\psi^3).
\end{align}
The $S$-anomaly is defined in terms of currents as
\begin{equation} \label{eq:AchiCurrents4c}
\mathcal{A}_{\eta}= \frac{1}{2} \I \gamma_5 \psi_\mu \< {J^R}^\mu\>+ \gamma_\mu \< {J^Q}^\mu\> \; .
\end{equation}
The suitable counterterm in the 4-component notation reads
\begin{align} \label{eqG:CounterGrav}
S_{\text{ct}}^R = \frac{1}{8 \pi^2}\left[3 (a-c) P_1 - 2 a P_2 \right]   \; ,
\end{align}
where
\begin{align} \label{eqG:Counter1}
P_1 & =  -\sqrt{2} \bar{\chi}^R \left( -\frac{1}{6}\gamma_{\rho \sigma} {W_{\mu \nu}}^{\rho \sigma}  + \I \gamma_5 F_{\mu \nu}^R \right) (\gamma^{[\mu} \phi^{\nu]}+ \mathcal{D}^{[\mu} \psi^{\nu]}) \nn\\
& \qquad - A_\mu^R \epsilon^{\mu \nu \rho \sigma}\left( \frac{4}{3}\bar{\phi}_\nu \mathcal{D}_{\rho} \psi_{\sigma}+ \frac{2}{3} \bar{\phi}_\nu \gamma_\rho \phi_\sigma \right)
\end{align}
and
\begin{align} \label{eqG:Counter2}
P_2 & =  -\sqrt{2} \bar{\chi}^R \left( 2 \I \gamma_5 F_{\mu \nu}^R - \tfrac{1}{4} {\epsilon_{\mu \nu}}^{\rho \sigma} F_{\rho \sigma}^R - P_{\mu \nu} + \tfrac{1}{6} R \, g_{\mu\nu} \right) \gamma^\mu \phi^\nu \nn\\
& \qquad - \I A_\nu^R \left( \sqrt{2} \mathcal{D}_\mu \bar{\chi}^R +  \bphi_{\mu} \right)  \left( \gamma_5 \gamma^\nu \phi^\mu - \gamma^\mu \gamma_5 \gamma^\nu \gamma_\rho \phi^\rho \right) + 2 A_\mu^R A_\nu^R P^{\mu \nu}
\; ,
\end{align}
where $P_{\mu \nu}= \frac{1}{2}(R_{\mu \nu}- \frac{1}{6} g_{\mu \nu} R)$ and we have omitted the terms proportional to $\mathcal{F}$.
In the above equations we have defined the Weyl tensor
\begin{equation}
W^{\rho \sigma \mu \nu}= R^{\rho \sigma \mu \nu}- \frac{1}{2}(R^{\rho \nu} g^{\sigma \mu}-R^{\rho \mu} g^{\sigma \nu}+ R^{\sigma \mu} g^{\rho \nu}-R^{\sigma \nu} g^{\rho \mu})+  \frac{1}{6} R ( g^{\rho \mu} g^{\sigma \nu} - g^{\rho \nu} g^{\sigma \mu}) \;
\end{equation}
and make an extensive use of the following identity
\begin{equation}
{\tilde{F}_\rho}^{\ \mu} \tilde{G}^{\nu \rho}= \frac{1}{2} F_{\rho \sigma} G^{\rho \sigma} g^{\mu \nu} + F_\rho^{\ \nu} G^{\mu \rho} \; ,
\end{equation}
with the convention
\begin{equation}
\tilde{G}^{\mu \nu}= \frac{1}{2} \epsilon^{\mu \nu \rho \sigma} G_{\rho \sigma} \; .
\end{equation}

\providecommand{\href}[2]{#2}\begingroup\raggedright\endgroup

\end{document}